\documentclass[11pt]{article} 
\usepackage[utf8]{inputenc}
\usepackage{multirow}
\usepackage{amsfonts}
\usepackage{amsmath}
\usepackage{amssymb}
\usepackage{amsthm}
\usepackage{mathtools}
\usepackage{dsdshorthand}
\usepackage{tikz}
\usetikzlibrary{snakes}
\usetikzlibrary{decorations}
\usetikzlibrary{trees}
\usetikzlibrary{decorations.pathmorphing}
\usetikzlibrary{decorations.markings}
\usetikzlibrary{external}
\usetikzlibrary{intersections}
\usetikzlibrary{shapes,arrows}
\usetikzlibrary{arrows.meta}
\usetikzlibrary{calc}
\usetikzlibrary{shapes.misc}
\usetikzlibrary{decorations.text}
\usetikzlibrary{backgrounds}
\usetikzlibrary{fadings}
\usetikzlibrary{tikzmark,calc,arrows,shapes,decorations.pathreplacing}
\tikzset{
        cross/.style={cross out, draw=black, minimum size=2*(#1-\pgflinewidth), inner sep=0pt, outer sep=0pt},
	branchCut/.style={postaction={decorate},
		snake=zigzag,
		decoration = {snake=zigzag,segment length = 2mm, amplitude = 2mm}	
    }}
\usepackage{caption}
\PassOptionsToPackage{dvipsnames}{xcolor}
\usepackage{xcolor}
    \definecolor{darkgreen}{rgb}{0,0.5,0}
    \definecolor{darkblue}{rgb}{0,0,0.6}
    \definecolor{purple}{rgb}{0.4,.2,0.7}
 \usepackage[margin = 2.5cm]{geometry}
\usepackage{graphicx}
\usepackage[hyperfootnotes = false, colorlinks = true, linkcolor = darkblue, citecolor = purple]{hyperref}
\usepackage{subcaption}
\usepackage{ytableau}
\renewcommand{\v}{\vspace{3mm}}

\usepackage{comment}

\newcommand{\ee}{\end{equation}}
\newcommand{\bea}{\begin{eqnarray}}
\newcommand{\eea}{\end{eqnarray}}
\def\la{\label}
\def\nref#1{(\ref{#1})}
\def\half{{1 \over 2 }}

\def\fft#1#2{{\frac{#1}{#2}}}

\begin{document}

\thispagestyle{empty}
\begin{center}
    ~\vspace{5mm}

  \vskip 2cm 
  
   {\LARGE \bf 
      The no boundary density matrix 
   }

    \vspace{0.5in}

 Victor Ivo$^1$, Yue-Zhou Li$^1$, and Juan Maldacena$^2$
  
    \vspace{0.5in}

  $^1$
{\it  Jadwin Hall, Princeton University,  Princeton, NJ 08540, USA }
   \\
   ~
   \\
  $2$
{\it   Institute for Advanced Study,  Princeton, NJ 08540, USA }


\end{center}

\vspace{0.5in}

\begin{abstract}
 
We discuss a no-boundary proposal for  a subregion of the universe.  In the classical approximation, this density matrix involves finding a specific classical solution of the equations of motion with no boundary. Beyond the usual no boundary condition at early times, we also have another no boundary condition in the region we trace out. We can find the prescription by starting from the usual Hartle-Hawking  proposal for the wavefunction on a full slice and  tracing out the unobserved region in the classical approximation. 
We discuss some specific subregions and compute the corresponding solutions. These geometries lead to phenomenologically unacceptable probabilities, as expected.  

We also discuss how the usual Coleman de Luccia bubble solutions can be interpreted as a possible no boundary contribution to the density matrix of the universe. These geometries lead to local (but not global) maxima of the probability that are phenomenologically acceptable. 
     
 \end{abstract}
 
\vspace{1in}

\pagebreak

\setcounter{tocdepth}{3}
{\hypersetup{linkcolor=black}\tableofcontents}

\section{Introduction and motivation }

\begin{figure}[h]
\centering \hspace{0mm}\def\svgwidth{130mm}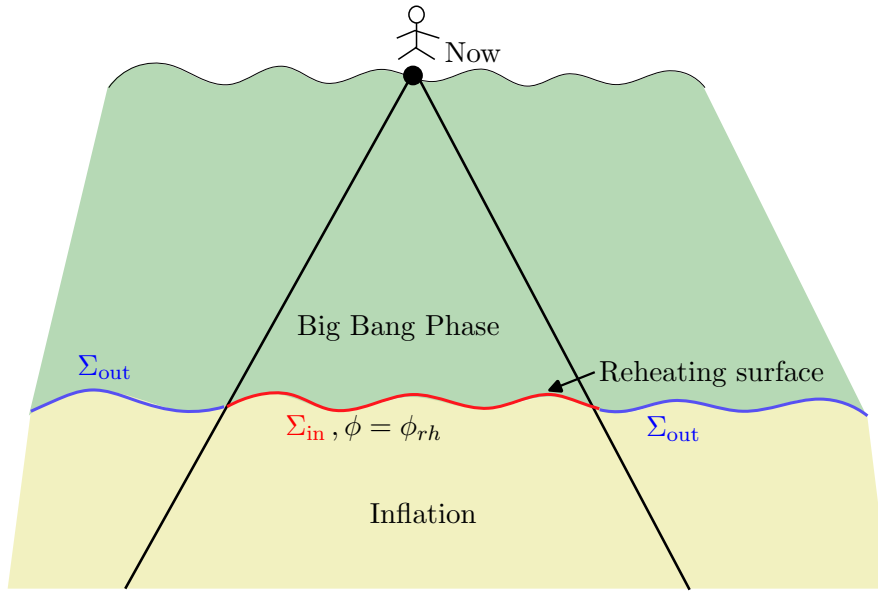
\caption { Sketch of a portion of the Penrose diagram for a cosmological spacetime highlighting different regions. We now receive signals only from a portion of the universe. For the purposes of this paper, we will think of the inflationary region of the universe as being quantum (light green) and the later one as a ``measurement apparatus" (dark green) that measures a region of the early universe, say a region of the reheating surface that sits at $\phi=\phi_{rh}$ (red). }
 \label{RegionObs}
 \end{figure}

The no boundary wavefunction of the universe is a theoretically well motivated proposal \cite{Hartle:1983ai}. 
In the usual formulation, it assumes we can make observations on a whole spatial slice, $\Sigma_3$.
However, we, as observers, only look at a part of a (probably) bigger universe, see figure \nref{RegionObs}. For this reason, we are most interested in making predictions for the portion of the spatial slice that we actually observe. 

\begin{figure}[h!]
    \centering
    \begin{subfigure}[b]{0.48\textwidth}
        \centering
       \hspace{0mm}\def\svgwidth{155mm}
\begingroup%
  \makeatletter%
  \providecommand\color[2][]{%
    \errmessage{(Inkscape) Color is used for the text in Inkscape, but the package 'color.sty' is not loaded}%
    \renewcommand\color[2][]{}%
  }%
  \providecommand\transparent[1]{%
    \errmessage{(Inkscape) Transparency is used (non-zero) for the text in Inkscape, but the package 'transparent.sty' is not loaded}%
    \renewcommand\transparent[1]{}%
  }%
  \providecommand\rotatebox[2]{#2}%
  \newcommand*\fsize{\dimexpr\f@size pt\relax}%
  \newcommand*\lineheight[1]{\fontsize{\fsize}{#1\fsize}\selectfont}%
  \ifx\svgwidth\undefined%
    \setlength{\unitlength}{366.30130798bp}%
    \ifx\svgscale\undefined%
      \relax%
    \else%
      \setlength{\unitlength}{\unitlength * \real{\svgscale}}%
    \fi%
  \else%
    \setlength{\unitlength}{\svgwidth}%
  \fi%
  \global\let\svgwidth\undefined%
  \global\let\svgscale\undefined%
  \makeatother%
  \begin{picture}(1,0.202379)%
    \lineheight{1}%
    \setlength\tabcolsep{0pt}%
    \put(0,0){\includegraphics[width=\unitlength,page=1]{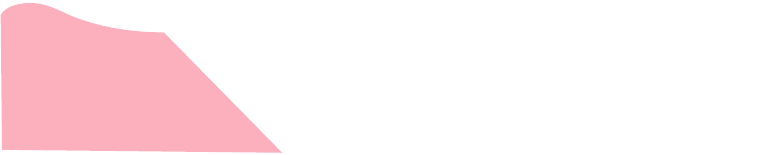}}%
    \put(0.11339172,0.18500319){\color[rgb]{0,0,0}\makebox(0,0)[lt]{\lineheight{1.25}\smash{\begin{tabular}[t]{l}$\Sigma_{\rm in}^-$\end{tabular}}}}%
    \put(0,0){\includegraphics[width=\unitlength,page=2]{traceout_data1.pdf}}%
    \put(0.27786113,0.17293927){\color[rgb]{0,0,0}\makebox(0,0)[lt]{\lineheight{1.25}\smash{\begin{tabular}[t]{l}${\color{blue} \Sigma_{\rm out}}$\end{tabular}}}}%
    \put(0.38453413,0.14970838){\color[rgb]{0,0,0}\makebox(0,0)[lt]{\lineheight{1.25}\smash{\begin{tabular}[t]{l}${\color{blue} \phi^+=\phi^-}$\end{tabular}}}}%
    \put(0.38463662,0.11488046){\color[rgb]{0,0,0}\makebox(0,0)[lt]{\lineheight{1.25}\smash{\begin{tabular}[t]{l}${\color{blue} \dot{\phi}^+=\dot{\phi}^-}$\end{tabular}}}}%
    \put(0,0){\includegraphics[width=\unitlength,page=3]{traceout_data1.pdf}}%
    \put(0.06799663,0.1179701){\color[rgb]{0,0,0}\makebox(0,0)[lt]{\lineheight{1.25}\smash{\begin{tabular}[t]{l}$\Sigma_{\rm in}^+$\end{tabular}}}}%
    \put(0,0){\includegraphics[width=\unitlength,page=4]{traceout_data1.pdf}}%
  \end{picture}%
\endgroup%

        \caption{}
    \end{subfigure}
    \hfill
    \begin{subfigure}[b]{0.48\textwidth}
        \centering
       \hspace{0mm}\def\svgwidth{140mm}
\begingroup%
  \makeatletter%
  \providecommand\color[2][]{%
    \errmessage{(Inkscape) Color is used for the text in Inkscape, but the package 'color.sty' is not loaded}%
    \renewcommand\color[2][]{}%
  }%
  \providecommand\transparent[1]{%
    \errmessage{(Inkscape) Transparency is used (non-zero) for the text in Inkscape, but the package 'transparent.sty' is not loaded}%
    \renewcommand\transparent[1]{}%
  }%
  \providecommand\rotatebox[2]{#2}%
  \newcommand*\fsize{\dimexpr\f@size pt\relax}%
  \newcommand*\lineheight[1]{\fontsize{\fsize}{#1\fsize}\selectfont}%
  \ifx\svgwidth\undefined%
    \setlength{\unitlength}{299.18760285bp}%
    \ifx\svgscale\undefined%
      \relax%
    \else%
      \setlength{\unitlength}{\unitlength * \real{\svgscale}}%
    \fi%
  \else%
    \setlength{\unitlength}{\svgwidth}%
  \fi%
  \global\let\svgwidth\undefined%
  \global\let\svgscale\undefined%
  \makeatother%
  \begin{picture}(1,0.22495732)%
    \lineheight{1}%
    \setlength\tabcolsep{0pt}%
    \put(0,0){\includegraphics[width=\unitlength,page=1]{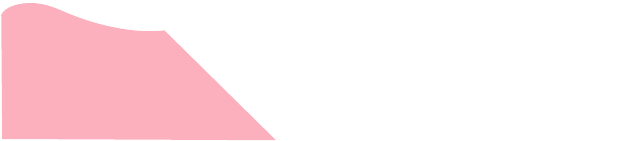}}%
    \put(0.13882772,0.20368377){\color[rgb]{0,0,0}\makebox(0,0)[lt]{\lineheight{1.25}\smash{\begin{tabular}[t]{l}$\Sigma_{\rm in}^-$\end{tabular}}}}%
    \put(0,0){\includegraphics[width=\unitlength,page=2]{traceout_data2.pdf}}%
    \put(0.08324962,0.12161383){\color[rgb]{0,0,0}\makebox(0,0)[lt]{\lineheight{1.25}\smash{\begin{tabular}[t]{l}$\Sigma_{\rm in}^+$\end{tabular}}}}%
    \put(0,0){\includegraphics[width=\unitlength,page=3]{traceout_data2.pdf}}%
    \put(0.36178903,0.16457895){\color[rgb]{0,0,0}\makebox(0,0)[lt]{\lineheight{1.25}\smash{\begin{tabular}[t]{l}{\color{blue} Trace-out} \\{\color{blue} boundary}\\{\color{blue} conditions}\end{tabular}}}}%
    \put(0.4081456,0.05534407){\color[rgb]{0,0,0}\makebox(0,0)[lt]{\lineheight{1.25}\smash{\begin{tabular}[t]{l}${\color{blue} \Sigma_{\rm out}}$\end{tabular}}}}%
    \put(0,0){\includegraphics[width=\unitlength,page=4]{traceout_data2.pdf}}%
  \end{picture}%
\endgroup%

        \caption{}
    \end{subfigure} 
    \caption{ (a) Given an observable region $\Sigma^{\pm}_{\rm in } $, we add a region $\Sigma_{\rm out} $  to complete each sheet   into full spatial slices $\Sigma_3^\pm $. We identify the fields and their derivatives on $\Sigma_{\rm out}$. We call these ``trace out'' boundary conditions.   There is significant freedom on the placement of $\Sigma_{\rm out}$. It can be taken all the way down to the past light cone of $\Sigma_{\rm in}^{\pm}$, as in (b).}
       \label{SigmaInOut}
\end{figure}

In this paper, we provide a simple extension of the proposal for the case that we observe only a finite  region of the universe, and we trace out over the rest of the universe. For this reason, it is naturally a proposal for the density matrix for the observable region of the universe. 
The process that traces out the unobserved region of the universe imposes an effective boundary condition that implies that we can forget about the unobserved region. In a sense, we can think of these boundary conditions as another ``no-boundary'', see figure \ref{SigmaInOut}. 
We will call these ``trace-out'' boundary conditions. They are the same type of boundary conditions one puts in the future when we perform computations in real time using 
a  Schwinger-Keldysh contour.  At early times, we impose the usual Hartle-Hawking like boundary conditions which say that the geometry is allowed to make an excursion into imaginary time in the direction that suppresses high energy contributions.  The  density matrix of an observable region of the universe was discussed in  \cite{Chandrasekaran:2022cip,Witten:2023xze,Chen:2024rpx,Kudler-Flam:2024psh} from a closely related perspective and we will attempt to discuss later the connection with the discussion here. 

The implications of the no boundary wavefunction for an observer who only does local observations was previously discussed in \cite{Hartle:2010dq,Hartle:2016tpo}, and our discussion is conceptually similar. We try state in a clearer way how the density matrix is to be computed from a classical solution.  Other aspects of the density matrices and the no boundary proposal were discussed in several other papers, starting with \cite{Page:1986vw,Hawking:1986vj} and most recently in  \cite{Fumagalli:2024msi}. 

The final geometry has a single physical boundary at the two copies of the region where we make observations, one copy for the bra and one for the ket. These two copies are connected at the boundary of the observable region, which has codimension two in the full spacetime. The interior spacetime has no other physical boundary and it can be viewed as a complex spacetime that everywhere obeys the Kontsevic-Segal-Witten criterion \cite{Kontsevich:2021dmb,Witten:2021nzp}. For the simple cases we will consider in this paper this spacetime has the topology of a sphere with a slit. 

Let us explain the general approach and the methods we will use. 
We view the no boundary proposal in the context of semiclassical gravity, where we first start with a classical geometry and then consider quantum corrections using gravity as an effective field theory. This appears to be a well defined procedure in this context. More precisely, it is well defined to the extent that the effective field theory approach is well defined. In other words, we say that the wavefunctional of the universe for a given field configuration $\Phi(\vec x)$ is 
\be 
\Psi[\Phi(\vec x ) ] \propto  \exp\left( i S[\Phi(\vec x,\tau)]  \right) ~,~~~~~~~~~ \Phi(\vec x ) \sim ( \Sigma_3 , \phi(\vec x) ) \la{NBWF}
\ee 
where $\Phi(\vec x) $ denotes the 3-geometry $\Sigma_3$ and other fields on this three geometry, such as a scalar field $\phi$. Similarly,  $\Phi(\vec x,\tau)$ is a full four dimensional spacetime with boundary set by $\Phi(\vec x)$. Here $\vec x$ denotes a point in $\Sigma_3$. The four dimensional spacetime is complex and has no boundary in the past. The deformation into the complex domain obeys a certain condition that we specify later, designed to dampen the fluctuations \cite{Hartle:1983ai}.  
In principle, we could consider quantum corrections around this classical geometry and systematically improve the right hand side \nref{NBWF} within the context of gravity as an effective field theory.
 
As it is well known, the no-boundary proposal appears to give results in contradiction with observations, for a review see \cite{Maldacena:2024uhs}. This might seem a good reason to abandon this proposal. However, the proposal seems so well motivated theoretically and so closely related to ideas that work very well in other contexts, such as black holes or anti-de-Sitter space \cite{Kapustin:2009kz,Drukker:2010nc}, that it seems worth exploring it further, with the hope that some day we will understand how to resolve the apparent  contradiction with observations. 

\begin{figure}[h!]
    \centering   
    \begin{subfigure}[b]{0.48\textwidth}
        \centering
       \hspace{0mm}\def\svgwidth{80mm}
\begingroup%
  \makeatletter%
  \providecommand\color[2][]{%
    \errmessage{(Inkscape) Color is used for the text in Inkscape, but the package 'color.sty' is not loaded}%
    \renewcommand\color[2][]{}%
  }%
  \providecommand\transparent[1]{%
    \errmessage{(Inkscape) Transparency is used (non-zero) for the text in Inkscape, but the package 'transparent.sty' is not loaded}%
    \renewcommand\transparent[1]{}%
  }%
  \providecommand\rotatebox[2]{#2}%
  \newcommand*\fsize{\dimexpr\f@size pt\relax}%
  \newcommand*\lineheight[1]{\fontsize{\fsize}{#1\fsize}\selectfont}%
  \ifx\svgwidth\undefined%
    \setlength{\unitlength}{614.21359806bp}%
    \ifx\svgscale\undefined%
      \relax%
    \else%
      \setlength{\unitlength}{\unitlength * \real{\svgscale}}%
    \fi%
  \else%
    \setlength{\unitlength}{\svgwidth}%
  \fi%
  \global\let\svgwidth\undefined%
  \global\let\svgscale\undefined%
  \makeatother%
  \begin{picture}(1,0.72959734)%
    \lineheight{1}%
    \setlength\tabcolsep{0pt}%
    \put(0,0){\includegraphics[width=\unitlength,page=1]{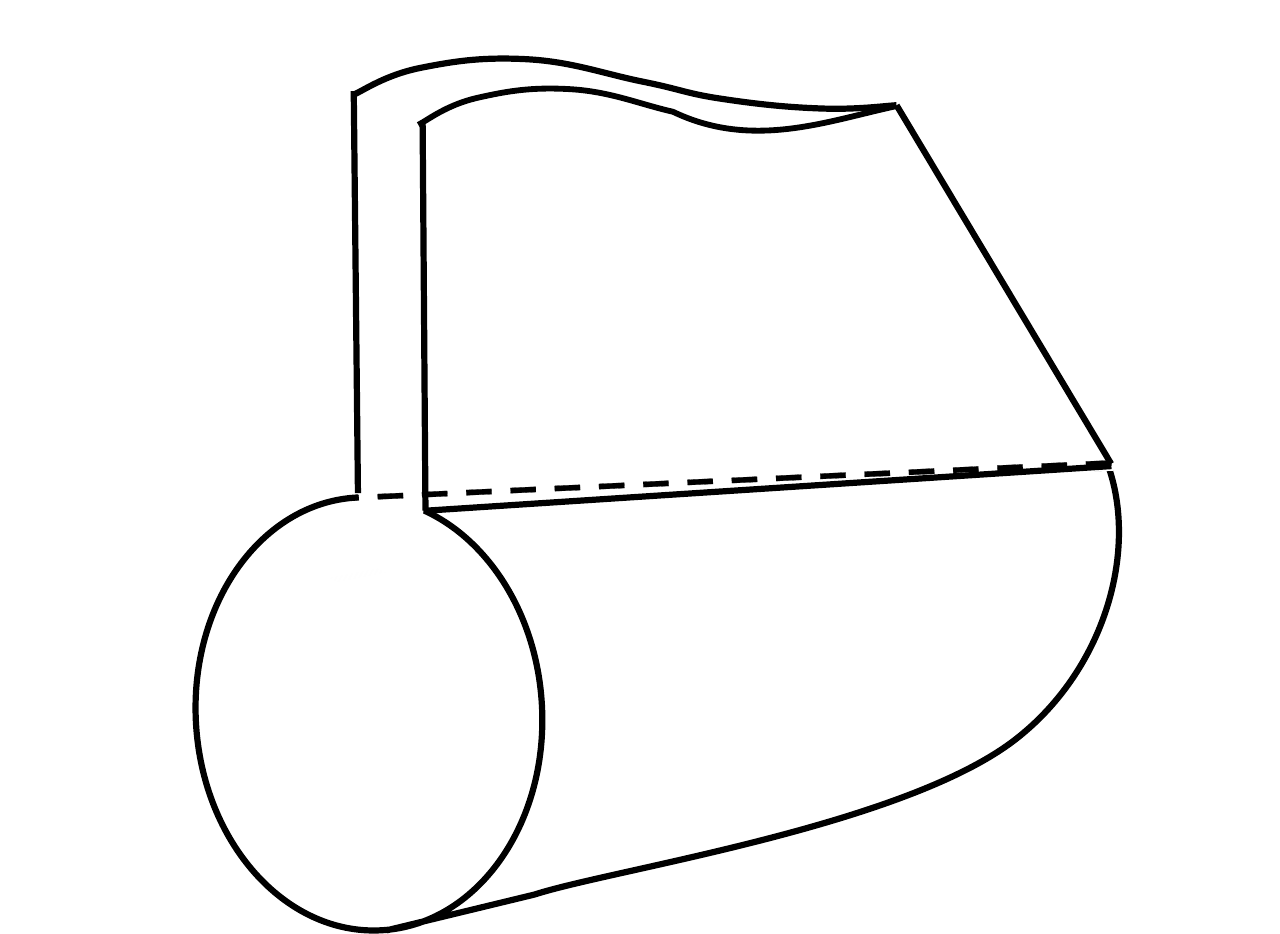}}%
    \put(0.46238543,0.58709248){\color[rgb]{0,0,0}\makebox(0,0)[lt]{\lineheight{1.25}\smash{\begin{tabular}[t]{l}$\Sigma_{\rm in}^+$\end{tabular}}}}%
    \put(0.41970546,0.70521522){\color[rgb]{0,0,0}\makebox(0,0)[lt]{\lineheight{1.25}\smash{\begin{tabular}[t]{l}$\Sigma_{\rm in}^-$\end{tabular}}}}%
    \put(0.36389802,0.18080268){\color[rgb]{0,0,0}\makebox(0,0)[lt]{\lineheight{1.25}\smash{\begin{tabular}[t]{l}$\gamma$\end{tabular}}}}%
    \put(0.19565028,0.49815759){\color[rgb]{0,0,0}\makebox(0,0)[lt]{\lineheight{1.25}\smash{\begin{tabular}[t]{l}$\uparrow \tau$\end{tabular}}}}%
    \put(0.03229023,0.30821231){\color[rgb]{0,0,0}\makebox(0,0)[lt]{\lineheight{1.25}\smash{\begin{tabular}[t]{l}$r_{S^2}=0$\end{tabular}}}}%
    \put(0.84794125,0.569071){\color[rgb]{0,0,0}\makebox(0,0)[lt]{\lineheight{1.25}\smash{\begin{tabular}[t]{l}Trace-out \end{tabular}}}}%
    \put(0,0){\includegraphics[width=\unitlength,page=2]{dm_geo1.pdf}}%
  \end{picture}%
\endgroup%

        \caption{}
    \end{subfigure}
    \hfill
    \begin{subfigure}[b]{0.48\textwidth}
        \centering
       \hspace{0mm}\def\svgwidth{80mm}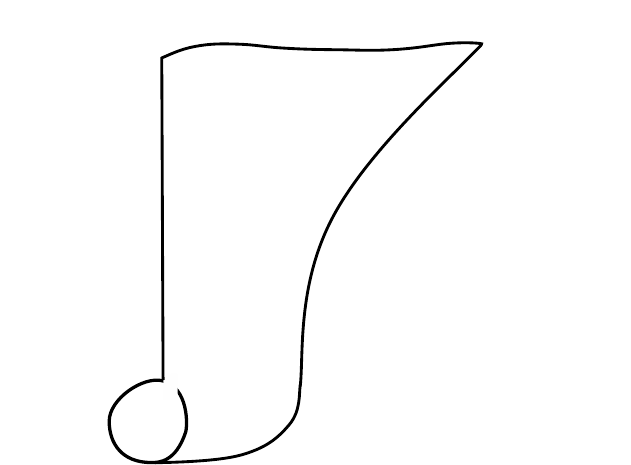
        \caption{}
    \end{subfigure} 
    \caption{ Sketch of the geometries that contribute to the density matrix. (a) Penrose-like diagram. (b) Diagram indicating the proper spatial size of the region.  $\tau$ indicates Lorentzian time and $\gamma$ is Euclidean time.   The left edge of the diagrams is where the radius of the $S^2 $ shrinks to zero is a smooth way. 
   }
 \label{Geometries}
 \end{figure}

In any theory, if we know the full wavefunction, we can easily compute the density matrix in a subregion. The no boundary proposal gives us the wavefunction, so one can simply trace out the unobserved region and get the density matrix of a subregion. 
In our context, this means dividing the boundary slice $\Sigma = \Sigma_{\rm in } \cup \Sigma_{\rm out}$. Here $\Sigma_{\rm in} $ is the region where we are making observations and $\Sigma_{\rm out}$ is the region outside where we are not making any observations. Then the boundary data splits into $\Phi_{\rm in}$ and $\Phi_{\rm out}$ which are the three geometries and fields in the $\Sigma_{\rm in} $ and $\Sigma_{\rm out} $ regions, see figure \ref{SigmaInOut}.  More precisely, we have two $\Sigma_{\rm in}^\pm $ regions corresponding to the bra and the ket entries of the density matrix or  two $\Phi^\pm_{\rm in}$ sets of field configurations. 
Starting with the density matrix of the full  state, given by $\Psi^* \Psi$, we can compute the density matrix of the observed region as 
 \be \la{TrOut}
 \rho[ \Phi^-_{\rm in}(\vec x) , \Phi^+_{\rm in}(\vec x ) ] = \int {\cal D } \Phi_{\rm out} 
 \Psi^*[ \Phi^-_{\rm in},\Phi_{\rm out}]\Psi[ \Phi^+_{\rm in},\Phi_{\rm out}] \propto  \Psi^*[ \Phi^-_{\rm in},\Phi^s_{\rm out}]\Psi[ \Phi^+_{\rm in},\Phi^s_{\rm out}]\,,
 \ee 
 where $\Phi^\pm_{\rm in}$ is the data on the bra and ket sides of the density matrix and we are tracing out over the data $\Phi_{\rm out}(\vec x)$ in the unobserved region. The main point is that, in the classical approximation,  this   functional integral can be evaluated via a saddle point approximation and therefore the final answer corresponds to evaluating the original density matrix at a particular field configuration,  $\Phi^s_{\rm out} $,  in the outside region. $\Phi^s_{\rm out}$  is a solution of some classical equations whose boundary data involves $\Phi^\pm_{\rm in}$. These classical equations have the usual no boundary conditions in the past.  More precisely, we can think of the full manifold ${\cal M}$ as the union of two manifolds ${\cal M } = {\cal M}^+ \cup {\cal M}^-$, which lead to four dimensional field configurations $\Phi^\pm(\vec x, \tau ) $ defined on ${\cal M}^\pm$ respectively. In the unobservable region we should impose the ``trace-out'' boundary conditions 
 \be \la{TrOutEq}
 \Phi_{\rm out}(\vec x)  \equiv  \Phi^+( \vec x,\tau_r ) = \Phi^-(  \vec x,\tau_r) ~,~~~~~~~~~\Pi^+( \vec x,\tau_r) = \Pi^-(  \vec x,\tau_r)
 ~,~~~~~~~~{\rm for }~~~~~~~ \vec x \in \Sigma_{\rm out} \,,
 \ee 
 for $\vec x$ in the outside region and  $\tau_r$ denotes the slice where $\Sigma $ lies in the full 4-manifold.  Here $\Pi^\pm$ are the canonically conjugate momenta. For a scalar field  they involve $\partial_\tau \phi^\pm(\vec x,\tau_r)$.  For the 3-metric  they involve the extrinsic curvature. 
The first equation arises because in \nref{TrOut} we are setting the entries of the original density matrix and the second equation in \nref{TrOutEq} arises from the saddle point condition 
\be \la{TrOutAct}
0 = i { \delta S^+ \over \delta \Phi^+}- i { \delta S^- \over \delta \Phi^-} ~~~~\to ~~~~~\Pi^+( \vec x,\tau_r) = \Pi^-(  \vec x,\tau_r)\,.
\ee 
Notice that,  after imposing \nref{TrOutEq} on $\Sigma_{\rm out}$, the equations of motion imply that $\Phi^+(\vec x, \tau) = \Phi^-(\vec x, \tau)$ in the whole domain of dependence of $\Sigma_{\rm out}$. This is the whole spacetime region that is determined by initial data on $\Sigma_{\rm out}$. 
This means that we can push the slice where we impose \nref{TrOutEq} all the way down to the past light-cone of the $\Sigma^\pm_{\rm in}$ surfaces and we can talk about a geometry of the kind represented in figure \ref{Geometries}. Note that $\Phi_{\rm out}$ is not part of the boundary data, it is simply that actual value of $\Phi^\pm$ once we find the solution. 
We can interpret \nref{TrOutEq} as a type of no-boundary condition, since it is identifying the solutions on the plus and minus parts of the contour and, furthermore, it is making the whole domain of dependence of $\Sigma_{\rm out}$ irrelevant. 

 Note that for the same reason that \nref{NBWF} is not properly normalized, the ``density matrix'' 
 \nref{TrOut} is not properly normalized either.  In fact, in situations where the scalar potential becomes zero or negative it is not even normalizable. We will not worry about this, imagining that we will apply selection criteria (or projection operators) that will remove this problem.

 This type of geometry is familiar from computations of correlators in thermal black hole backgrounds. In that case, one can use a Schwinger Keldysh contour which only includes the region outside the horizon, as well as a Euclidean geometry preparing the thermal state, see e.g. \cite{Son:2002sd,Skenderis:2008dg,Jana:2020vyx}.  
 Here we simply apply a similar idea to cosmology, see figure \ref{Geometries}.   

 In this paper,  we will compute such semiclassical solutions in a couple of simple contexts. 
 
 The first is closely related to the original context in which the no boundary proposal was discussed \cite{Hartle:1983ai,Halliwell:1984eu}, see also \cite{Hartle:2007gi,Hartle:2008ng,Hartle:2010vi}. We study an inflationary situation and consider a subregion $\Sigma_{\rm in}$ which sits at some constant value of the scalar field, say $\phi= \phi_{rh}$, and has an arbitrary three-geometry. We can then compute the density matrix for various values of this three-geometry. In fact, we will restrict to the scalar mode in the geometry, or the overall scale factor, so our three geometries will be conformally flat. The portion of the geometry under consideration could have a scalar curvature with any sign. 
 We study this as a purely mathematical exercise in order to understand the type of classical solutions that are involved in the computation of the density matrix. In fact, 
 the density matrix computed in this context displays a large ``probability pressure'' for the universe to become small, of the order of Hubble size at the minimum value $\phi_r$. This is a well known problem with the no-boundary proposal \cite{Vilenkin:1987kf}, as reviewed in \cite{Maldacena:2024uhs}\footnote{ 
 Other authors have stated the problem in other related forms.  }.

 In fact, the problem with the Hartle Hawking wavefunction in the inflationary context is related to   the fact that the bulk solutions are complex. Let us explain why. Consider the diagonal components of the density matrix 
 \be 
 \log \rho[ \Phi(\vec x), \Phi(\vec x)] \sim \log 
  {\Big [ }  \Psi^*[\Phi(\vec x ) ]\Psi [\Phi(\vec x)]   {\Big ] } 
 \sim  i S^+[\Phi^+(\vec x, \tau ) ] - i S^-[\Phi^-(\vec x, \tau )]\,.
\la{RhoNB} \ee 
 Taking the derivative with respect to $\Phi$ we get 
 \be \la{MomMax}
   { \delta  \log \rho[ \Phi(\vec x), \Phi(\vec x)] \over \delta \Phi(\vec x ) }  \approx      i  { \delta S^+ \over \delta \Phi(\vec x ) }  - i { \delta S^- \over \delta \Phi(\vec x ) }    \sim  i   \Pi^+(\vec x) - i \Pi^-(\vec x) = - 2 \, {\rm Im}\left(\Pi^+(\vec x)\right)\,.
  \ee 
For real boundary values, we expect that the two solutions are complex conjugates of each other 
\be 
\Phi^+(\vec x , \tau ) = \left( \Phi^-(\vec x , \tau ) \right)^* \,.
\ee 
If they are complex, this means that the conjugate momenta $\Pi^\pm(\vec x)$ are not equal. Then, \nref{MomMax} means that the probability is not stationary with respect to variations of $\Phi(\vec x)$. In other words, we are not evaluating the density matrix or the wavefunction at a maximum of the probability. 

   Since we are considering classical solutions, the probability pressure for going to a different value of $\Phi(\vec x) $ is very large. This will be the case for all configurations discussed in our first set of examples in section \ref{HHDM}. Our purpose for exploring this problem is just to formulate and learn how to solve the classical equations that compute the density matrix, but we make no claim that these geometries are phenomenologically important.

 On the other hand, if we had a real solution, namely a solution with $\Phi^+(\vec x , \tau) = \Phi^-(\vec x , \tau) $,     then this means that we evaluating the probability at a stationary point, a point where the first derivative vanishes (in the classical approximation).   Whether that is a maximum or a minimum, would need to be worked out by changing $\Phi$ away from the value where the solution is real. These are the type of geometries that are more likely to be relevant for phenomenological reasons.

Our second example involves geometries that are an analytic continuation of the 
Euclidean solutions  studied by Coleman and de Luccia \cite{Coleman:1980aw},  which are normally interpreted in terms of bubble nucleation and have been discussed in the context of open inflation starting from \cite{Bucher:1994gb}.  Here we will interpret them as geometries contributing to the density matrix of the universe. These two interpretations are closely related.  
In this case, there is a real solution which represents a local maximum in the probability. 
These geometries are known to give predictions that are compatible with observations in our universe, if we have sufficient ordinary slow roll inflation after the nucleation event,  and they were discussed in the context of open inflation \cite{Bucher:1994gb,Sasaki:1994yt,Linde:1995rv,Yamamoto:1996qq,Tanaka:1998mp,Garriga:1998he,Yamauchi:2011qq}.  
They are usually viewed in the context of eternal inflation,  and as an alternative to the no boundary proposal. However,  they can also be interpreted as a contribution to the no boundary density matrix of the universe.

For all geometries we consider in this paper, the geometry that prepares the density matrix for a subregion of the universe and the one that computes the wavefunction on the full slice are intimately related, due to \nref{TrOut}. However, the fact that the geometry that prepares the density matrix can be restricted to smaller subregion of the four dimensional manifold seems to give a more economical description. Moreover, it is a description that that makes it clear that IR effects from the unobserved region are irrelevant. This might turn out to be a significant advantage in the eternal inflation context.

 All our discussions are classical, and we leave the computation of one loop determinants for the future. However, for gaussian fields, this classical approximation does indeed contain the full wavefunctional and we expect a close relation to the recent discussions in \cite{Chen:2024rpx,Kudler-Flam:2024psh}.

\section{Density matrix for a massless scalar field in de-Sitter}

\label{masslesssc}

In this section, as a warm up problem,  we discuss the density matrix for a massless scalar field in a rigid, non-gravitating de-Sitter space. It is important to note that the computation of density matrices for subregions in quantum field theory is a well-developed subject, and the discussion here is just a special case, see \cite{Casini:2022rlv} for a review. 
Our objective in this section is to show how the density matrix can be derived using the semiclassical approximation of the path integral. 
Since we are considering a free field theory, the wavefunction is Gaussian so that the classical solution gives the exact answer, up to a field independendent prefactor that should arise from the one loop correction. 

We set the radius of de Sitter to one ($H=1$)  and consider the global slicing
\be \la{dSSthr}
ds^2 = -d\tau^2 + \cosh^2 \tau d\Omega_3^2 \,,\quad ~~~~~ d\Omega_3^2 = d\theta^2 + \sin^2 \theta d\Omega_2^2 \,.
\ee 

\begin{figure}[h]
    \begin{center}
    \begin{tikzpicture}[decoration={markings, mark= at position 0.52 with {\arrow{stealth}}}]
    \node (a) at (-2,2) {$\tau$};
	\draw (a.north west) -- (a.south west) -- (a.south east);
    \draw[->] (-3.5, 0) -- (3.5, 0) node[right] {$\tau$};
   \draw[thick,postaction={decorate}] (0,2) -- (0,0.1)  node at (0.7,1) {$\tau=i\gamma$};
   \draw[] node at (0.4,2) {$i\fft{\pi}{2}$};
   \draw[] node at (0.5,-2) {$-i\fft{\pi}{2}$};
    \draw[thick,postaction={decorate}] (0,0.1) -- (2,0.1)
node at (2.6,0.26) {$\varphi_b^+(\tau_r)$}; 
     \draw[thick,postaction={decorate}] (2,-0.1) -- (0,-0.1) ;
     \draw[thick,postaction={decorate}] (0,-0.1) -- (0,-2)
     node at (2.6,-0.24) {$\varphi_b^-(\tau_r)$};
\draw[thick, decoration={brace, amplitude=10pt}, decorate] (3.3, 2) -- (3.3, 0.15) node at (4.6,1.1) {$\text{ket}, ~\mathcal{M}^+$};
\draw[thick, decoration={brace, amplitude=10pt}, decorate] (3.3, -0.15) --(3.3, -2)  node at (4.6,-1.1) {$\text{bra}, ~\mathcal{M}^-$};
\end{tikzpicture}
    \end{center}
    \caption{Countour of integration for the time coordinate in the spacetime metric \nref{dSSthr}.  }
    \label{TimeContour}
\end{figure}
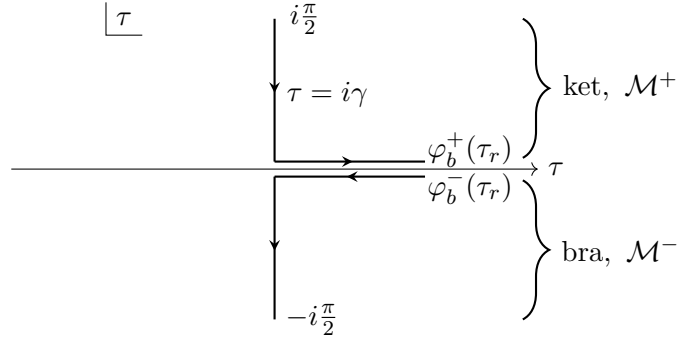

\begin{figure}[h]
\centering \hspace{0mm}\def\svgwidth{85mm}
\begingroup%
  \makeatletter%
  \providecommand\color[2][]{%
    \errmessage{(Inkscape) Color is used for the text in Inkscape, but the package 'color.sty' is not loaded}%
    \renewcommand\color[2][]{}%
  }%
  \providecommand\transparent[1]{%
    \errmessage{(Inkscape) Transparency is used (non-zero) for the text in Inkscape, but the package 'transparent.sty' is not loaded}%
    \renewcommand\transparent[1]{}%
  }%
  \providecommand\rotatebox[2]{#2}%
  \newcommand*\fsize{\dimexpr\f@size pt\relax}%
  \newcommand*\lineheight[1]{\fontsize{\fsize}{#1\fsize}\selectfont}%
  \ifx\svgwidth\undefined%
    \setlength{\unitlength}{264.33868889bp}%
    \ifx\svgscale\undefined%
      \relax%
    \else%
      \setlength{\unitlength}{\unitlength * \real{\svgscale}}%
    \fi%
  \else%
    \setlength{\unitlength}{\svgwidth}%
  \fi%
  \global\let\svgwidth\undefined%
  \global\let\svgscale\undefined%
  \makeatother%
  \begin{picture}(1,0.93109306)%
    \lineheight{1}%
    \setlength\tabcolsep{0pt}%
    \put(0,0){\includegraphics[width=\unitlength,page=1]{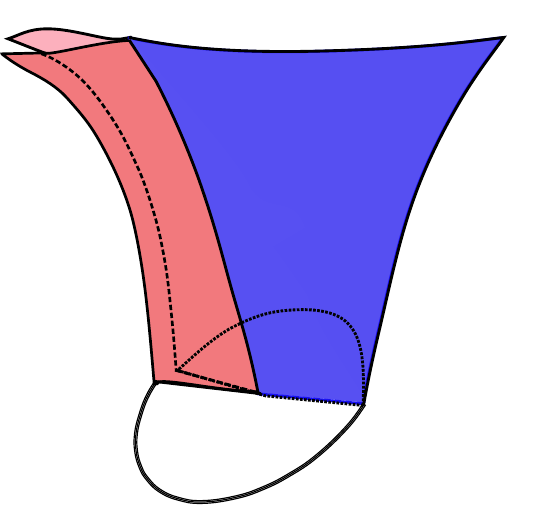}}%
    \put(0.18413946,0.79765385){\color[rgb]{0,0,0}\makebox(0,0)[lt]{\lineheight{1.25}\smash{\begin{tabular}[t]{l}$\Sigma_{\rm in}^+$\end{tabular}}}}%
    \put(0.03379519,0.9055957){\color[rgb]{0,0,0}\makebox(0,0)[lt]{\lineheight{1.25}\smash{\begin{tabular}[t]{l}$\Sigma_{\rm in}^-$\end{tabular}}}}%
    \put(0.41336994,0.86632057){\color[rgb]{0,0,0}\makebox(0,0)[lt]{\lineheight{1.25}\smash{\begin{tabular}[t]{l}$\Sigma_{\rm out}$\end{tabular}}}}%
    \put(0.44048795,0.58010141){\color[rgb]{0,0,0}\makebox(0,0)[lt]{\lineheight{1.25}\smash{\begin{tabular}[t]{l}Trace-out  $\mathcal{M}_{\rm out}$\end{tabular}}}}%
    \put(0,0){\includegraphics[width=\unitlength,page=2]{traceout.pdf}}%
    \put(0.0781021,0.45401532){\color[rgb]{0,0,0}\makebox(0,0)[lt]{\lineheight{1.25}\smash{\begin{tabular}[t]{l}${\rm dS}_4$\end{tabular}}}}%
    \put(0,0){\includegraphics[width=\unitlength,page=3]{traceout.pdf}}%
    \put(0.61422083,0.01698174){\color[rgb]{0,0,0}\makebox(0,0)[lt]{\lineheight{1.25}\smash{\begin{tabular}[t]{l}$S^4$\end{tabular}}}}%
  \end{picture}%
\endgroup%

\caption{ Sketch of the geometry for the computation of the density matrix for a subregion of de-Sitter space that lives on a late time slice. In blue, we see the region we trace out, and $\Sigma_{\rm out} $ could be any slice in that region. As we go to Euclidean time, we need to consider a Euclidean space which is a 4-sphere with a slit, where we attach the two lorentzian sheets.  }
 \label{deSitterSlit}
 \end{figure}

The action for the scalar field is  
\begin{equation}
\label{masslessscact}
I=\int_{\mathcal{M}}d^{4}x\sqrt{-g}\left[ - \frac{1}{2}(\nabla \varphi)^{2} \right] \,.
\end{equation}
The manifold ${\cal M } = {\cal M}^+ \cup {\cal M}^-$ is depicted in figure \ref{deSitterSlit}. 
${\cal M}^+$ involves the Lorentzian region $\tau>0$ together with the Euclidean region $\tau = i \gamma $, $\gamma \in [ 0, {\pi \over 2} ] $. ${\cal M}^-$ involves a second Lorentzian region $\tau>0$ together with the Euclidean region $\tau = i \gamma $, $\gamma \in [ - { \pi \over 2} , 0] $.
In the Lorentzian regions, the action $I$ in \nref{masslessscact} has a different overall sign for ${\cal M}^\pm$. 
This can be understood as originating from the opposite orientation of the Schwinger-Keldysh contour, see figure \ref{TimeContour}. Alternatively, we can say that we can parametrize the contour in terms of a real coordinate, say $\hat t$, and view the metric as being complex, see \cite{Kontsevich:2021dmb,Witten:2021nzp}.

The classical equations of motion from \nref{masslessscact} are
\begin{equation}
\label{eommasslesssc}
\nabla^{2}\varphi=0\,.   
\end{equation}
The boundary conditions are that the solution should be smooth at the $\tau =\pm i \pi/2$ endpoints of the contour in figure \ref{TimeContour}, and that the field has values $\varphi_b^\pm $ at time $\tau =\tau_r$. 
 
Once we have the solution, we can evaluate the action. After integrating by parts, and  using \nref{eommasslesssc}, we find 
\bea \la{AcFiFS}
\log \rho[\varphi_b^-,\varphi_b^+]  &= & \log \left[ \Psi^*[\varphi_b^-] \Psi[\varphi_b^+] \right] =  iI = \fft{1}{2} \int_{\partial \mathcal{M}}d^{3}x\sqrt{h} \,  \varphi\, i n^\mu \partial_{\mu} \varphi 
\cr &=&\fft{1}{2} \int_{\Omega_3} d\Omega_3  \, \cosh^3 \tau_r (  \varphi_b^+\, i \partial_\tau \varphi^+ - \varphi_b^-\, i \partial_\tau \varphi^- ) |_{\tau = \tau_r} 
\eea 
where $n^\mu$ is the normal vector to the spacelike surface on which we measure the observables, and $h$ is the induced metric on this surface. We have also indicated the answer for a global slice at constant $\tau$ in \eqref{dSSthr}.

\subsection{The wavefunction and density matrix on the whole slice}
\la{WfdSWh}

In this subsection, we review the more standard computation of the wavefunctional on the whole spatial slice \cite{Halliwell:1984eu}. 

We solve the wave equation \nref{eommasslesssc} by the separation of variables as 
\be \la{vphiExp}
\varphi = \sum_{\ell,m_1,m_2 } \hat c_{\ell,m_1,m_2} f_{\ell}(\tau) Y_{\ell,m_1,m_2}(\vec \Omega) \,,
\ee 
where these mode functions obey 
 \begin{equation} \la{eomSep}
\frac{1}{\cosh^{3}\tau}\partial_{\tau}(\cosh^{3}\tau \partial_{\tau}f_\ell)+\frac{\ell (\ell+2)}{\cosh^{2}\tau} f_\ell=0 \,,\quad \nabla^{2}_{S^{3}}Y_{\ell,m_1,m_2}=-\ell(\ell+2)Y_{\ell,m_1,m_2}\,,
\end{equation}
where $\nabla_{S_{3}}^{2}$ the Laplacian on the three sphere and the $Y_{\ell,m_1m_2} $ are the spherical harmonics. 

For computing the wavefunction (or its conjugate), 
it is necessary to find solutions $f_\ell$ that are regular at $\tau = i \pi/2$ (or $\tau = - i \pi/2$), see the contour in figure \ref{TimeContour}. 
The solutions are given by 
\begin{equation}
f_{\ell}^{\pm}(\tau)=\bigg(\frac{\ell+\cosh^{2}\tau \pm i \ell \sinh \tau}{\cosh^{2}\tau}\bigg)e^{\mp 2 i \ell \arctan e^{-\tau}}\,,\label{eq: f_l dS}
\end{equation}
which are   normalized so that $\lim_{\tau \to \infty} f_\ell =1$. The $+$ sign corresponds to the ket, used to compute $\Psi$, and the $-$ to the bra, used to compute $\Psi^*$.  

As a side comment,   we can consider the coordinates that are useful for the Penrose diagram of de-Sitter
\be \la{ConfCo}
ds^2 =  { 1 \over \sin^2 \eta }(- d\eta^2 + d \Omega_3^2 ) ~,~~~~~~~~~ ~~~~~~\cosh  \tau =   { 1 \over \sin(- \eta) }\,,
\ee 
where $\eta \in [ -\pi, 0]$ and the asymptotic future corresponds to $\eta \to 0^-$. Then, 
the wavefunctions in \nref{eq: f_l dS} become particularly simple\footnote{This simplicity is related  to the connection between this problem and the problem of a scalar field with a quartic action that is Weyl invariant, see section 2 of \cite{Maldacena:2011mk}. After a Wely transformation to $ R \times S^3$ this quartic action describes a set of harmonic oscillators. }

\begin{equation}  
f_{\ell}^{\pm}(\eta)=\bigg[\bigg(1+\frac{\ell}{2}\bigg)e^{\pm i \ell \eta}-\frac{\ell}{2}e^{\pm i (\ell+2)\eta}\bigg] = 
\left[  \cos \eta - \sin \eta \partial_\eta \right] e^{ \pm i ( \ell +1) \eta }\,.\label{eq: fell}
\end{equation}
Note that in these coordinates they have purely positive or purely negative frequencies in $\eta$. 
 
We can expand the boundary conditions as 
\be \la{VarPhivExp} 
\varphi_b^\pm = \sum_{\ell,m_1,m_2 } c^\pm_{\ell,m_1,m_2}   Y_{\ell,m_1,m_2}(\vec \Omega)\,.
\ee
and from now on we always take $\tau_{r}\gg 1$, while leaving the finite $\tau_r$ case to the appendix \ref{FiniteTrApp}. Equating $ \varphi^\pm (\tau_r, \vec \Omega) = \varphi_b^\pm(\vec \Omega)$, and using \nref{vphiExp}, we find that 
\be 
\hat c^\pm_{\ell,m_1,m_2}  = c^\pm_{\ell,m_1,m_2} \,,
\ee 
which determines the solution everywhere in spacetime. 
Then the wavefunction, for large $\tau_r$,  is 
\begin{align} 
\Psi[ \varphi_b^+] & 
\propto \exp( i I^+ ) = \exp\left( i \half \int d\Omega_3 \cosh^3 \tau_r \varphi_b^+ \partial_\tau \varphi^+ \Big|_{\tau=\tau_r}\right) \la{WaFu1} =\\
&= \exp\left( \sum_{\ell,m_1,m_2} c^+_{\ell,m_1,m_2} c^+_{\ell,-m_1,-m_2} 2\pi^2 \left[ - i { e^{\tau_r} \over 4}  \ell(\ell+2)  - \half \ell(\ell+1)(\ell+2) + \mathcal{O}(e^{ -\tau_r}) \right] \right)\,,
\label{WaFu}
\end{align}

We can then compute the density matrix in the full space 
\be \la{DensFull}
\rho[ \varphi^-_b, \varphi_b^+] \sim \exp( i I^+ - i I^- )  \,. 
\ee 
We see that the off diagonal terms contain a rapidly oscillating piece coming from the $e^{\tau_r}$ term in \nref{WaFu}. Such rapidly oscillating terms depend in a local way on the boundary conditions. In fact, they involve the Laplacian acting on the boundary values of the field.  On the other hand, the diagonal terms do not depend on the cutoff $\tau_r$. For the diagonal component, we set $\varphi^+_b = \varphi^-_b = \varphi_b$ and obtain 
\be \la{DenMaF}
\log \rho[ \varphi_b , \varphi_b ]= \log \left| \Psi[ \varphi_b]\right|^2 =  - 2 \pi^2 \sum_{\ell,m_1,m_2}   \ell(\ell+1)(\ell+2)   |c_{\ell,m_1,m_2}|^2  \,.
\ee 
(The finite $\tau_r$ version can be found in \eqref{ActFifintau}.)

\subsection{Computing the density matrix  of a subregion}

We now divide the total spatial $S^3$ slice into two subregions $\Sigma_{\rm in}$ and $\Sigma_{\rm out}$, $S^3 = \Sigma_{\rm in} \cup \Sigma_{\rm out}$. 
A straightforward method for computing the density matrix for a subregion is to start from \nref{DensFull}, set $\varphi^+_b = \varphi_b^- = \varphi_{b, {\rm out}}$ in $\Sigma_{\rm out}$ and integrate over this function. 
Using the saddle point approximation for this integral (which happens to be exact for a Gaussian integral), we get the following boundary conditions
\be \la{TrOutSF}
\varphi^+(\tau_r, \vec \Omega) = \varphi^-(\tau_r, \vec \Omega) ~,~~~~~~~~~\partial_\tau \varphi^+(\tau_r, \vec \Omega) = \partial_\tau \varphi^-(\tau_r, \vec \Omega) ~,~~~~~~~~{\rm for }  ~~~~~~ \vec \Omega \in \Sigma_{\rm out}\,.
\ee 
Note that the fields are not restricted here, but they are restricted to be equal in the $\pm $ sheets. So, we can say that the two sheets have been identified. In fact, we can run this identification along the backward lightcone of the region $\Sigma_{\rm in}$, so that, for the purposes of this computation, we have just a smaller region of the de-Sitter spacetime. 

Along $\Sigma_{\rm in}$ we have a true physical boundary and we can set arbitrary boundary conditions 
\be \la{InBCFF}
\varphi^{\pm}(\tau_r, \vec \Omega ) = \varphi_{b, {\rm in}}^\pm (\vec \Omega) 
 ~,~~~~~~~~{\rm for }  ~~~~~~ \vec \Omega \in \Sigma_{\rm in}\,.
\ee 

After setting all these boundary conditions \nref{InBCFF} \nref{TrOutSF}, together with the conditions that the field should be smooth in the Euclidean region, we get a single solution of the equations of motion. Then we can evaluate the final action using \nref{AcFiFS}. 
The only difference is the nature of the boundary conditions we used to obtain the classical solutions. In other words, now the $ c^\pm_{\ell, m_1,m_2} $ are determined in terms of \nref{InBCFF} and \nref{TrOutSF}, as we discuss in the next subsection. 
 
\subsection{Density matrix on a spherical ball}
\la{DMCap}

In order to give a more explicit example, we can consider the simple case where $\Sigma_{\rm in}$ is a spherical ball. Namely, we consider a region on the $S^3$ such that $ 0 \leq \theta \leq \theta_0$ which sits at some value $\tau_r \gg 1$.  

For simplicity, we will also evaluate the density matrix on profiles which are $SO(3) $ symmetric under rotations of the $S^2$ in \nref{dSSthr}. This implies that the functions only depend on $\theta$,  but not on the angles of $S^2$. This means that we should consider a particular SO(3) invariant combination of the spherical harmonics 
\begin{equation} \la{YHarm}
Y_{\ell}(\theta)\equiv\frac{\sin[(\ell+1)\theta]}{\sin \theta}  \propto \sum_{m} Y_{\ell,m,-m}\,,
\end{equation}
and the expansion of the fields is  
\begin{equation}
\label{homexp}
\varphi^{\pm}=\sum_{\ell}c_{\ell}^{\pm}Y_{\ell}(\theta)f_{\ell}^{\pm}\,.(\tau)\,, 
\end{equation}
Let us say that we fix a profile $\varphi_{b}(\theta)$ for $0\leq \theta\leq\theta_{0}$ in both bra and ket sides,  and for $\theta_{0}< \theta \leq \pi$ the bra and the ket are glued using trace-out boundary conditions.  
Since we are focusing on the diagonal components of the density matrix, we have that 
 $\varphi^{+}=\varphi^{-}$ in the entire $\tau=\tau_{r}$ slice, which implies that  $c_{\ell}^{+}=c_{\ell}^{-}$. Furthermore, for diagonal matrix elements we have that $\varphi^+ = (\varphi^-)^*$, so that the $c_\ell$ are real. 
  %
 For $\theta>\theta_{0}$ we must also impose that $\partial_{\tau}\varphi^{+}-\partial_{\tau}\varphi^{-}=0$. Using
\begin{equation}
i \cosh^{3}\tau_{r} \bigg(\frac{\partial_{\tau}f_{\ell}^{+}(\tau_r)}{f_{\ell}^{+}(\tau_r)}-\frac{\partial_{\tau}f_{\ell}^{-}(\tau_r)}{f_{\ell}^{-}(\tau_r)}\bigg)  =- 2   \ell(\ell+1)(\ell+2)\,.
\end{equation} 
we arrive at the final form of the equations that determine $c_\ell$ at $\tau_r\gg 1$
\begin{align}
\label{masslesscond}
& \sum_{\ell=0}^\infty c_{\ell}Y_{\ell}(\theta)=\varphi_{b}(\theta)\,,\quad ~~~~~~~\text{for}\,\, 0 \leq \theta\leq\theta_{0}\,,\nonumber\\
& \sum_{\ell=0}^\infty\ell(\ell+1)(\ell+2)c_{\ell}Y_{\ell}(\theta)=0\,,\quad \text{ for}\,\,  \theta_{0} < \theta \leq \pi  \,,
\end{align}
 (The equations for finite  $\tau_r$   can be found in \eqref{masslesscondfintau}.)

This system of equations can be solved numerically or  analytically, as discussed in appendix \ref{Numerical}. This gives a unique solution. 

We get the density matrix for this configuration by computing the action \nref{WaFu1}. 
\begin{equation} \la{ActFi}
\log \rho[ \varphi_b(\theta), \varphi_b(\theta)] \sim -2 \pi^{2} \sum_{\ell}\ell(\ell+1)(\ell+2)c_{\ell}^{2}\,,
\end{equation}
where the $c_\ell $ depend on $\varphi_b(\theta) $ through the equations in \eqref{masslesscond}. 
This is the same expression as we had in \nref{DenMaF}. This is to be expected, since the tracing out procedure is done via saddle point, so that in the end we are evaluating the same action on a particular configuration, namely the solution of \nref{masslesscond}. 

It is also possible to write \nref{ActFi} explicitly in terms of the boundary profile as, see appendix \ref{AnalyticSolution} 
\be 
\log \rho[ \varphi_b(\theta), \varphi_b(\theta)] \sim  \int_0^{\theta_0}  d\theta \int_0^{\theta_0 } d \theta' G(\theta, \theta') \varphi_b(\theta) \varphi_b(\theta') \,,\label{eq: bilinear action}
\ee 
where $G(\theta, \theta')$ is some function, see \eqref{eq: shape G}. This makes the dependence on $\varphi_b(\theta)$   manifest. 

Note that \nref{ActFi} is independent of $c_0$, as expected, and it has a maximum where $c_\ell =0$ for $\ell> 0$, which is the solution of  \nref{masslesscond} when $\varphi_b = \text{constant}$, which also implies that $\varphi^+ = \varphi^-$ is that same constant everywhere. 
This agrees with our discussion around \nref{MomMax} where we pointed out that extrema of the probability arise where the solutions are real.

\section{No boundary density matrix in slow roll inflation}
\la{HHSection}

We now turn to the case where we have dynamical gravity. It is convenient to focus on the case of single field inflation, both because of its cosmological interest, and because the time dependent scalar field gives us a notion of time. 
 
\subsection{Generalities about slow roll inflation }
\label{SetupHH}

We consider the standard single field slow roll inflationary which is described by the action\footnote{Here $M_{pl}$ is the reduced Planck mass defined via   $M^2_{pl} \equiv 1/(8 \pi G_N)$. } 
\begin{equation}\la{InflAct}
I=\frac{1}{2} M_{pl}^{2}\bigg(\int_{\mathcal{M}}d^{4}x\, \sqrt{-g}R-2 \int_{\partial{\mathcal{M}}}d^{3}y \sqrt{h} K\bigg)+\int_{\mathcal{M}}d^{4}x\, \sqrt{-g}\bigg(-\frac{1}{2}(\nabla \phi)^{2}-V(\phi)\bigg)\,,
\end{equation}
where the potential obeys the slow roll approximation 
\be \la{SLRPa}
 \epsilon \equiv \half M_{pl}^2 { {V'}^2 \over V^2 } \ll 1 ~,~~~~~~~~\eta \equiv M_{pl}^2 { V'' \over V } \ll 1 \,. 
 \ee 

\begin{figure}[h!]
    \centering
    
    \begin{subfigure}[b]{0.46\textwidth}
        \centering
       \hspace{0mm}\def\svgwidth{82mm}
\begingroup%
  \makeatletter%
  \providecommand\color[2][]{%
    \errmessage{(Inkscape) Color is used for the text in Inkscape, but the package 'color.sty' is not loaded}%
    \renewcommand\color[2][]{}%
  }%
  \providecommand\transparent[1]{%
    \errmessage{(Inkscape) Transparency is used (non-zero) for the text in Inkscape, but the package 'transparent.sty' is not loaded}%
    \renewcommand\transparent[1]{}%
  }%
  \providecommand\rotatebox[2]{#2}%
  \newcommand*\fsize{\dimexpr\f@size pt\relax}%
  \newcommand*\lineheight[1]{\fontsize{\fsize}{#1\fsize}\selectfont}%
  \ifx\svgwidth\undefined%
    \setlength{\unitlength}{238.83111668bp}%
    \ifx\svgscale\undefined%
      \relax%
    \else%
      \setlength{\unitlength}{\unitlength * \real{\svgscale}}%
    \fi%
  \else%
    \setlength{\unitlength}{\svgwidth}%
  \fi%
  \global\let\svgwidth\undefined%
  \global\let\svgscale\undefined%
  \makeatother%
  \begin{picture}(1,0.65862737)%
    \lineheight{1}%
    \setlength\tabcolsep{0pt}%
    \put(0.90792712,0.02842709){\color[rgb]{0,0,0}\makebox(0,0)[lt]{\lineheight{1.25}\smash{\begin{tabular}[t]{l}$\phi$\end{tabular}}}}%
    \put(0.07564815,0.63436217){\color[rgb]{0,0,0}\makebox(0,0)[lt]{\lineheight{1.25}\smash{\begin{tabular}[t]{l}$V(\phi)$\end{tabular}}}}%
    \put(0,0){\includegraphics[width=\unitlength,page=1]{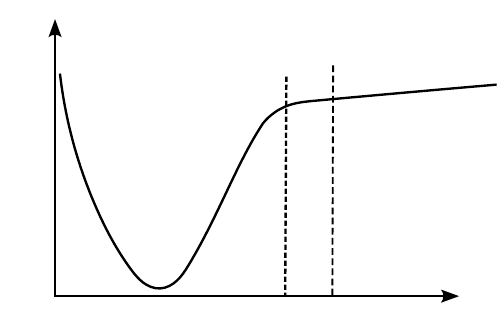}}%
    \put(0.55883326,0.0279192){\color[rgb]{0,0,0}\makebox(0,0)[lt]{\lineheight{1.25}\smash{\begin{tabular}[t]{l}$\phi_{rh}$\end{tabular}}}}%
    \put(0.654132,0.02711583){\color[rgb]{0,0,0}\makebox(0,0)[lt]{\lineheight{1.25}\smash{\begin{tabular}[t]{l}$\phi_{r}$\end{tabular}}}}%
  \end{picture}%
\endgroup%

        \caption{}
    \end{subfigure}
    \hfill
    \begin{subfigure}[b]{0.46\textwidth}
        \centering
       \hspace{0mm}\def\svgwidth{82mm}
\begingroup%
  \makeatletter%
  \providecommand\color[2][]{%
    \errmessage{(Inkscape) Color is used for the text in Inkscape, but the package 'color.sty' is not loaded}%
    \renewcommand\color[2][]{}%
  }%
  \providecommand\transparent[1]{%
    \errmessage{(Inkscape) Transparency is used (non-zero) for the text in Inkscape, but the package 'transparent.sty' is not loaded}%
    \renewcommand\transparent[1]{}%
  }%
  \providecommand\rotatebox[2]{#2}%
  \newcommand*\fsize{\dimexpr\f@size pt\relax}%
  \newcommand*\lineheight[1]{\fontsize{\fsize}{#1\fsize}\selectfont}%
  \ifx\svgwidth\undefined%
    \setlength{\unitlength}{238.83111668bp}%
    \ifx\svgscale\undefined%
      \relax%
    \else%
      \setlength{\unitlength}{\unitlength * \real{\svgscale}}%
    \fi%
  \else%
    \setlength{\unitlength}{\svgwidth}%
  \fi%
  \global\let\svgwidth\undefined%
  \global\let\svgscale\undefined%
  \makeatother%
  \begin{picture}(1,0.65862737)%
    \lineheight{1}%
    \setlength\tabcolsep{0pt}%
    \put(0.90792712,0.02842709){\color[rgb]{0,0,0}\makebox(0,0)[lt]{\lineheight{1.25}\smash{\begin{tabular}[t]{l}$\phi$\end{tabular}}}}%
    \put(0.07564815,0.63436217){\color[rgb]{0,0,0}\makebox(0,0)[lt]{\lineheight{1.25}\smash{\begin{tabular}[t]{l}$V(\phi)$\end{tabular}}}}%
    \put(0,0){\includegraphics[width=\unitlength,page=1]{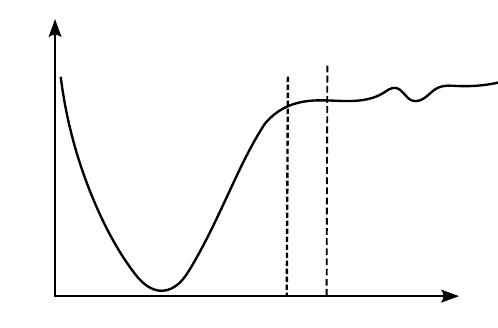}}%
    \put(0.56167715,0.02511287){\color[rgb]{0,0,0}\makebox(0,0)[lt]{\lineheight{1.25}\smash{\begin{tabular}[t]{l}$\phi_{rh}$\end{tabular}}}}%
    \put(0.64363368,0.02387696){\color[rgb]{0,0,0}\makebox(0,0)[lt]{\lineheight{1.25}\smash{\begin{tabular}[t]{l}$\phi_{r}$\end{tabular}}}}%
    \put(0,0){\includegraphics[width=\unitlength,page=2]{potential_bub.pdf}}%
    \put(0.74737028,0.34761507){\color[rgb]{0,0,0}\makebox(0,0)[lt]{\lineheight{1.25}\smash{\begin{tabular}[t]{l}small bump\end{tabular}}}}%
  \end{picture}%
\endgroup%

        \caption{}
    \end{subfigure} 
    \caption{Qualitative shape of the potentials we will consider in this paper. (a) An inflationary potential with a slow roll region for $\phi> \phi_{rh}$. (b) A potential with an additional bump   which can lead to Coleman de Luccia bubble solutions. Here we have denoted by $\phi_{rh}$ the value where inflation ends. The value $\phi_r$ denotes a value of $\phi$ further into inflationary region up to which we can approximate the potential as linear, see \nref{LinPot}. }
    \label{PotentialShape}
\end{figure}
 
When we evaluate the density matrix we will think of $\phi$ as a clock and view the 3-geometry as describing the dynamical variable. Just to be more definite, we imagine that we fix $\phi $ to some value $\phi_{rh}$ which is just slightly prior to the reheating region, just before the end of inflation. Then we are going to be interested in the probabilities (or density matrix) for various shapes for the three geometry of this surface. For simplicity, we will only explicitly discuss the scalar mode, namely the overall scalar curvature of the three surface, ignoring the tensor modes. Including the tensor modes in the discussion should be straightforward. 

The no boundary geometries we will consider look as follows. Starting from $\phi=\phi_{rh}$ and going backwards in time, we have a long period of Lorentzian evolution preceded, in the past, by an excursion into Euclidean signature. We find that the part that contributes significantly to the probability arises from the region where we have an excursion into Euclidean signature.
This is an important point that can be argued as follows.

\subsection{Argument that the probability does not depend on the superhorizon slice }
\label{superhorizon}

A property of the inflationary solution is that it is an attractor, so that slightly different initial conditions lead to essentially the same inflationary trajectory. To be specific, for spatially homogeneous solutions we can view the scale factor $a$ of the three geometry as a function of the time $\phi$. So we write $a(\phi)$. When we semiclassically evaluate the wavefunction of the universe at some value $\Psi(a_1,\phi_1)$ we will need to consider a solution $a_1^+(\phi)$ such that $a_1^+(\phi_1) = a_1$, in addition $a^+_1(\phi)$ should obey a regularity condition at early times that involves an excursion into complex geometries. This implies that generically the solution $a_1^+(\phi)$ is complex, even though $a_1$ and $\phi_1$ are real. When we evaluate $\Psi^*$, we have a solution $a^-_1(\phi)$ which generically will be different than $a^+_1(\phi)$ because the prescription for going into Euclidean time is different. 

However, at leading order in slow-roll,  if one solves the equations of motion neglecting the effects of spatial curvature there is a unique classical solution for $a$, call it $a_{f}(\phi)$, that is equal to $a_{1}$ at $\phi=\phi_{1}$, this is the usual slow-roll inflationary trajectory in flat slices. Then, because of the attractor property,  at late times both solutions $a^{\pm}(\phi)$ will be equal to $a_{f}(\phi)$ up to subleading corrections 
\begin{equation}
\label{LatExP2}
a_{1}^{\pm}(\phi)=a_{f}(\phi)\bigg[1+O\bigg(\frac{1}{a^{2}}\bigg)\pm i O\bigg(\frac{1}{a^{3}}\bigg)\bigg]\,.
\end{equation}

Note that the imaginary part of $a^{\pm}$ is smaller than the leading real part by a factor of $a^{-3}$. Now, we will argue that under these circumstances we will get the same value for the on shell action, or the same value for the density matrix,  as long as we evolve it along this classical flat slice solution $a_{f}(\phi)$. A similar point was previously discussed in \cite{Hartle:2007gi}. We argue this as follows. Under a small change of boundary conditions the action changes as 
\begin{equation}
\begin{gathered}
\label{rhochange}
\delta \log \rho=i[(\partial_{a}I^{+}-\partial_{a}I^{-})\delta a_{1}+(\partial_{\phi}I^{+}-\partial_{\phi}I^{-})\delta \phi_1]\\
=i[(p_{+}(\phi_{1})-p_{-}(\phi_{1}))\delta a_{1}-(H_{+}(\phi_{1})-H_{-}(\phi_{1}))\delta \phi_{1}]\,,
\end{gathered}
\end{equation}
where we used the usual Hamilton Jacobi formulas to write the derivatives of the action in terms of the conjugate momentum $p_{\pm}$ of $a$ and its Hamiltonian $H_{\pm}$ in the ket and bra respectively. Using an expansion like \nref{LatExP2} we note that $p^\pm = p \pm   p_I$,  where  $p$ is real and $p_I$ is purely imaginary. $p_I$ is small,   $p_{I}\sim a^{-1}$. Then we find that   
 \begin{equation}
H(a_{1},p_{+},\phi_{1})-H(a_{1},p_{-},\phi_{1})=2 p_{I} \partial_{p}H(a_{1},p,\phi_{f})+O\bigg(\frac{\dot{a}_{f}}{a_{1}^{3}}\bigg)=2 p_{I}\frac{d a_{f}}{d\phi}+O\bigg(\frac{\dot{a}_{f}}{a_{1}^{3}}\bigg)\,.
\end{equation}

The first equality results from Taylor expanding the Hamiltonian difference in $p_{I}$. The second one arises from using the Hamiltonian equation of motion,  dropping spatial curvature terms of order $a^{-2}$, writing $\partial_{p}H={ d {a}_{f} \over d \phi}\big(1+O(a^{-2})\big)$ and using the inflaton solution  in flat slices. Therefore, \nref{rhochange} implies
\begin{equation}
\label{rhochangefinal}
\delta \log \rho=i 2 p_{I} \big(\delta a_{1}-\frac{da_{f}}{d\phi}\delta \phi_{1}\big)+O\bigg(\frac{\dot{a}_{f}\delta \phi_{1}}{a_{1}^{3}}\bigg)\,.
\end{equation}
This means that if we choose $\delta a_{1}$ and $\delta \phi_{1}$ to be related as they are for the classical flat slice solutions,   $\frac{\delta a_{1}}{\delta \phi_{1}}={ d {a}_{f} \over d \phi} $, then the density matrix does not change at leading order, up to an error of order $a_{1}^{-2}$, since ${ d {a}_{f} \over d \phi} \sim O(a_{1})$. 

More precisely, in our case we have a full field 
$ a( \vec x , \phi) $, but the argument is the same, as long as we are considering fields that vary over length scales larger than the horizon size at the time $\phi$ where we evaluate the action. In that case \nref{LatExP2} is valid.  In particular, this implies that we vary the superhorizon slice where we evaluate the action. In other words, we can choose different spatial slices, even slices which sit at a time $\phi$ that depends on $\vec x$, and the total density matrix will still have the same value. 

As an aside, we note that a somewhat similar argument is usually applied in AdS, in the context of the classical limit of holographic renormalization \cite{deBoer:1999tgo,Bianchi:2001kw,Heemskerk:2010hk}.  There it is important to take the local terms in the action explicitly into account, by cancelling them with suitable counterterms, for example. 
For the diagonal elements of the density matrix that we have been considering, the argument is simpler since such local terms cancel between the bra and ket sides. As a side comment, note that these local and purely imaginary terms in the wavefunction, or off diagonal components of the density matrix,
are important for enforcing the leading order equations of motion for the bulk fields.

\subsection{A simplified approximation for the potential in the relevant region}

\begin{figure}[h!]
   \begin{center}
   \includegraphics[scale=.4]{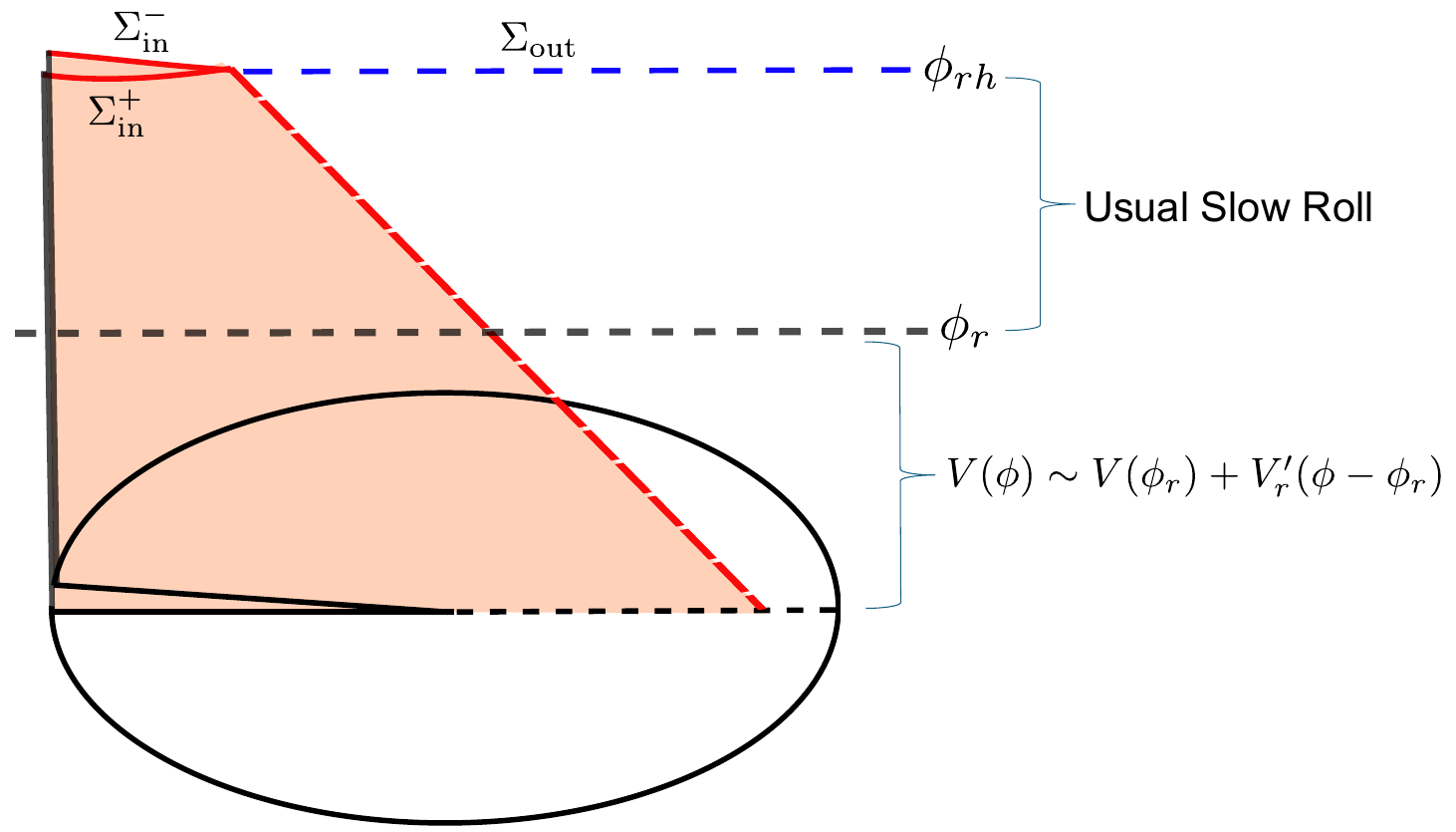}
    \end{center}
    \caption{Geometry relevant to the inflationary computation. We evaluate the action on a surface at $\phi_r$, up to which we can approximate the potential as being linear.
   }
    \label{InflationFig}
\end{figure}

In principle, we are interested in evaluating the density matrix at $\phi_{rh}$, at the end of inflation. As we go back in time, the actual solutions we consider track fairly closely a real inflationary trajectory up to a time we call  $\phi_* $ where they mainly go into Euclidean space. In order to analyze the solution, it is convenient to expand the potential up to linear order around that time. After a few efolds the solution will again closely track the inflationary trajectory. For simplicity, we can evaluate the action at an earlier time, say a time $\phi_r$ which is a few efolds after $\phi_*$ but close enough that we can still use the linear potential approximation. As we explained in section \ref{superhorizon}, we get the same answer as we would have obtained in we evaluated the action at $\phi_{rh}$, see figure \ref{InflationFig}. This assumes that we are considering fluctuations of the geometry of $\Sigma^\pm_{\rm in}$ that correspond to superhorizon distances at $\phi_r$. 

 Then it is convenient to expand the potential as  
\be \la{LinPot}
V = V_r + V'_r ( \phi - \phi_r ) + \cdots ~,~~~~~~~~~~V_r \equiv V(\phi_r)\,.
\ee 
In addition,  the metric can be approximated as being the standard de-Sitter metric with Hubble constant $H_r$, with $ 3 M_{pl}^2 H_r^2  = V_r$. Here we are imagining that $\phi_r $ is further up the potential than $\phi_{rh}$, see figure \ref{InflationFig}.

With all these approximations taken into account, we are now ready to write a simplified version of the action
\bea 
I&=&\frac{1}{2} { M_{pl}^{2} \over H_r^2} \bigg(\int_{\mathcal{M}}d^{4}x\, \sqrt{-g}( R - 6 ) -2 \int_{\partial{\mathcal{M}}}d^{3}y \sqrt{h} K\bigg) +
\cr \la{AcSca}
& ~& +  { 2 \epsilon_r M_{pl}^2 \over H_r^2  }  
\int_{\mathcal{M}}d^{4}x\, \left[ - \half (\nabla \varphi)^2 - 3 \varphi  \right] ~,~~~~~~
{\rm with } ~~~~~\varphi \equiv  { ( \phi - \phi_r) \over M_{pl} \sqrt{ 2 \epsilon_r } }\,.
\eea
  This metric differs from 
 the one in \nref{InflAct} by a factor of $H_r^2$. In writing \nref{AcSca} we have neglected a possible $\varphi^2$ term inside the square backet whose coefficient is proportional to the $\eta$ slow roll parameter   \nref{SLRPa}\footnote{The term is $-{ 3\over 2 } \eta \varphi^2 $ inside the square bracket in \nref{AcSca}.     }. We will also assume that $\phi =\phi_r$ is the surface where we evaluate the action, this means that $\varphi=0$ at the boundary, see figure \ref{InflationFig}. 
 
 The equations of motion for the action in the first line of \nref{AcSca} imply that we have a de Sitter space with unit radius
\be \la{UnDs}
ds^2 = - d\tau^2 + \cosh^2 \tau d\Omega_3^2 ~,~~~~~~~~~d\Omega_3^2 = d \theta^2 + \sin^2 \theta d\Omega_2^2  \,.
\ee  
The equation  of motion for $\varphi $  is
\be \la{Eomvf}
- \nabla^2 \varphi + 3 =0\,, 
\ee 
where the Laplacian is over a unit radius $dS_4 $ \nref{UnDs}.

 We also assumed that the solutions of \nref{Eomvf}  do not backreact on the metric,  even for a non-zero $\varphi$ solution. Let us explain why we can do this. 
 We first start with a purely de-Sitter metric and find a solution of  \nref{Eomvf} that obeys the boundary conditions. We consider boundary conditions that involve order one profiles for the field $\varphi$. Their contribution to perturbations to the metric away from $dS_4$ will be then of order $\epsilon_r$, let us call these fluctuations $\delta g_{\mu \nu}$. 
When we insert them into the action, the fact that the de Sitter background solved the classical equations  for the first line in \nref{AcSca} means that the action does not change to first order in $\delta g_{\mu \nu}$. For this argument, it is important to  note that the original solution was obeying the boundary conditions and therefore when we solve for $\delta g_{\mu \nu} $ we will set it to zero at the boundary. 
The conclusion is that we can ignore the deformations away from the purely de-Sitter metric, for the purposes of evaluating the probabilities.  

It turns out that the (inverse of) the coefficient of the second line  \nref{AcSca} determines the overall size of the  scalar fluctuations in inflation. If this coefficient is smaller than of order one, it means that we are in the slow roll eternal inflation regime. Since we will be evaluating actions classically here we will assume it is larger than one,   $ \frac{\epsilon_{r} M_{pl}^{2}}{H_{r}^{2}} \gg 1$. As explained in \cite{Creminelli:2008es}, one can still use the properties of the free field $\varphi$ to make interesting statements for eternal inflation.

\subsection{Quick review of the spatially homogeneous solution}

Before turning to more complicated configurations, let us discuss the simple spatially homogeneous solutions, independent of the coordinates of $S^3$  \cite{Maldacena:2024uhs}, see also \cite{Janssen:2020pii},
\begin{equation}
\label{phic}
\varphi^\pm_{c}(\tau)=\frac{1}{1\mp i \sinh \tau}-\ln(1\mp i \sinh \tau)-(\tau \leftrightarrow \tau_{r})\,.
\end{equation}
These solutions obey that $\phi = \phi_r $, or $\varphi =0$, at $\tau = \tau_r$. $\tau_r$ is just a parameter that says how big the $S^3$ is at time $\phi = \phi_r$, namely $a_r = \cosh \tau_r$. 
These solutions, \nref{phic}, are smooth at $\tau = \pm i { \pi \over 2} $ respectively. These   conditions  amount to saying that the solutions are smooth on either the north or south pole of the $S^4$. Since the condition is different for the $+$ or $-$ solutions we find that  $\varphi^\pm$ are not equal to each other.  

We now compute the action for these solutions. For the solutions we will consider in our paper, we will imagine that, {\it by hand}, we evaluate the probability for a very large $a_r$, or very large $\tau_r $, at some value $\phi_r$.
The first line in \nref{AcSca}   gives a constant contribution corresponding to the de-Sitter entropy with cosmological constant $V_r$, $ S_r = 24\pi^2 M_{pl}^4/V_r $.  We then get, for $\tau_r\gg 1$, 
\be \la{AcLat}
\log \rho_c  = i(I_+ - I_-) = S_r  + { 2 \epsilon_r M_{pl}^2 \over H_r^2 } 8 \pi^2 \left[ - \tau_r+2 \log 2 - { 1 7 \over 12} \right] ~,~~~~~~S_r \equiv     24\pi^2 {M_{pl}^4 \over V_r}\,,
\ee  
where the second term comes from inserting \nref{phic} into the action \nref{AcSca} and integrating $\tau$ from $\pm i \pi/2$ to $\tau_r$. We first note that if we vary $\tau_r $ in order to search for the most probable solution, we are driven to small $\tau_r$.\footnote{ In fact, from \nref{Acco}  (which is the expression valid for any $\tau_r $)  we find that the most probable solution sits at $\tau_r=0$.} Of course, this reproduces the well known fact that the no-boundary proposal tries to have as little inflation as possible. Since there we are putting the future boundary at $\phi_r$, it just simply says we have no inflation and we go directly  into Euclidean space. Of course, the term linear in $\tau_r $ in \nref{AcLat} tells us how unlikely it is to have a scale factor of about $\tau_r $ e-folds greater than the Hubble scale at the time $\phi_r$. In other words,
\be\la{UnProba}
\rho_c \propto \exp\left( - { 2 \epsilon_r M_{pl}^2 \over H_r^2 } 8 \pi^2 \Delta {\cal  N } \right) \,,
\ee 
where $\Delta {\cal N } \sim \tau_r$ is the number of efolds up to time $\tau_r$, or the number of Hubble size regions at time $\tau_r$. 

As we said in the introduction, the fact that the solutions are complex is intimately related to the fact that we are not evaluating the probability at a maximum. Of course, this is {\it not} the most probable configuration, but we will continue studying such configurations to explore the answers that the no-boundary proposal gives us. 

Alternatively, we can think that we are looking at the Hartle Hawking state but we are evaluating observables that are strongly peaked around $\tau= \tau_r$, that involve a very large spatial slice. They have to be peaked strongly enough to overwhelm the exponential suppression. We can certainly mathematically consider such observables, but their relevance to the physics we measure is not clear.  Similarly strongly peaked observables were discussed in \cite{Chen:2024rpx,Kudler-Flam:2024psh} for the same reason.

\subsection{Quick review of the spatially inhomogeneous case }

Here we simply consider a problem similar to the one we considered for a massless scalar in de Sitter in section \ref{WfdSWh}  where we set the boundary conditions 
\be 
\varphi^\pm(\vec \Omega, \tau_r)   = \varphi_b(\vec \Omega) \,,
\ee 
with $\varphi_b(\vec \Omega)$ a  given function,  which is the argument of the diagonal components of the density matrix. 

The only difference relative to the computation in section \ref{WfdSWh} is that we have a linear term in the action for $\varphi$ \nref{AcSca}.
But we can simply write
\be \la{phiDec}
\varphi^\pm = \varphi^\pm_c + \tilde \varphi^\pm  ~,~~~~~~~~~~ \nabla^2 \tilde \varphi =0\,,
\ee 
where $\varphi^\pm_c$ is the solution discussed  in \nref{phic} which obeys \nref{Eomvf}, implying that $\tilde \varphi$ obeys the second equation in \nref{phiDec}. 
The evaluation of the action then involves 
\be 
\int_{\cal M } \left[\half (\nabla \varphi)^2 +  3 \varphi \right] = \int_{\cal M } \left[\half (\nabla \varphi_c)^2 +  3 \varphi_c \right]  +  \int_{\Sigma } \left[ \half \varphi_b \partial_n \tilde \varphi + \varphi_b \partial_n \varphi_c \right]\,,
\ee 
where we did some integration by parts and used the equations of motion. 
The first term reproduces the second term in 
\nref{AcLat}. Then, we  obtain, for the diagonal part of the density matrix, 
\be 
\log \rho = \log \rho_c + { 2 \epsilon_r M_{pl}^2 \over H_r^2 } \int_{S^3 } d\Omega_3 \left[ \half  \cosh^3 \tau_r \varphi_b ( i \partial_\tau \tilde \varphi^+ - i \partial_\tau \tilde \varphi^-)  - 4 \tilde \varphi_b \right]\,,
\ee 
where the integral is over a unit radius $S^3$. We  used that 
\be  \la{DervFc}
- i \cosh^3\tau ( \partial_\tau \varphi_c^+ -  \partial_\tau \varphi_c^-) = 4 \,.
\ee 
We can now expand $\  \varphi_b$ in terms of spherical harmonics, as in \nref{VarPhivExp} and $\tilde \varphi^\pm$ as in   \nref{homexp}  to obtain 
\be 
\log \rho = \log \rho_c  - { 2 \epsilon_r M_{pl}^2 \over H_r^2 }  2 \pi^2 \left[    4   c_0  + \sum_{\ell, m_1,m_2 } \ell ( \ell +1) (\ell + 2)   |  c_{\ell, m_1,m_2}|^2  \right]\,. \la{LateRho}
\ee 
In appendix \ref{InHomApp}, we discuss the action for general  $\tau_{r}$  in \nref{FinRho}, and setting $\tau_r =0$ we recover the wavefunction in the form discussed in \cite{Chen:2024rpx}.

\subsection{The wavefunction in terms of the scalar curvature}

In the previous subsection, we computed the wavefunction as a function of the scalar field on a given spatial slice with constant curvature, set by the choice of $\tau_r$. From the inflationary point of view it is better to think about the wavefunction on a given value of the scalar field, say $\varphi = 0 $ (or $\phi = \phi_r$) as a function of the curvature or the geometry of the three dimensional slice. 
When we think about the scalar mode, we are simply talking about the overall Weyl factor in the metric. We express this in terms of deviations from the round metric  
\be \la{3Geom}
ds^2_{\Sigma } = e^{ 2 \omega} \cosh^2 \tau_r d\Omega_3^2 \,.
\ee 
where $\omega= \omega(\vec \Omega)$ is an arbitrary function of the coordinates on the three sphere. And $\tau_r$ is just a parameter which sets the overall scale of the metric\footnote{This parameter is  redundant, since we could absorb it by shifting $\omega$. Nevertheless, we still keep it because we want to take the limit $\tau_r \to \infty$ keeping $\omega $ fixed.}.   

In the inflationary region,  these two choices (spatially constant $\phi$ or spatially constant scale factor of the metric) differ only by a gauge choice. In particular, we can find the metric 
\nref{3Geom} as a slice of un-deformed   de-Sitter space \nref{UnDs} by picking a suitable slice. This is particularly easy to do at superhorizon scales\footnote{It is simpler in this case because we can neglect the part of the induced metric coming from 
$d\tau^2$. }, or for $\tau_r \gg 1$, keeping $\omega(\vec \Omega)$ fixed. The slice in $dS_4$ with metric \nref{3Geom} sits at  
\be \la{SliTau}
\tau = \tau_r + \omega \, ~~~~~{\rm for } ~~~~~ \tau_r \gg 1.
\ee 

The scalar curvature (Ricci scalar) of \nref{3Geom} is 
\be \la{CurvSli}
R^{(3)} = { e^{ - 2 \omega } \over \cosh^2 \tau_r } \left[   6 - 2 (\nabla \omega)^2 - 4 \nabla^2 \omega \right]\,,
\ee 
where $\nabla$ denotes the gradient on an $S^3$ with unit radius. 
This means that if we are interested in the amplitude to find  a surface at $\varphi =0$ with scale factor set by $\omega(\Omega)$, we can equivalently look for the amplitude to find a profile for 
$\varphi = \omega $ on a surface with $\omega=0$. 
The reason is the following. 
The  general  solutions of \nref{Eomvf} for $\varphi $  are linear in $\tau$ for large $\tau$ 
\be \la{varphiLa}
\varphi = - (\tau -\tau_r) + f( \vec \Omega_3) + \mathcal{O}(e^{ - 2 \tau })\,,
\ee 
where $f$ is function on the spatial three sphere of $dS_4$ \nref{UnDs}. 
This means that the surface where $\varphi =0 $ occurs at $\tau - \tau_r = f(\vec \Omega) $. Therefore,   \nref{SliTau} implies that  
\be \la{vflp}
\omega(\vec \Omega)  =f(\vec \Omega)\,.
\ee 

This means that the wavefunctions or density matrices can be simply related among these two gauges. 
\be 
\rho[ \omega(\vec \Omega) , \omega(\vec \Omega) ; \varphi =0] = \rho[ \varphi_b(\vec\Omega)  , \varphi_b(\vec\Omega) ;   \omega =0] ~,~~~~~~~~~~~~{\rm with } ~~~~~~~ \varphi_b(\vec \Omega) = \omega(\vec \Omega)\,.
 \ee

 Note that we are evaluating the action of two physically distinct slices which are related by time evolution over superhorizon distances. We used section \ref{superhorizon} to conclude that the action is the same evaluated on these two slices. 
 
In the inflationary region we can pick either of these two gauges for describing the fluctuations. However, in order to track the physics after we exit inflation, it is better to think in terms of the curvature because it is conserved on superhorizon scales through the later evolution of the universe (in the single field inflation case).  

Note that we can trust our approximations \nref{AcSca} for evaluating order one fluctuations in $\varphi$ which translate in order one fluctuations of $\omega$ and, therefore, order one fluctuations in \nref{CurvSli}. In fact, we can consider situations where the curvature is negative, for example.  Note that we are not linearizing the expression for the curvature   \nref{CurvSli}.  Here we are considering classical solutions, but the same point was useful for the quantum theory in the analysis of eternal inflation in 
  \cite{Creminelli:2008es}.

\subsection{The density matrix for a subregion}

\label{dmsubregion}
 
In this section, we discuss the problem of computing the density matrix for a subregion.
We are interested in considering a portion of a spatial slice, $\Sigma_{\rm in}$, or more precisely two portions, $\Sigma_{\rm in}^\pm$, one for the bra and one for ket entries of the density matrix. 

Conceptually, we are interested in finding a no-boundary geometry that computes the classical approximation to the density matrix. 

One possible no boundary geometry is one where we extend the spatial slice by adding an unobserved piece, 
$\Sigma_{\rm out}$, which is the same for the bra and the ket and we look for a complex geometry with no boundary in the past, by deforming with the usual prescription. 
The process of tracing out, classically sets the fields to be equal on the $+$ and $-$ sheets of the geometry at $\Sigma_{\rm out}$. 
We would like to say that this ``trace-out'' boundary condition at $\Sigma_{\rm out} $ is physically and mathematically similar to saying that we have no physical boundary there. It boils down to a precription for connecting the fields on the two sheets, and we can place $\Sigma_{\rm out} $ anywhere.

The conclusion, is that the final mathematical problem is the following. We are given the scale factor of the geometry $\omega(\vec \Omega) $ on some portion of the sphere $\Sigma_{\rm in}$. We extend this to a full slice covering the full $S^3$ by adding $\Sigma_{\rm out}$. 
Then we look for a solution of the equations of motion for $\varphi$ \nref{Eomvf} on a two sheeted spacetime ${\cal M } = {\cal M}^+ \cup {\cal M^-}$ which both end on the full three sphere. ${\cal M}^+$ is half of de-Sitter joined with a half $S^4$ and similarly for ${\cal M}^-$. The solutions should be smooth in the corresponding half $S^4 $ regions obtained by analytic continuation. 

The boundary conditions in the future slice are 
\bea 
& ~& \varphi_+ =f_+  (\Omega)= \omega_+(\Omega) ~,~~~~~~~ \varphi_- = f_-(\vec \Omega)= \omega_-(\Omega) ~,~~~~~~{\rm for } ~~~~~~\tau = \tau_r ~,~~~~\vec \Omega \in \Sigma^{\pm}_{\rm in}\,,\la{EqnsTo1}
\\
&~& \varphi_+ = \varphi_- ~,~~~~~~~~~~~~~~~~~~~~~~~ \partial_\tau \varphi_+ = \partial_\tau \varphi_- ~,~~~~~ ~~~~~~~~
{\rm for }~~~~~~~~~\tau = \tau_r ~,~~~~\vec \Omega \in \Sigma_{\rm out} \,.\la{EqnsTo2}
\eea 
Let us explain again the logic for  \nref{EqnsTo2}, which are the ``trace-out'' boundary conditions. The condition $\varphi^+ =\varphi^- $ comes  because we are tracing out this region,  and the condition on the time derivative comes from a saddle point approximation for the integration over all possible profiles of the field $\varphi$ on $\Sigma_{\rm out}$, see \nref{TrOut} \nref{TrOutEq}. 

Equivalently, we can say we have a single complex manifold ${\cal M}$ with boundary at $\Sigma^\pm_{in}$, where the Lorentzian metric is obtained  by changing the phase of the  $g_{\hat t \hat t }$ (initially  Euclidean)  component of metric by $e^{ \pm i \pi(1-\epsilon) }$ for the $\pm$ parts of the boundary \cite{Kontsevich:2021dmb,Witten:2021nzp}.  

For most of this paper we will consider the {\it diagonal} components of the density matrix, where $\omega_+ = \omega_-$ and therefore the boundary conditions $f_+ = f_-$. We will still have that $\varphi_+ \not = \varphi_- $ away from the boundary due to the different prescriptions for going into Euclidean time. But, under these circumstances, we will find that $   \varphi_+ = (\varphi_-)^*  $.  
  And, as we explained above, we will also have that $\varphi_+ = \varphi_- $ in the whole domain of dependence of $\Sigma_{\rm out}$. 

Once we solve the equations \nref{EqnsTo1} \nref{EqnsTo2}, the final value of the density matrix is given by the same expression as the density matrix evaluated on the whole spatial slice. 
In other words it is given by \nref{LateRho}. 
The only difference is that we are not directly given the values of the $c_{\ell,m_1,m_2}$, we need to find them in terms of $\omega(\vec \Omega) $ in region $\Sigma_{\rm in} $ by solving the problem \nref{EqnsTo1} \nref{EqnsTo2}.  

Let us give more details on how we can determine the solution to \nref{EqnsTo1} \nref{EqnsTo2}. 
We write the field $\varphi = \varphi_c + \tilde \varphi $ as in \nref{phiDec} and expand $\tilde \varphi$ 
  in spherical harmonics,  as in \nref{homexp},   
\begin{equation} \la{Expvft}
\tilde{\varphi}^{\pm}=\sum_{\ell=0}^{\infty}   c_{\ell,m_{1},m_{2}}Y_{\ell,m_{1},m_{2}}(\Omega_{3}) f_{\ell}^{\pm}(\tau)\,,
\end{equation}
where we have used that, for the diagonal elements of the density matrix,  \nref{EqnsTo2} imply that $\varphi^+ = \
\varphi^- $ on the whole spatial slice, so that     $ c_{\ell,m_{1},m_{2}}^{\pm}$  are equal (and we dropped the $\pm$ index in the $  c$'s).
The term in the trace-out boundary condition \nref{EqnsTo2} involving the derivative becomes 
\be 
 i \cosh^3 \tau_r ( \partial_\tau \tilde \varphi^+ - \partial_\tau \tilde \varphi^+ ) = -  i \cosh^3 \tau_r ( \partial_\tau  \varphi^+_c - \partial_\tau   \varphi^+_c ) = 4\,, 
 \ee 
 after using \nref{DervFc}.
 This implies that the problem \nref{EqnsTo2} becomes a problem for the coefficients $  c_{\ell,m_1,m_2}$  
\begin{equation}
\begin{gathered}
\la{LargeEqc}
\sum_{\ell,m_{1},m_{2}}  c_{\ell,m_{1},m_{2}}Y_{\ell,m_{1},m_{2}}(\vec \Omega)=\omega (\vec \Omega ), ~~~~~\text{ for } ~~~~~~\vec{\Omega}  \in \Sigma_{\rm in}\,,\\
\sum_{\ell,m_{1},m_{2}}^{\infty} \ell(\ell+1)(\ell+2)c_{\ell,m_{1},m_{2}}Y_{\ell,m_{1},m_{2}}(\vec \Omega)=-2 ,~~~~~~~~\text{ for }~~~~~~~~\vec{\Omega} \in \Sigma_{\rm out} \,.
\end{gathered}
\end{equation}
We can view this equation as an equation purely for the profile of $\tilde \varphi(\vec \Omega)$ at the boundary. By using the expansion \nref{Expvft} we solved the equation of motion and imposed the correct boundary conditions in the past. Solving \nref{LargeEqc} we determine the profile of $\tilde \varphi$ in $\Sigma_{\rm out}$. These equations are the final equations that determine the saddle point value of  $\varphi_{\rm out}^s$ in the procedure discussed around \nref{TrOut}. 

The final action remains \nref{LateRho}, evaluated in a solution of \nref{LargeEqc}. 
In  the large $\tau_r $ limit this is  
\begin{equation}
\begin{gathered}
\label{OnshellactHH}
\log\rho =S_{r}-\frac{8 \pi^{2}\epsilon_{0} M_{pl}^{2}}{H_{r}^{2}}\bigg(2 \tau_{r}+2 c_{0}-4 \ln 2+\frac{17}{6}+\frac{1}{2}\sum_{\ell,m_{1},m_{2}}^{\infty} \ell(\ell+1)(\ell+2)|c_{\ell,m_{1},m_{2}}|^{2}\bigg)\,,
\end{gathered}
\end{equation}

In the following subsections, we solve these equations in a series of examples. 
In those examples, we will consider regions and boundary data that is $SO(3)$ symmetric, as in section \ref{DMCap}, so that we will expand the field using the $Y_\ell(\theta) $ functions defined in \nref{YHarm}. 

The dependence on the mode $c_0$ in \nref{OnshellactHH} is the same as what was found in \cite{Chen:2024rpx,Kudler-Flam:2024psh}
using different methods.

\subsection{The density matrix when we fix the area of a  surface}
\label{FixArea}

A simple example for the  $\Sigma_{\rm in} $ region is a tiny strip around $\theta = {\pi \over 2 } $, in the limit where the interval in the $\theta $ direction becomes very small. This means that we are only fixing boundary conditions at   $\theta={\pi \over 2} $, where we set $\varphi =0$,  and we have trace-out boundary conditions at all other values of $\theta$. 

Though here we take $\tau_{r}\gg 1$ , we also consider this problem for finite $\tau_r$  in appendix \ref{Finitetauarea}. Notice that $\tau_r$ is the value of $\tau$ where we impose the boundary condition, so that choosing $\tau_r$ is equivalent to choosing the radius of this sphere. 


In this case, the equations 
\eqref{FinEqc} become 
\begin{equation} \la{EqnSe}
\sum_{\ell} c_\ell Y_{\ell}\left({ \pi \over 2} \right)=0\,,\quad  ~~~~\sum_{\ell}^{\infty} c_{\ell}\ell(\ell+1)(\ell+2)Y_{\ell}(\theta)=-2+b \delta\left(\theta-{ \pi \over 2 } \right)\,.
\end{equation}
where $b$ is a constant to be determined. This is simply the statement that \nref{EqnSe} should hold everywhere except at $\theta = {\pi \over 2 }$. 
Since the left hand side  of \nref{EqnSe} has no $\ell=0$ harmonic, the integral of the right hand side against $\sin^{2}\theta$ should be zero, implying $b=\pi$. Integrating \nref{EqnSe} against the other harmonics yields
\begin{equation} \la{clpos}
c_{\ell,\rm fa} \coloneqq c_{\ell}=\frac{2 Y_{\ell}(\frac{\pi}{2})}{\ell(\ell+1)(\ell+2)}\,,~~~~~~{\rm for }~~~~~ \ell > 0\,.
\end{equation}
Finally,  $c_{0}$ is determined by requiring that $\varphi=0$ at $\theta=\frac{\pi}{2}$, leading to
\begin{equation} \la{clzero}
c_{0,\rm fa} \coloneqq c_{0}=2\log 2-\frac{3}{2}\,.
\end{equation}
The expression for the action can be simplified in this problem by using that either $\varphi=0$ or $\dot{\varphi}^{+}=\dot{\varphi}^{-}$ at $\tau=\tau_{r}$, which implies
\begin{equation}
0=\int_{\Omega_{3}}\varphi(\partial_{\tau}\varphi^{+}-\partial_{\tau}\varphi^{-})=\frac{16 i}{3}2 \pi^{2}\bigg(2 c_{0}+\sum_{\ell=1}^\infty c_{\ell}^{2}\ell(\ell+1)(\ell
+2)\bigg)\,.\label{eq: fixA c0 eq}
\end{equation}
Then,  the final action can be written in terms of $c_{0}$ alone and it gives  
\begin{equation}
\label{fixaaction}
\log \rho_{\rm fa} = S_r -\frac{8 \pi^{2}\epsilon_{r} M_{pl}^{2}}{H_{r}^{2}}\bigg(2 \tau_{r}+\frac{4}{3}-2\log 2\bigg)\,.
\end{equation} 
Note that the probability density still depends on $\tau_r$, which sets the size of the sphere. Furthermore, it is exponentially small for large $\tau_{r}$, which implies that it is a very unlikely configuration. But since the field is unconstrained away from the two sphere at $\theta=\frac{\pi}{2}$ and $\tau=\tau_{r}$, this configuration is the most likely among those that have a two-sphere of this size. In addition, we see from the $\tau_r$ term in \nref{fixaaction} that the probability will also increase if we make this two-sphere smaller.

\begin{figure}[h]
    \centering
    \includegraphics[width=0.5\linewidth]{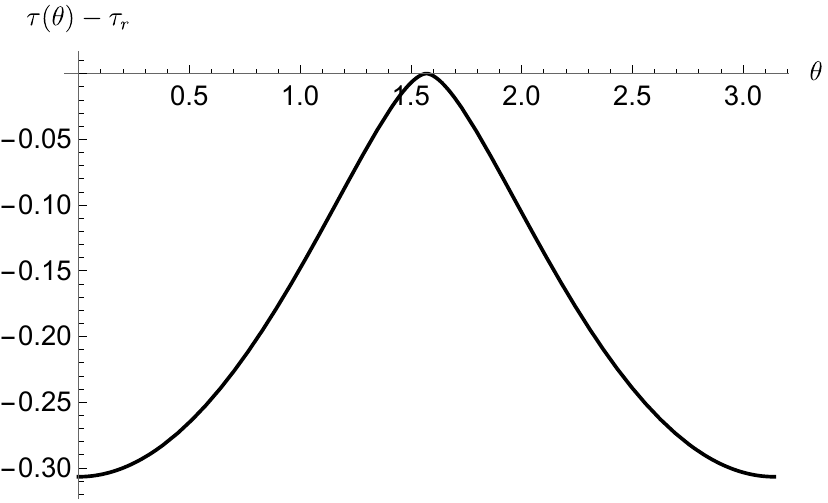}
    \caption{$\tau(\theta)-\tau_{r}$ of the surface with $\varphi=0$ in the problem where we only fix the area at $\theta = \pi/2$. Note that this is negative, indicating that we have a smaller number of e-folds away from the region where we are fixing the metric (away from $\theta =\pi/2$). 
    }
    \label{zerophifixedarea}
\end{figure}

Having solved for the coefficients $c_{\ell}$ \nref{clpos} \nref{clzero}, it is straightforward to do the sum and find the shape $\varphi(\theta)$ to be
\begin{equation}
\label{phifa}
\varphi_{\rm fa}(\theta) \coloneqq \varphi(\theta)=\frac{1}{\sin \theta}\bigg[\frac{(1-\sin \theta)}{2}\log\bigg(\frac{1-\sin \theta}{2}\bigg)\\-\frac{(1+\sin \theta)}{2}\log\bigg(\frac{1+\sin \theta}{2}\bigg)\bigg]\,.
\end{equation}
We can also solve for $\varphi$ everywhere in $\tau$, we still take $\tau_{r}\gg 1$ but allow $\tau$ to be finite. We find for $\varphi^{\pm}$
\begin{equation}
\begin{gathered}
\label{fixedasurf}
\varphi^{\pm}(\tau,\theta)=\tau_{r}-\ln(2\cosh \tau)+\frac{1}{\sin \theta}\bigg[\frac{(\tanh \tau-\sin \theta)}{2}\log\bigg(\frac{\tanh \tau-\sin \theta \pm i 0^+}{2}\bigg)\\-\frac{(\tanh \tau+\sin \theta)}{2}\log\bigg(\frac{\tanh \tau+\sin \theta}{2}\bigg)\bigg]\,,
\end{gathered}
\end{equation}
where terms exponentially small in $\tau_{r}$ were neglected. The $i0^+$ prescription tells us how to continue the logarithm when  $ \tanh \tau < \sin \theta $. This is the region lying in the past light cone of $\theta= \pi/2$, which is where $\Sigma_{\rm in}$ is located. Outside this region, for $\tanh \tau > \sin \theta$,  the two solution are equal, as expected.\footnote{Note that in the coordinates \nref{ConfCo}, $\tanh \tau = \cos \eta $. } 

\begin{figure}[h!]
    \centering
    \includegraphics[width=0.5\linewidth]{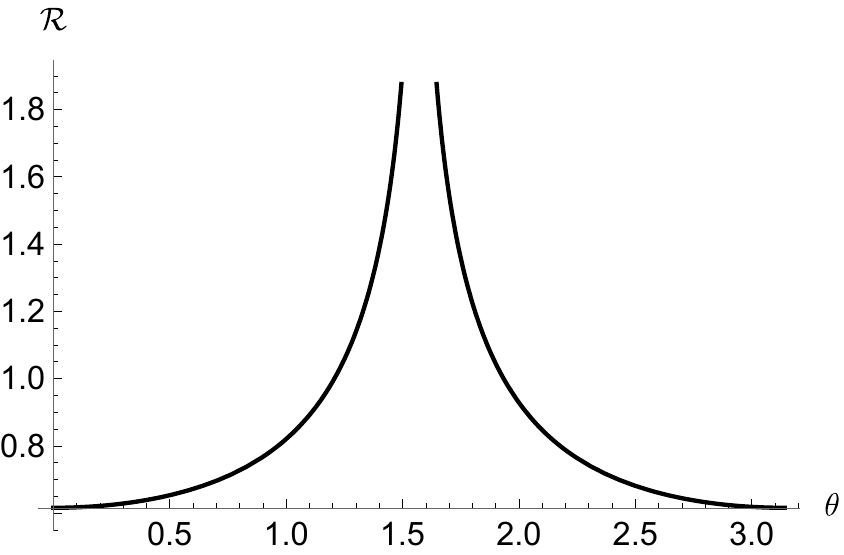}
    \caption{Plot of the normalized intrinsic curvature \nref{req} of the $\varphi=0$ surface in the problem where we only fix the area of the hemisphere.}
    \label{intriscfixedarea}
\end{figure}

 
Therefore, \nref{fixedasurf} equivalently fixes the shape of the $\varphi=0$ surface for this problem, as shown in figure  \ref{zerophifixedarea}.
At large $\tau$, this surface is continuous at $\theta=\frac{\pi}{2}$ and has continuous first derivative, but the second derivative of $\tau(\theta)$ is divergent there. In particular, this implies that the intrinsic curvature $R$ of the surface 
diverges at $\theta=\pi/2$, see figure \ref{intriscfixedarea}.
Note that since we took the large $\tau_{r}$ limit, the surface curvature is suppressed as $e^{-2 \tau_{r}}$, so it is more meaningful to plot instead the ratio of the curvature of this surface to the curvature $R^{(2)}_{r}$ of the two sphere whose area are kept fixed. 
We then define and plot the following dimensionless curvature variable ${\cal R}$,\footnote{The factor of 3 was inserted so that when all spheres have unit radius this ratio is one.}
\begin{equation}
\label{req}
{\cal R}=\frac{R^{(3)}}{3 R_{r}^{(2)}}\,,
\end{equation}
which will appear often in the next subsection.

\la{HHDM}

\subsection{The density matrix for subregions  with fixed curvature}
\label{FixCurv}

Using the setup from section \ref{dmsubregion} one can compute the probability for subregions with various 3-geometries at $\phi=\phi_r$ slice. Since we are considering scalar deformations, we can fix various values of the scalar curvature. As a simple example, let us consider a subregion $\Sigma_{\rm in}$ with the topology of a solid ball, with various constant values for the scalar curvature inside and a fixed radius surface (the boundary of $\Sigma_{\rm in}$). 

We can embed the boundary of $\Sigma_{\rm in}$ as the sphere at $\theta=\pi/2$ and $\tau = \tau_r$. Again, $\tau_r$ parametrizes the size of this sphere. It is then convenient to parametrize the scalar curvature of the interior of $
\Sigma_{\rm in }$ as the ratio ${\cal R} $ defined in \nref{req}. 

\begin{figure}[h!]
\begin{subfigure}[h]{0.7\linewidth}
\centering
    \includegraphics[width=\linewidth,height=0.4\linewidth]{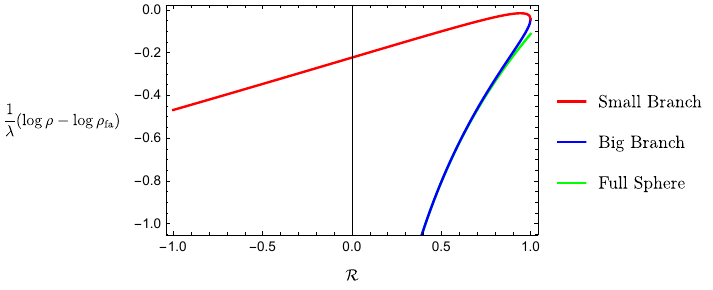}
\caption{}
\end{subfigure}
\begin{subfigure}[h]{0.2\linewidth}
\centering 
\includegraphics[height=\linewidth]{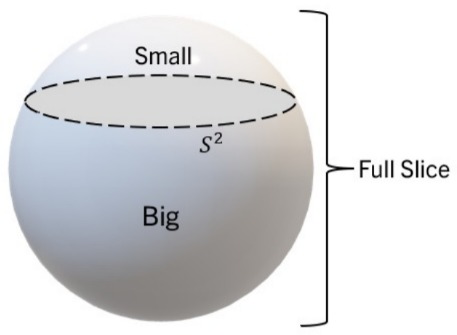}
\caption{}
\end{subfigure}
    \caption{a) The logarithm of the probability density as a function of the curvature ${\cal R}$. We subtracted the same quantity for the fixed area solution \nref{fixedasurf}. We also divided this difference by $\lambda\equiv \frac{8 \pi^{2} \epsilon_{0}M_{pl}^{2}}{H_{r}^{2}}$.  There is a maximum for the small branch at ${\cal{R}} \approx 0.94$. Note the Big Branch is always subleading, since large volumes are suppressed, as we saw already for full slices in \nref{AcLat}.  The curve was computed using the method in \nref{Numerical} with $n_{max}=1000$. b) Lower dimensional cartoon of how to split the $S^{3}$ into a small and big part by the cut of a $S^{2}$(which is a $S^{1}$ in the lower dimensional cartoon). Note that because of the non-trivial embedding \nref{omegaeq} the angle of this sphere is not simply the $\theta$ in global slices.}
    \label{plotactcurv}
\end{figure}

It is a simple matter to find the embedding of such constant curvature slices inside de-Sitter\footnote{For example, we could use coordinate systems where spatial slices have positive or negative curvature.}. The expressions for these slices are particularly simple for large $\tau_r $ 
\be 
\label{omegaeq}
\tau = \tau_r  + \omega\,,\quad \omega =  -\log\bigg(1+\sqrt{1-{\cal R}}\cos \theta\bigg)\,,
\ee 
where we set $\omega =0$ at $\theta=\pi/2$. For the positive curvature case, there is a second branch where we change the sign in front of the square root in \nref{omegaeq}. This is so because by cutting $S^{3}$ via a $S^{2}$ one is left with two portions, one that shrinks to zero as the $S^{2}$ goes to zero, while the other goes to the full $S^{3}$. We call these the small and big branch respectively, which correspond to the $\pm$ signs in front of the square root, and are related by $\theta \leftrightarrow \pi-\theta$, see fig \ref{plotactcurv}b.

As explained in section \ref{dmsubregion},  we need to solve the equations \nref{LargeEqc}, with 
$c_{\ell, m_1,m_2} Y_{\ell, m_1, m_2 } \to c_{\ell} Y_{\ell}(\theta) $ due to the spherical symmetry. 
One can solve this system of equations either numerically, as in \nref{Numerical}, or using the analytical methods of \nref{AnalyticSolution}.   In figure \ref{plotactcurv}, we plot the logarithms of the probability densities as a function of ${\cal R}$. 

An interesting point is that, for a fixed size boundary sphere, there is a value of ${\cal R}$ that maximizes the probability density. Numerically, this is   ${\cal R} \approx 0.94$. Moreover, the action for this solution is  still smaller than the action for the fixed area solution  \nref{fixaaction}. The reason is that when we fix the curvature we are putting a constraint on the whole slice, while the fixed area solution had a constraint only on the sphere. So if we liberate the bulk constraint, then the solution will relax to the fixed area solution \nref{fixedasurf} and the probability will increase.  
We can see this explicitly from the action \nref{OnshellactHH} as follows.  Setting $\varphi=0$ at $\theta=\frac{\pi}{2}$ fixes $c_{0}=-\sum_{\ell=1}^{\infty}c_{\ell}Y_{\ell}(\frac{\pi}{2})$. We plug this back into the action \nref{OnshellactHH} and complete the squares to find
\begin{equation}
\begin{gathered}
\label{sumofsqract}
\log \rho=S_{r} -\frac{8 \pi^{2}\epsilon_{0}M_{pl}}{H_{r}^{2}}\bigg[2 \tau_{r}+\frac{4}{3}-2 \ln 2+\frac{1}{2}\sum_{l=1}^{\infty}\ell(\ell+1)(\ell+2)\bigg(c_{\ell}-\frac{2 Y_{\ell}(\frac{\pi}{2})}{\ell(\ell+1)(\ell+2)}\bigg)^{2}\bigg]\,.
\end{gathered}
\end{equation}
This implies that the probability will be strictly smaller than the one for the fixed area solution \nref{fixedasurf}, where all squares vanish \nref{clpos}. 
In fact, if we define  $\delta c_{\ell}=c_{\ell}-c_{\ell,fa}$, which is the deviation of the angular momentum coefficients from the ones in \nref{clpos} and \nref{clzero}, the mathematical problem of finding $\delta c_{\ell}$ also becomes conceptually simpler since it is that of a massless scalar in \nref{masslesscond} with proper adjustments 
\begin{equation}
\begin{gathered}
\sum_{\ell}\delta c_{\ell}Y_{\ell}(\theta)=-\log\big(1+\sqrt{1-{\cal R}}\cos \theta \big)-\varphi_{\rm fa}(\theta)\,,\quad  \text{for $\theta<\frac{\pi}{2}$}\,,\\
\sum_{\ell}\ell(\ell+1)(\ell+2)\delta c_{\ell}Y_{\ell}(\theta)=0\,,\quad \text{for $\theta>\frac{\pi}{2}$}\,,
\label{eq: equiprobFixcurv}
\end{gathered}
\end{equation}
where $\varphi_{\rm fa}$ can be found in \eqref{phifa}. Note that the last sum in the action in \nref{sumofsqract} becomes that of the free scalar in \nref{ActFi}. 
This 
means that finding the solutions for this fixed curvature problem is essentially the same as finding the solutions for the free scalar field problem.   

\begin{figure}[h!]
\begin{subfigure}[b]{0.45\linewidth}
    \centering
    \includegraphics[width=0.9\linewidth]{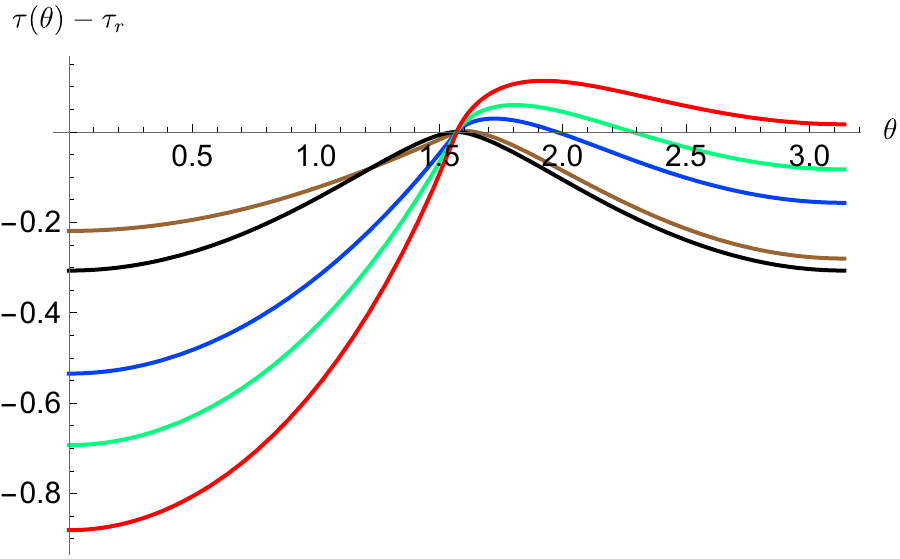}
    \caption{}
\end{subfigure}
\begin{subfigure}[b]{0.5\linewidth}
    \centering
    \includegraphics[width=1.1\linewidth]{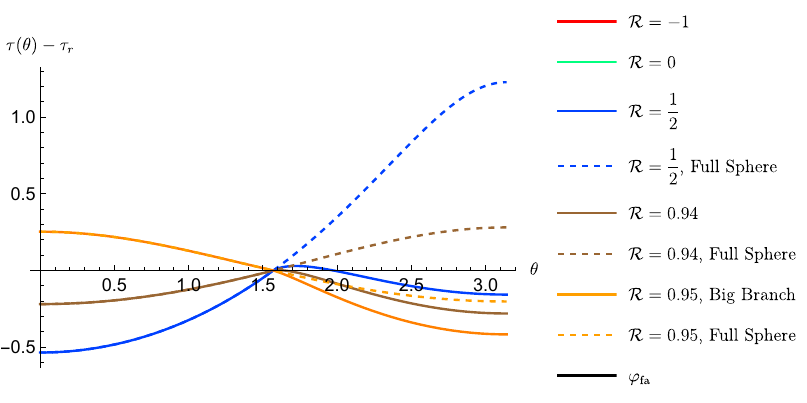}
    \caption{}
\end{subfigure}
    \caption{a) The solid colored curves, are surfaces with a fixed value of ${\cal{R}}$ for $\theta<\frac{\pi}{2}$, but unconstrained everywhere else. The black curve is the $\varphi=0$ surface from the problem in section \nref{FixArea} where we only fix $\varphi$ at $\frac{\pi}{2}$, plotted here for comparison.  
    The solution with constant ${\cal{R}}$ that maximizes the probability in figure \ref{plotactcurv} is shown in brown, and it sits close to the black curve, as expected.
   Note, that the two curves have different scalar curvatures, one is constant and the other is not, see figure  \ref{intriscfixedarea}. b) The dashed lines of the same color are surfaces constrained to have that constant value of curvature everywhere. Note the dashed lines are above the solid lines. This means that the surfaces shrink  for $\theta > \pi/2$, where we relax the curvature constraint.  The orange line corresponds to the big branch solution discussed below \nref{omegaeq}, of the given curvature.}
    \label{plotzerophi}
\end{figure}

It is also interesting to plot the $\varphi=0$ surfaces from these solutions for different values of ${\cal R}$. This tells us how these surfaces are embedded in de-Sitter, see figure \ref{plotzerophi}.  

\begin{figure}[h!]
    \centering
    \includegraphics[width=0.5\linewidth]{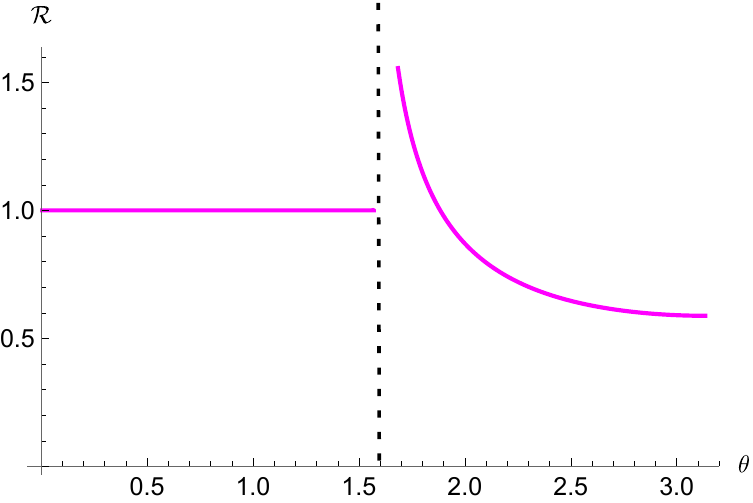}
    \caption{Curvature ${\cal{R}}$ versus $\theta$ for the problem where the curvature is fixed to ${\cal{R}}=1$ for $\theta<\frac{\pi}{2}$. At exactly the transition point $\theta=\frac{\pi}{2}$ the curvature blows up because its shape $\tau(\theta)$ has discontinuous second derivative there. For $\theta>\frac{\pi}{2}$ the curvature is positive but finite, only becoming very big near the transition point. }
    \label{curvtheta}
\end{figure}

It is interesting to understand how they change when the curvature constraints are modified. We can consider two solutions. One is where we fix the curvature on the whole global slice to a value set by ${\cal R}$ in \nref{omegaeq}. The other is one where fix it only inside the two sphere, as we are discussing in this section. In this second case, we find that the $\varphi=0$  surface (or $\phi = \phi_r$) is driven to lower values of the scale factor, see figure \ref{plotzerophi}b, 
reflecting the fact that there is a probability pressure to smaller size. This change is rather fast at the location of the two sphere, so that the curvature blows up there, see figure \ref{curvtheta}. 

Something that should be stressed again is the following. While the calculations performed here answer possible questions about the probability to observe various surfaces, this probability is phenomenologically unreasonable. There is a large probability pressure for the universe to become smaller than the sizes of the surfaces where we are evaluating the probabilities. So, we are computing the probabilities of very unlikely events, at the tail of the probability distribution. In particular, in all these computations we are fixing the area of the boundary of $\Sigma_{\rm in}$ to be very large (many Hubble radii). But the probability becomes larger if we made it smaller. 
The reason we are performing the computation is to explain that the computation of the density matrix is (classically) well defined and one can find the explicit solutions. We are not making any claim that the probabilities discussed here fit the observations in our universe.

\section{Density matrix from  bubble geometries }
\label{sec: bubble}

\subsection{General Discussion and Setup}

In this section, we consider a geometry that arises when we have a more complicated potential than we have been considering so far. We want a potential that does not obey the slow roll condition for some values of the field $\phi$ and therefore can support Euclidean bubble   geometry like the ones considered by Coleman and de Luccia \cite{Coleman:1980aw}. For a qualitative potential of the kind we have in mind see figure \ref{PotentialShape}(b).  

\begin{figure}[h!]
    \centering
    \includegraphics[width=0.7\linewidth]{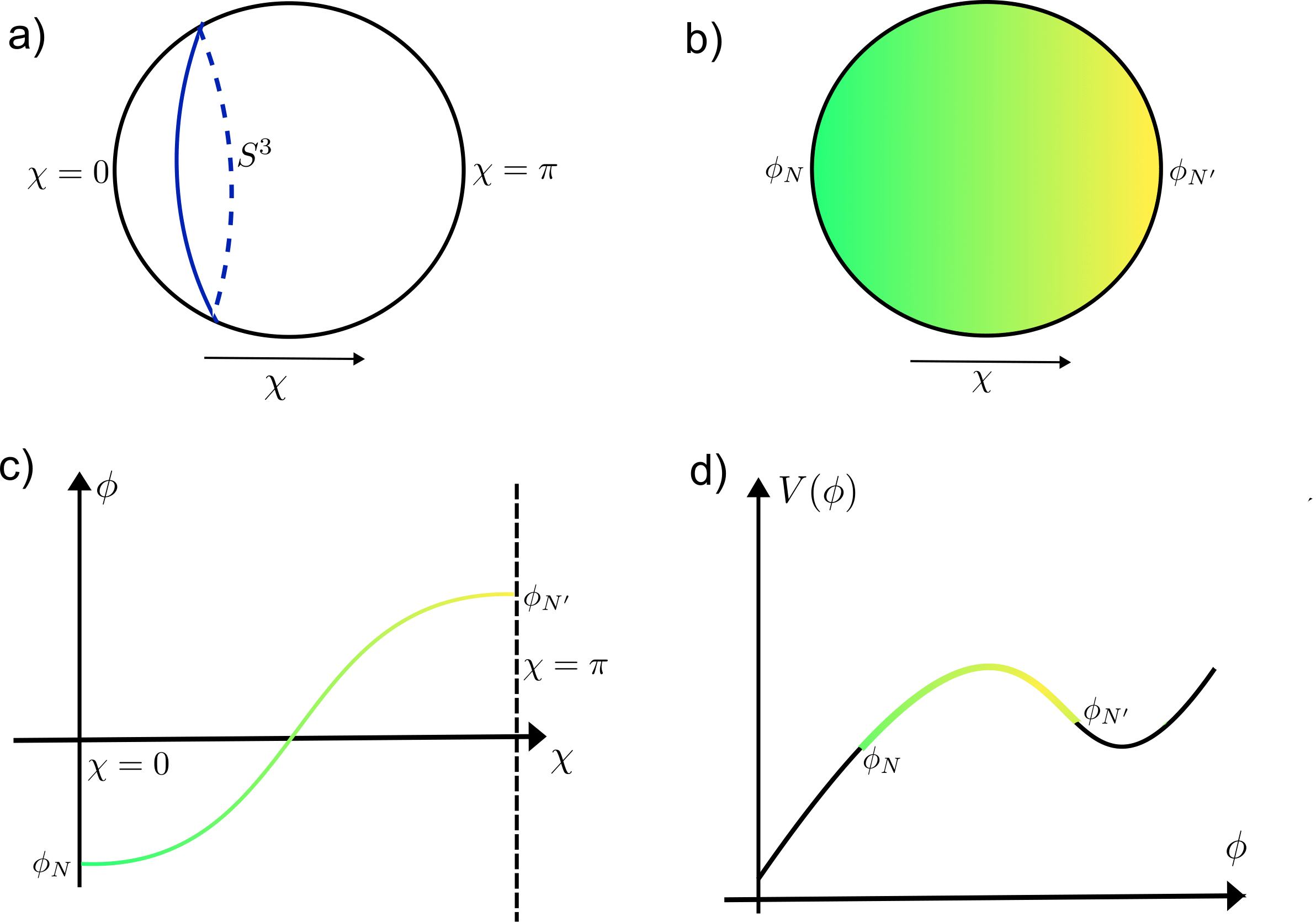}
    \caption{a) Sketch of the Euclidean Bubble geometry, where $\phi$ only changes through the polar coordinate $\chi$. Each surface of fixed $\chi$ is a $S^{3}$, that goes to zero size in the endpoints. b) Bubble solution in the sphere, the gradient from green to yellow parameterizes $\chi$, e.g, the flow from $\phi_{N}$ to $\phi_{N'}$. c) Plot of background Bubble solution $\phi(\chi)$ in the full range d) Bubble potential, with region transversed by the Euclidean bubble colored by the same gradient from b) and c).}  
    \label{BubbleGeosug}
\end{figure}

\begin{figure}[h!]
    \centering    \includegraphics[scale=0.6]{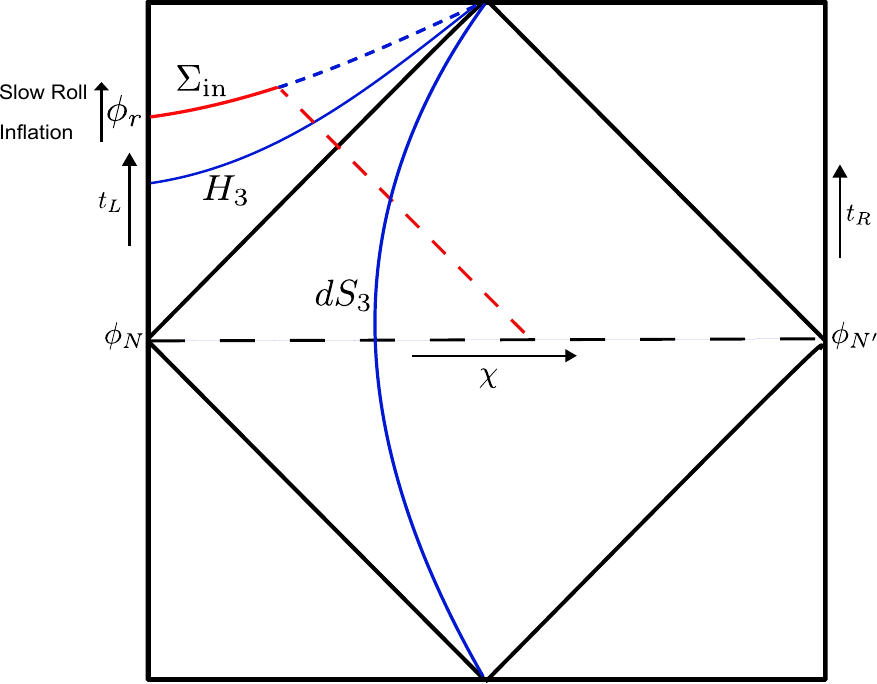}
    \caption{Penrose Diagram of the $dS_{4}$ from the bubble solution. $\Sigma_{\text{in}}$ is the region observed in the density matrix which is at late $t_{L}$ such that the scalar is under slow-roll regime. In principle for the density matrix calculation we can ignore the region outside the light cone of $\Sigma_{\text{in}}$ in the Lorentzian section.}\label{BubbleLor}
\end{figure}

In this case, we have a Euclidean solution with a non-constant value of the field. This is a solution with $SO(4)$ symmetry, a smaller symmetry than the $SO(5)$ symmetry of a round $S^4$, see figure \ref{BubbleGeosug}. These geometries have  round $S^3$ slices. There are two points where these slices shrink smoothly to zero, let us call these two points $N$ and $N'$. We can call the values of the scalar field $\phi_{N}$ and $\phi_{N'}$ at these two points. 
We can continue these solutions to Lorentzian signature where the $SO(4) $ symmetry becomes $SO(3,1)$ and the geometry has both $H_3$ and $dS_3$ slices, see figure \ref{BubbleLor}. 
The forward light cone of the points $N$ or $N'$ can be sliced with $H_3$ slices. The region that is spacelike separated from these two points has $dS_3$ slices. In the thin wall approximation, the bubble wall would sit along one of these $dS_3$ slices. 

We consider potentials where the Lorentzian evolution to the future of the point $N$ involves a period of slow roll inflation. This will happen for potentials of the qualitative form depicted in figure \ref{PotentialShape}(b). We are not particularly interested in what happens to the future of the point $N'$, it could asymptote to a de Sitter space, a flat space, or collapse into a singularity, as it would   if the potential becomes negative in that region. 

Of course, this type of bubble solutions were discussed before as possible descriptions of the early universe in the context of open inflation \cite{Bucher:1994gb,Linde:1995rv,Sasaki:1994yt,Yamamoto:1996qq,Tanaka:1998mp,Garriga:1998he,Yamauchi:2011qq}  or as processes involving tunneling between different vacua of the a possibly very complicated landspace \cite{Lee:1987qc,Blau:1986cw}.
Of course, we could have similar solutions involving many scalar fields. For simplicity, we focus on the case of a single scalar field. 
Our point is simply a reinterpretation of these solutions as contributions to the no-boundary proposal for the density matrix of the universe.   

We are imagining we make observations in the inflationary region after the bubble nucleation.  So we  consider  
 a value of the scalar field, $\phi_r $,  which is further down the potential compared to the value  $\phi_N$, so that we had at least several e-folds of inflation between $\phi_N$ and $\phi_r$. 
We now want to compute the probability that we observe different shapes for the geometry at constant $\phi = \phi_r$. More precisely, we want to consider a finite ball shaped region and ask about the probability for various 
values of the geometry there. In other words, if our universe were to have arisen from this process we would now be observers looking at a finite region of these spatial slices and we want to compute probabilities (or density matrices) for these finite portions, see figure \ref{BubbleLor}. 

Among the different geometries that this ball can take, there is a very special one which is the geometry that this ball takes on the classical solution in figure \nref{BubbleLor}. This is the geometry of a ball shaped portion of the $H_3$ slice that sits at  $\phi =\phi_r$. This slice has a special value of the curvature determined by the classical solution. This classical solution can be continued to the bubble at early times, and is therefore an admissible no boundary geometry.
The no boundary philosophy assigns a probability of the form 
\be 
\rho \sim e^{ - I_{\rm bubble}^E }\,,
\ee 
where $I_{\rm bubble}^E  $ the action of the Euclidean bubble solution in figure \ref{BubbleGeosug}. The full Lorentzian plus Euclidean geometry in this case looks like the one depicted in figure \ref{Geometries}. 

 Since this Lorentzian solution is real, we know from the discussion around \nref{MomMax} that this 3-geometry is an extremum of the probability distribution. In addition, this  probability does not depend on the size of the region or the area of the two sphere bounding the portion of hyperbolic space. Furthermore,  the extremum of the probability distribution is indeed a local maximum. This follows simply from the analysis  in \cite{Yamamoto:1996qq}, which studied the spectrum of primordial fluctuations in open inflation. They found that the spectrum is well defined, with all modes being described by ordinary gaussian distributions. This means that we have a local maximum in the probability distribution. 

Therefore, in this case, we get a phenomenologically acceptable distribution,  as long as we have a sufficiently large number of e-folds of inflation after the Euclidean geometry so that the negative curvature is smaller than the current bounds.  

The ``only'' problem is that the solutions discussed in section \ref{HHSection} actually dominate over the ones we discuss here. So we are not really solving the main problem with the no boundary proposal.

\begin{figure}[h!]
   \begin{center}
    \includegraphics[scale=.6]{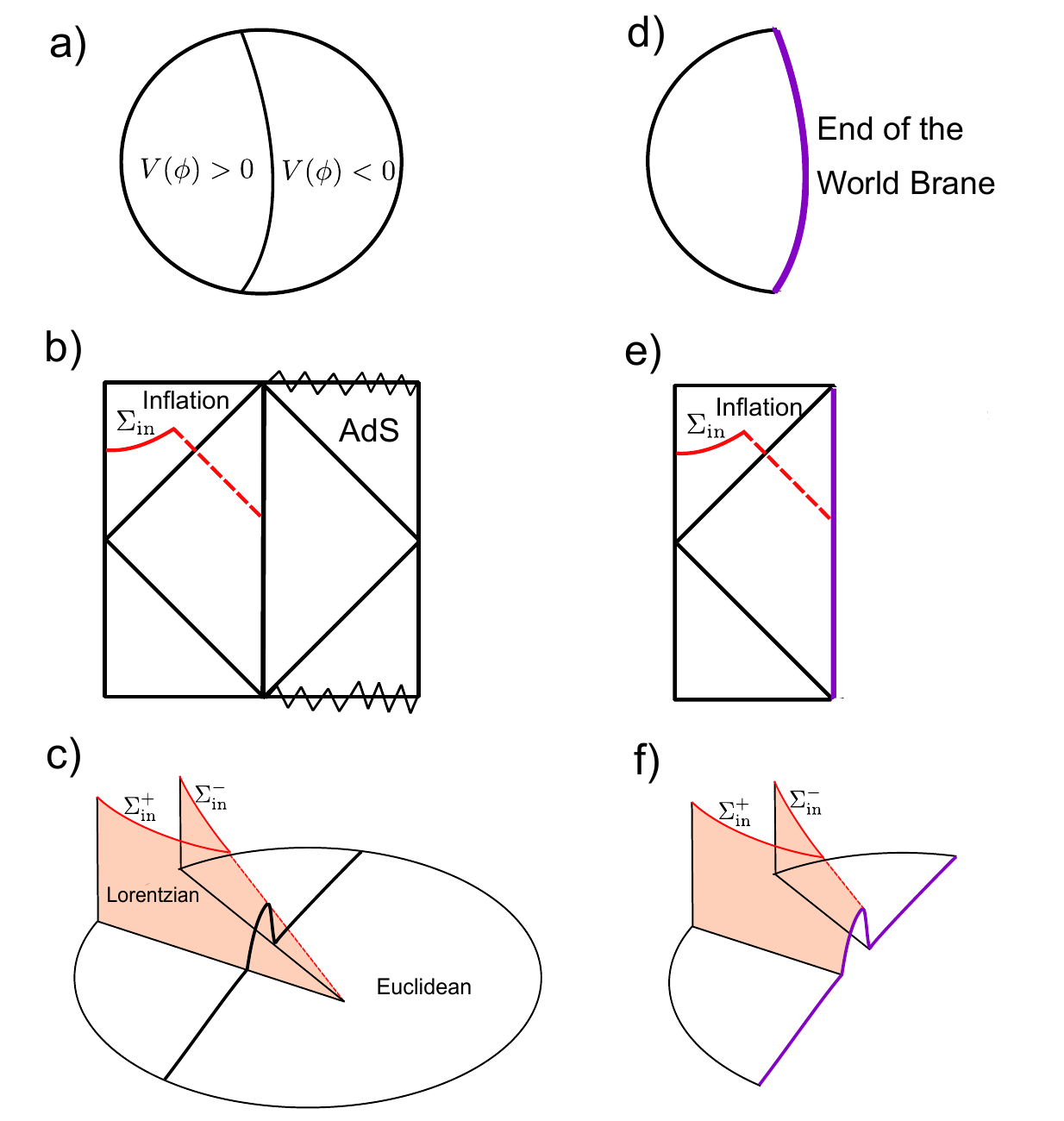}
    \end{center}
    \caption{ 
    a) A bubble solution that interpolates from $dS$ to $AdS$, with the black curve in the middle representing the bubble wall. b) The Penrose diagram for the setup in figure a, the red dashed line is the light cone of an observed region $\Sigma_{\text{in}}$. c) The geometry relevant for computing the density matrix for the region $\Sigma_{\rm in}$, showing both the Lorenzian and Euclidean regions.    d) Bubble solution for de Sitter with an end of the world brane.  e) The Penrose diagram for the Lorentzian geometry resulting from analytic continuation from the Euclidean  geometry in figure d. f) The full Lorenzian and Euclidean geometry that computes the density matrix for region $\Sigma_{\rm in } $.  }
    \label{OtherBubblessug}
\end{figure}

Some of the bubble geometries that we could consider arise when a metastable de-Sitter decays into the inflationary universe we observe. We want to emphasise that we can also consider bubble geometries that do not have such an interpretation. For example, in the thin wall approximation, we can have  an anti-de-Sitter spacetime on the other side. Alternatively we can have an end of the world brane that has nothing on the other side. These cases also contribute to the density matrix of the universe, as long as we get our inflationary universe in the interior of the Lorentzian bubble solution, see figure \ref{OtherBubblessug}.\footnote{As a side comment, the two cases discussed in figures \ref{OtherBubblessug}a and \ref{OtherBubblessug}d can be related to each other if we think that there is a CFT living at the end of the world brane whose dual is an AdS space \cite{Maldacena:2010un}.} 

In the case of the end of the world brane geometry discussed in figure \ref{OtherBubblessug}def, we could add an index to the end of the world brane, as it was done for a similar model in \cite{Penington:2019kki}. If the number of indices, $k$, is sufficiently large (larger than  the exponential of the difference in actions)  then this solution would dominate over the usual Hartle-Hawking like saddle. This might be an interesting mechanism to enhance the dominance of the bubble nucleation solution and should be explored further. 
One potential issue is that the same geometry could be reinterpreted as a geometry that mediates  the decay of the inflationary solution into end of the world branes,  and this process would happen with probability one due to the large number of end of the world branes.  
 
In the following subsections,  we describe in more detail some examples. The results are simply describing the open inflation results \cite{Yamamoto:1996qq}, but in the language of the no boundary proposal, in the same way that 
\cite{Halliwell:1984eu} is restating the results of the usual Bunch Davies wavefunctions in terms of the no boundary proposal.

\subsection{Specific example of a potential with an analytic solution }

\begin{figure}[h]
\begin{subfigure}[h]{0.5\linewidth}
\includegraphics[scale=0.4]{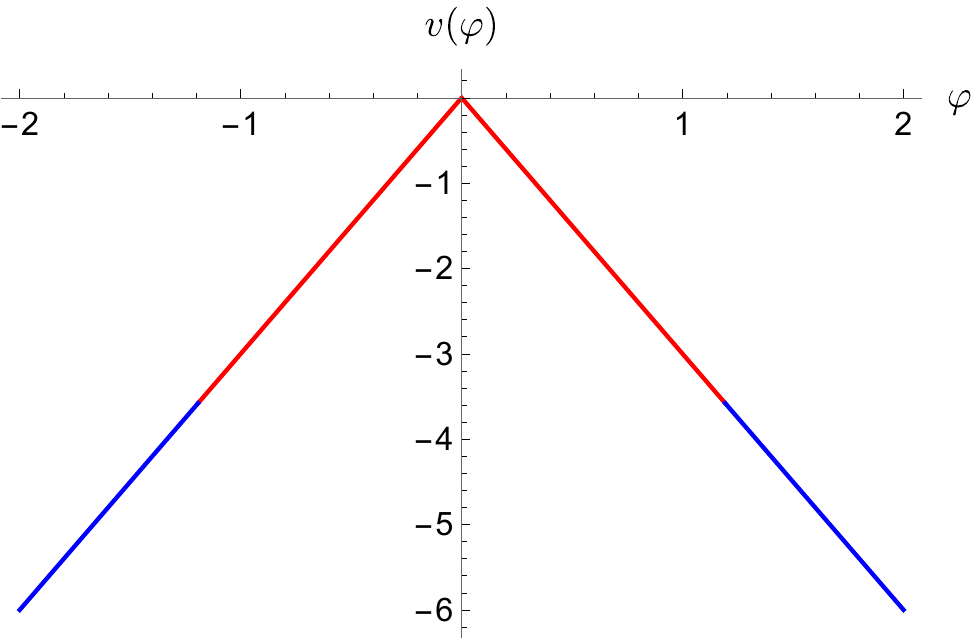}
\caption{}
\end{subfigure}
\begin{subfigure}[h]{0.5\linewidth}
\includegraphics[scale=0.4]{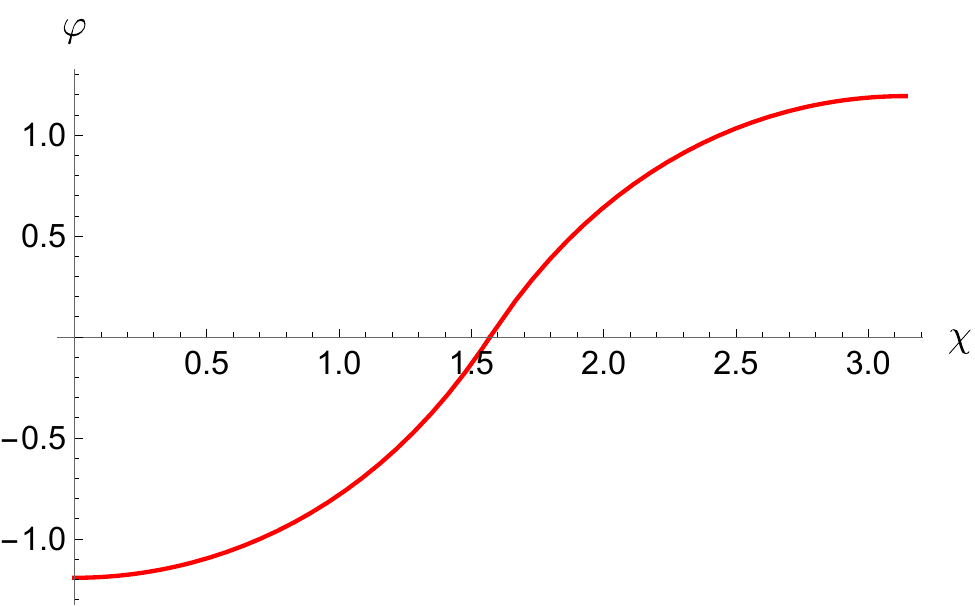}
\caption{}
\end{subfigure}
    \caption{   (a) Simple potential \nref{SimPotLin}. In red we show the region explored by the Euclidean bubble solution \nref{SolLin}. (b) We plot the solution \nref{SolLin}.}
    \label{BubbleLinear}
\end{figure}

In this section, we want to describe an example in more detail.  
We can work in the same approximation that was described around \nref{AcSca} where the metric was very close to the de-Sitter metric but we change the second line of \nref{AcSca} to 
\be 
\label{bubblefirstact}
{M_{pl}^2 \over H_0^2 } 2 \epsilon_0 \int \left[ - \half (\nabla \varphi)^2 - v(\varphi) \right]\,,
\ee 
where $v(\varphi) $ is a  potential  that is asymptotically linear $v \sim  3 \varphi $ for small enough $\varphi$ but it can have some order one feature near $\varphi \sim 0 $, see figure \nref{BubbleLinear} for an example. We have also been imagining that in the zeroth order approximation we have some constant $V_0$ leading to a de-Sitter space with Hubble scale $H_0$. 

In fact, as a particular solvable example can can consider 
\be \la{SimPotLin}
v(\varphi) = - 3  |\varphi | \,.
\ee 

In this case we see that the slow roll approximation is violated at $\varphi =0$. It also goes beyond the thin-wall approximation.
We can explicitly find the bubble solution. The geometry is a four sphere 
\be \la{Sfour}
ds^2 = d\chi^2 + \sin^2 \chi d\Omega_3^2 \,.
\ee 
And the scalar field has the profile 
\bea 
\varphi_{\rm bu} &=& -  { \cos \chi  \over 1 + \cos \chi } - \log (1 + \cos \chi ) ~,~~~~~~~~~~{\rm for } ~~~\chi< { \pi \over 2 }\,,
\cr 
\varphi_{\rm bu} &=&   - { \cos \chi \over 1 - \cos \chi } + \log (1 - \cos \chi ) ~,~~~~~~~~~~{\rm for } ~~~\chi >  { \pi \over 2 }\,,\la{SolLin}
\eea 
which is smooth at both $\chi=0, \pi$ and has the right discontinuity in the second derivative at $\chi = \pi $, where $\varphi_{\rm bu} =0$.

We can compute the action of this bubble  
\be \la{Acib}
i I_b = -I_{b}^E= S_0 +{M_{pl}^2 \over H_0^2 } 2 \epsilon_0  2\pi^2 ( \log 16 - { 5 \over 3 } ) \,,  
\ee 
where $S_0$ is the entropy for a de-Sitter space with Hubble constant $H_0$. 
In potentials with a barrier, we can also consider the Hawking-Moss instanton \cite{Hawking:1981fz}, which is an $S^4$ with a constant scalar field set at the top of the potential barrier (a local maximum). In this case the maximum is at $v=0$ so that the Hawking Moss action is
$iI_{HM} = S_0$. We see that  $iI_b > iI_{HM}$ so that the bubble dominates. 

The continuation of this solution \nref{Sfour} \nref{SolLin} to Lorentzian signature, in the bubble interior region, is obtained by setting $t = - i \chi $. It gives  the spacetime 
\be \la{ClaInBu}
ds^2 = - dt^2 + \sinh^2 t ds^2_{H_3} ~,~~~~~~~~ \varphi_{\rm bu} = { \cosh t \over 1 + \cosh t } - \log (1 + \cosh t ) \,.
\ee 
We see that this has a structure similar to \nref{phic}, but the crucial difference is that now $\varphi$ is real. Therefore $\varphi^+ =\varphi^- $ for this solution. 

More precisely, let us discuss what problem this solution is solving. Let us pick some value  of $\varphi_r$ that is sufficiently negative so that we have a few e-foldings after the Euclidean region. We then attempt to compute the probability that the surface where $\varphi= \varphi_r$ contains an observable region $\Sigma_{\rm in}$ 
 with constant negative curvature, where the geometry of  the three surface is  
\be 
 ds^2 = e^{ 2 \omega } ds^2_{H_3} ~,~~~~~~~~~~ds^2_{H_3} = d\rho^2 + \sinh^2\rho \,d\Omega_2^2\,,
 \ee 
 for $\rho < \rho_r$. This is a region with the topology of a solid ball with constant negative curvature. 

In principle, as in the discussion in section \ref{FixCurv}, we can pick any constant value of $\omega$ that we want. 
It is interesting to consider the special value which is the one on the classical solution \nref{ClaInBu}
  \be \la{RhoZ}
  e^{ 2 \omega_r} = \sinh^2 t_r ~,~~~~~~~~{\rm with } ~~~~~\varphi_r = \varphi_{\rm bu}(t_r) =   { \cosh t_r \over 1 + \cosh t_r } - \log (1 + \cosh t_r )\,,
  \ee 
  where the second equation is determining the time $t_r$ where $\varphi_{\rm bu}(t) = \varphi_r $ according to the classical solution \nref{ClaInBu}. The first equation in \nref{RhoZ} is then setting $\omega$ to the special value $\omega_r$ that the solution \nref{ClaInBu} has at $t=t_r$. 
  
\begin{figure}
\begin{subfigure}[t]{0.32\linewidth}
    \includegraphics[width=\linewidth]{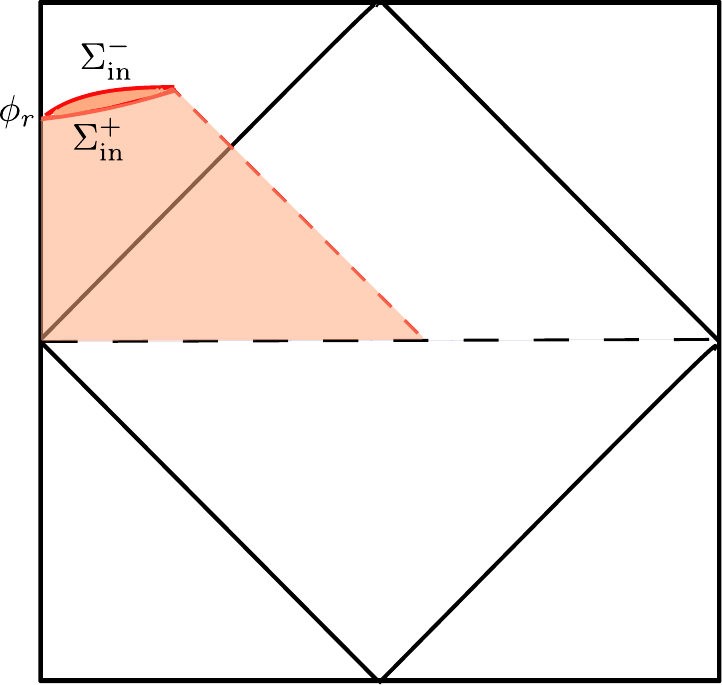}
    \caption{}
\end{subfigure}
\begin{subfigure}[t]{0.32\linewidth}
    \includegraphics[width=\linewidth]{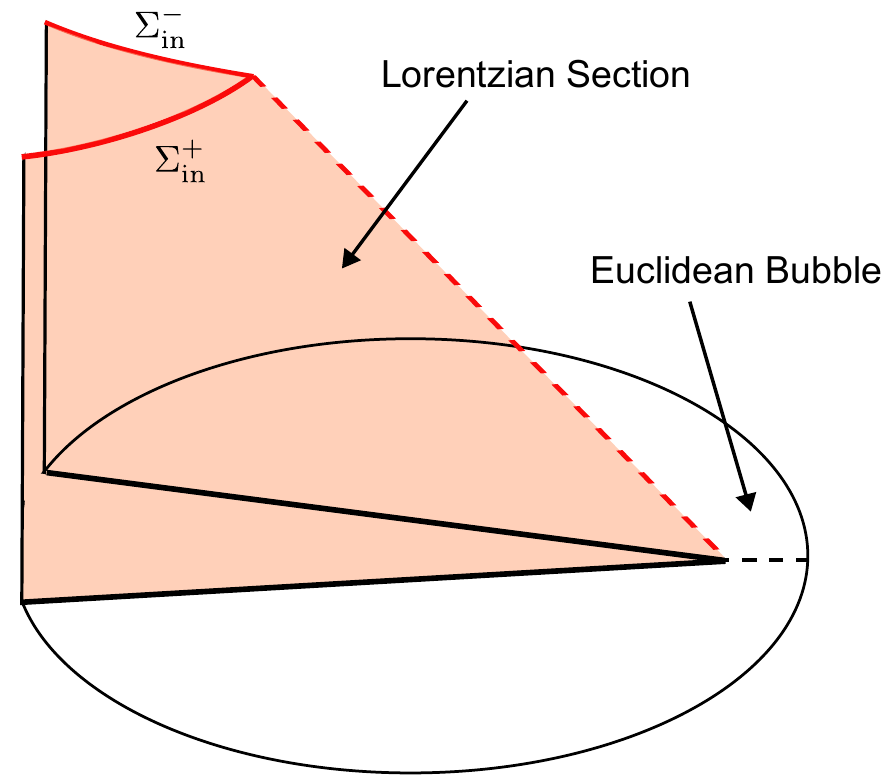}
    \caption{}
\end{subfigure}
\begin{subfigure}[t]{0.32\linewidth}
    \includegraphics[width=\linewidth]{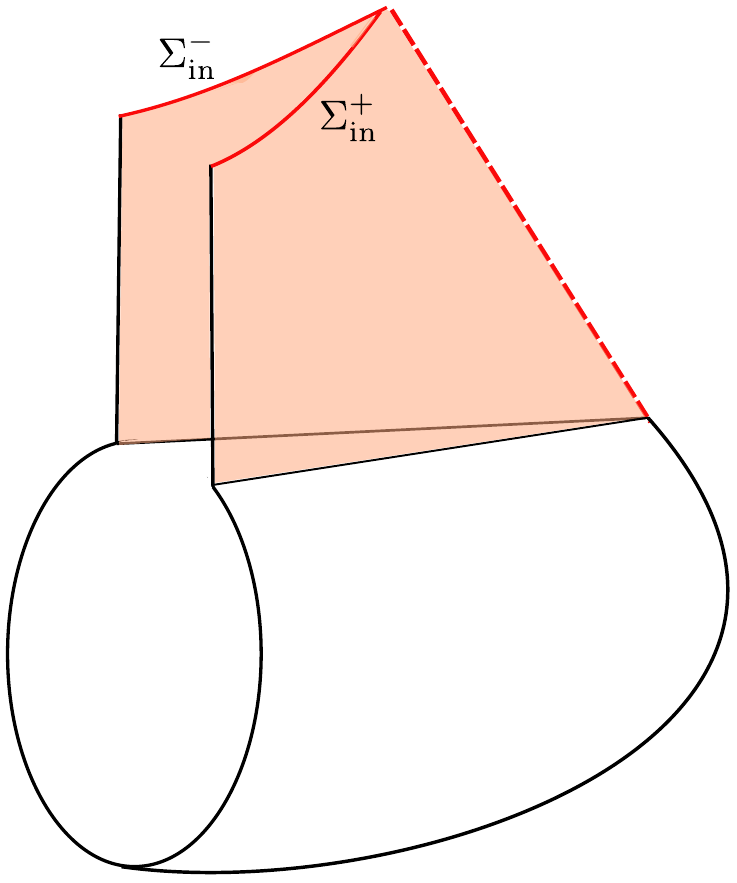}
    \caption{}
\end{subfigure}
\caption{(a) The full extended real solution.   (b) The part of the solution that really matters for the density matrix of $\Sigma_{\rm in }$. (c) Another representation of the solution which makes the connection with thermal-like states more manifest. In this representation the Euclidean deformations of the bra and ket geometries get connected in Euclidean time. }
    \label{fig:enter-label}
\end{figure}

Then the no boundary geometry that computes the (diagonal) component of the 
density matrix for this particular three geometry, is given by a portion of the Lorentzian geometry we discussed above, together with the Euclidean bubble, see figure \ref{fig:enter-label}. 
The action of this geometry is 
\be \la{DensOmr}
\rho( \omega_r , \omega_r) \sim e^{ - I_b^E}\,,
\ee 
where $I_b^E$ is the action of the Euclidean bubble.  The Lorentzian parts do not contribute because the contributions from the $+ $ and $-$ sheets cancel out. This type of cancellation is, of course, true in any real Lorentzian solution.  In particular, \nref{DensOmr} is independent of $\rho_r$, so that we get the same answer for any size of the ball (at least classically).  



So far we discussed the density matrix evaluated in a very particular 3-geommetry. We can change the 3-geometry and compute the corresponding solutions. A simple situation is when the 3-geometry differs from the one defined in \nref{RhoZ} by a small amount, $\omega = \omega_r + \delta \omega$. The answer to this is that the fluctuations $\delta \omega$ have a gaussian wavefunction centered on $\delta \omega =0$. This is the problem studied in \cite{Yamamoto:1996qq}. In the next subsection we explain in more detail how this can be done.

\subsection{Density matrix for fluctuations around the Lorentzian continuation of the bubble solution   }


\label{dmbub}

\begin{figure}
\begin{subfigure}[h]{0.5\linewidth}
    \centering
    \includegraphics[width=0.8\linewidth]{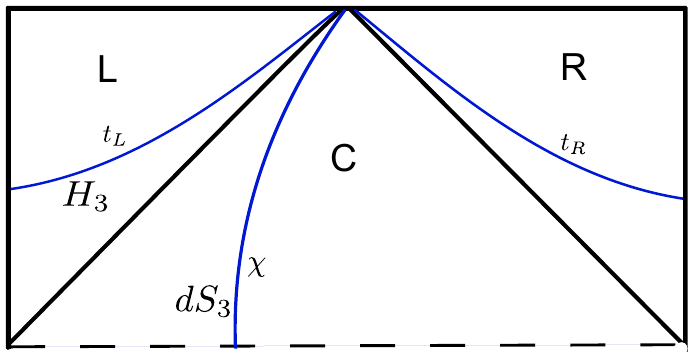}
    \caption{}
\end{subfigure}
\begin{subfigure}[h]{0.5\linewidth}
    \centering
    \includegraphics[width=0.8\linewidth]{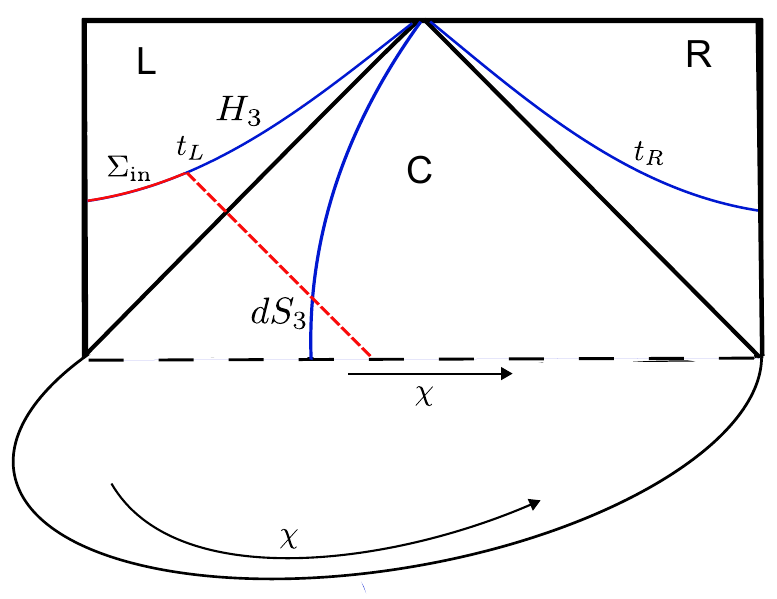}
    \caption{}
\end{subfigure}
    \caption{a) Penrose diagram of the Lorenztian sector of the manifold, the slices of fixed $t_{L}$, $t_{R}$ and $\chi$ are shown in the respective regions they are defined in. b) Penrose diagram of $dS_{4}$ with the addition of the Euclidean bubble that prepares the state for the Lorentzian section.}
    \label{CoordinateRegions}
\end{figure}

In this section we discuss the computation of the density matrix around the basic real solution.   Namely, we consider boundary conditions where the scale factor of the metric differs by a small amount relative to the one in \nref{RhoZ}, $\omega = \omega_r + \Delta \omega$. This can be analyzed by considering  small fluctuations around the original real solution, $\varphi = \varphi_{\rm bu} + \delta \varphi $. 
Expanding the action we find the action for the small fluctuations 
\be 
I = { 2 \epsilon_0 M_pl^2 \over H^2 } \int \left[ - \half (\nabla \delta \varphi ) - \half v''( \varphi_{\rm bu}) \delta \varphi^2 \right]\,.
\ee 
In the case that we neglect the backreaction on the geometry, it is convenient to use the following coordinates, see figure \ref{CoordinateRegions}, 
\bea 
ds^2 &=& -dt_L^2 + \sinh^2 t_L ( d\rho^2+ \sinh^2 \rho  d\Omega_2)^2\,,  ~~~~~~~{\rm in ~region ~L}
\cr 
ds^2 &=& d\chi^2 + \sin^2\chi  ( -d \hat \tau^2  + \cosh^2\hat \tau  d\Omega_2)^2 \,, ~~~~~~~{\rm in ~region ~C}
\cr 
ds^2 &=& -dt_R^2 + \sinh^2 t_R ( d\rho^2+ \sinh^2 \rho d\Omega_2)^2\,,  ~~~~~~~{\rm in ~region ~R}
\cr 
ds^2 &=& d\chi^2 + \sin^2 \chi ( d\theta^2 + \sin^2 \theta d\Omega_2)^2 \,, ~~~~~~~{\rm in ~the ~Euclidean~ region  }
\eea
These coordinates are connected as follows in the $\pm$ sheets 
\be \la{Sheets}
t_L = \pm i \chi ~,~~~~~~~t_R = \pm i (\pi - \chi) ~,~~~~~~~ \rho = \mp i \theta ~,~~~~~~~~\hat \tau = \pm i ({ \pi \over 2 } - \theta )\,.
\ee 

The last continuation is slightly unconventional, we do it like this so that the condition that the fields are smooth at $\hat \tau = \pm i {\pi \over 2} $ translates into the conditions that the  fields are smooth at $\theta=0$, see the last expression in \nref{YHarm}. We can imagine a contour in the complex plane that joints smoothly these solutions where around $t=0$ we deform it in a way that is consistent with \nref{Sheets}. 

In general, we also have angular momentum $l$ on $S^2$. For simplicity we will just discuss the case when the angular momentum on $S^2 $ is zero, $l=0$. 
This means that we consider spherically symmetric profiles in the observable region. The general case can be done without many changes. 

Then we can pick eigenfunctions 
\be \la{YHarm}
Y_p(\rho) = { \sin p \rho \over \sinh \rho } ~,~~~~~~~~~~ Y_p(\theta)  = { \sinh p\, \theta  \over \sin \theta } ~,~~~~~~~~
Y^\pm_p (\tau ) = { \sinh p ( { \pi \over 2 } \pm i \hat \tau) \over \cosh \hat \tau }\,,
\ee 
which obey 
\be \la{YHarmEq}
\nabla^2_{H_3} Y_p = - ( p^2 +1 ) Y_p ~,~~~~~~~~~\nabla^2_{S^3} Y_p =   ( p^2 +1 ) Y_p ~,~~~~~~\nabla^2_{dS^3} Y^\pm_p=  ( p^2 +1 ) Y_p^\pm\,,
\ee 
where $p$ can be related to the $S^3$ angular momentum by $\ell=-1\pm i p$. All eigenfunctions \nref{YHarm} continue into each other. 

The idea is to expand 
\be \la{Expansi}
\delta \varphi^\pm  = \int dp \,c^{ \pm , \dot a } (p) f_{p}^{\pm , \dot a }(t) Y_p(\rho) \,,
\ee 
where $\dot a = \dot \pm $ are two choices of solutions in each of the regions. For example, we can choose them to be solutions that behave as $t_L^{ -1 \dot \pm i p } $ near $t_L\sim 0$. Once they are defined there, they can be analytically continued through the other branches. 
More precisely, these functions obey the equations 
\bea 
\label{bubfluceom}
0&=&{ 1 \over \sinh^3  t}{ d \over d t } \left[ \sinh^3 t {d f_{p} \over d t } \right] + \left[ \delta v'' - { (p^2 +1) \over \sinh^2 t }\right]  f_p =0 ~,~~~ 
\cr 
0&=&{ 1 \over \sin^3  \chi}{ d \over d \chi } \left[ \sin^3 \chi {d f_{p} \over d \chi } \right] +\left[ - \delta v'' - { (p^2 +1) \over \sin^2 \chi }\right] f_p=0\,,
\eea 
in each of the regions. 
We demand that the solutions continue properly under \nref{Sheets}, this gives rise to the index $\pm$ in $f$.  There are two choices of solutions in each region that give rise to the index $\dot \pm $.

In \nref{Expansi} we have neglected the so called ``wall motion'' modes, see \cite{Hamazaki:1995dy,Yamamoto:1996qq}, since they do not contribute to fluctuations of the scalar curvature, but they are important for tensor fluctuations \cite{Garriga:1996pg,Garcia-Bellido:1995ruf}.  In particular, those wall modes are Goldstone bosons that arise from the spontaneous breaking of the symmetry from $SO(5)$ to $SO(4)$ by the bubble solutions. These modes belong to the type of massive (actually tachyonic in our case) Goldstone bosons discussed in \cite{Watanabe:2013uya}. Actually, we only get modes with angular momentum larger than two because the lowest modes corresponds to gauge symmetries.   See appendix \ref{app: wall mode} for further discussion.  

We start with four functions $f^{\pm, \dot \pm}$. 
Since we plan to trace out the right region, we can set the $\pm $ functions to be equal in the right region. This halves the number of functions.   In addition, if we plan to evaluate only the diagonal  density matrix elements of left region, then we can set the asymptotic expression of the $\pm $ fields in the left region also to be equal.   This then leaves only one independent function for each $p$. This procedure is detailed in appendix \ref{SpecBub}.

Expanding the solutions for large $t_L$ we find that 
\be 
\label{phibigsmall}
\delta \varphi^\pm  = \int dp c(p) Y_p(\rho) \left[ 1 \mp i \frac{8}{3}  \mu(p) e^{ - 3 t_L} \tanh(\pi p) p (p^2+1) \right ] \,,
\ee  
where we have neglected subleading terms that are equal for the $\pm$ indices. 
where $\mu(p) $ is obtained by solving the radial equation with the above boundary conditions, tracking carefully how the solution transforms in the $+$ vs the $-$ sheets. 

If we want to trace out a region, then the problem that we need to solve is by now familiar 
\bea  
&~ &  \int dp\, c(p) Y_p(\rho) = \delta \varphi^{\rm in}_b = \delta \omega^{\rm in }_b ~,~~~~~~~~{\rm for } ~~~~~~~ 0 \leq \rho \leq \rho_r \,,
 \cr 
 &~ &  \int dp\, c(p) \mu(p)  Y_p(\rho) =   0 ~,~~~~~~~~~~~~~~~~~{\rm for } ~~~~~~ \rho_r < \rho \,. 
 \la{EquSoBB} 
 \eea

 Once we find these $c(p) $ we determine $\delta \varphi $ on the whole slice. In other words, we find also $\delta \varphi_{\rm out} $. Once we find it, we can then insert this into the action which gives us  
 \be \la{RhoOI}
 \log \rho[ \omega^{\rm in}_b , \omega^{\rm in}_b] \sim  - 2\pi^{2}  \int dp\, c(p)^2 \mu(p) \tanh(\pi p) p (p^2 +1) \,.
 \ee 
for the values of $c(p) $ that solve \nref{EquSoBB}.

\section{Conclusions}

In this paper, we have discussed a no-boundary proposal for the density matrix of a subregion of the universe. 
The motivation was to phrase the no-boundary proposal in terms of quantities that are observable to a physical observer such as ourselves, see figure \ref{RegionObs}.  We asked questions about the nature of the surface of constant inflaton (or constant time) near the end of inflation. The Hartle Hawking measure wants this surface to be a Hubble size sphere at that time. On the other hand, we have put in a projector which selects the unlikely component when this surface contains a large number of Hubble patches at this time, so that we can talk about the  superhorizon   shape of the surface. We have imposed this by constraining the area of the surface of a large ball shaped region. Subject to this constraint, we have then computed the probability for observing various shapes for the geometry of such regions.  This is a somewhat unphysical question due to the presence of this constraint. This is an unfortunate feature of the Hartle-Hawking measure, which will hopefully be solved in the near future. 

We have also considered geometries of the bubble nucleation type where it is possible to find local maxima in the probability that allow for a large universe at the time of reheating, a scenario discussed in the open inflation context \cite{Bucher:1994gb,Linde:1995rv,Sasaki:1994yt,Yamamoto:1996qq,Tanaka:1998mp,Garriga:1998he,Yamauchi:2011qq}  as well as in the landscape \cite{Lee:1987qc,Blau:1986cw}. This a reasonable probability distribution except for the fact that the contribution discussed in the previous paragraph is also present and it appears to dominate.

We emphasized that we could impose the ``trace-out'' boundary conditions in the past lightcone of the subregion. However, when we performed explicit computations, we actually imposed such boundary conditions at a fixed $\tau$ slices. This was a purely technical step that enabled us,   to use the standard method of separation of variables. But the idea that the boundary conditions could be imposed along the past lightcone is more economical from the purely theoretical and conceptual point of view, since it eliminates the unobservable region of the universe from the discussion. This fact could be useful for discussions of the measure problem in eternal inflation \cite{Guth:2007ng}.

Though we discussed just the classical geometries,   in principle, we should also include the quantum corrections. For these quantum corrections, the sharp boundary condition at  $\Sigma_{\rm in} $ will pose problems. In fact, there are two types of problems. First, the concept of a wavefunctional in quantum field theory is subject to UV divergences due to the specification of the value of the fields at super short distances along $\Sigma_{\rm in}$. This problem can be cured by tracing out over short distances when we define the density matrix, so that we focus on sufficiently long distances  along the $\Sigma$ directions (superhorizon distances). The second problem is that we need to split the quantum field theory vacuum in two parts, at the boundary of the region $\Sigma_{\rm in}$.\footnote{A density matrix for a finite region of the universe was defined in section 3.1 of \cite{Dong:2020uxp}. They identified the bra and the ket with two complementary regions in the same universe. This is an example of a  density matrix that does not behave as expected near the edges of the interval, it has the twice the expected Unruh temperature.} We expect that the careful analysis of the von Neumann algebras in   \cite{Kudler-Flam:2024psh,Chen:2024rpx} addresses this issue.

Once we have the density matrix,  a natural object to compute is the entropy. We leave a full exploration of this question to the future. Of course, if we only think about the small fluctuations, the entropy computation will be the same as the entropy computation of quantum fields in a large superhorizon size region, such as what was discussed in \cite{Maldacena:2012xp}. More interestingly, one would like to understand more precisely what we should keep fixed in such computations. We could conceivably keep the area of $\Sigma_{\rm in}$ fixed as we compute the entropy. If there is no condition on the angle around this surface, then we expect that the $n^{th}$ replica is given by the same geometry that computes the single replica. In that case, the entropy will be given by the action  we have computed in the case that we only fixed the geometry of the boundary of $\Sigma_{\rm in}$, see \nref{fixaaction}. 
In addition, we should understand more precisely the effects of the quantum contributions to the density matrix. 

Let us briefly comment on the connection to the recent papers \cite{Chen:2024rpx,Kudler-Flam:2024psh} discussing density matrices in cosmology using algebraic field theory constructions. The first comment is that the trace that these papers define corresponds to what we would call the ``Hartle Hawking state''. In the linear potential approximation that we are using, this state is not a normalizable state. This lack of normalizability is due to the exponential preference for a smaller number of e-folds  \nref{UnProba}. \footnote{Of course, if the full potential has a positive minimum, then the Hartle Hawking state is indeed normalizable. However, we are now zooming into a region of the inflationary potential. So the non-normalizability simply means that we are driven away from this region. } The  papers \cite{Chen:2024rpx,Kudler-Flam:2024psh}  then consider states that are normalizable. A particular way to choose a state would consist of picking a projection operator in the algebra whose trace is finite. This is physically similar to our projection onto a configuration with a given area. It seems it should be possible to find a direct connection between our formalism and the one discussed in   \cite{Kudler-Flam:2024psh}. One point that we are making here is that the Hartle-Hawking philosophy, specially as articulated in \cite{Hartle:2007gi,Hartle:2008ng}, is that we make the observations in the future. From that point of view the ``state'' that appears in \cite{Chen:2024rpx,Kudler-Flam:2024psh} could be viewed as a manifestation of a selection condition on the Hartle-Hawking state.

As a side comment, suppose that we think that the observations at the reheating surface are given by a CFT (as in dS/CFT), or a suitable 
$T^2$ deformation, see e.g. \cite{Araujo-Regado:2022gvw,Batra:2024kjl}.   Then, computing the density matrix of a subregion, $\Sigma_{\rm in}$ should correspond to considering the two copies of this CFT, the one for the ket and the one for the bra, and joining them at the boundary of the region $\Sigma_{\rm in}$ with a suitable boundary condition. In this way, we only need to consider the dual field theory in the finite region $\Sigma_{\rm in}$ with a boundary, and not outside of this region.\footnote{ A related comment is that if we consider the state of the universe as being given by a MERA-like collection of isometries \cite{SinaiKunkolienkar:2016lgg,Bao:2017iye,Cotler:2022weg} (or unitaries acting on a tensor product of $|0\rangle $ state ancillas), then, when we consider only a subregion, we only need to consider the tensors that are in the so called ``causal-cone'' of the subregion \cite{Vidal:2008zz}.  }

A different aspect of the density matrix  in cosmology was discussed in \cite{Page:1986vw,Hawking:1986vj,Chen:2020tes,Fumagalli:2024msi}. These discuss possible geometric connections between the bra and the ket in the Euclidean past, even in the case where $\Sigma_{\rm in}$ is the full spatial slice. Of course, it would be interesting to understand whether such geometries are also important for our universe.

\subsection{Speculative comments about the case involving slow roll eternal inflation}

\begin{figure}[h!]
    \centering
    \begin{subfigure}[b]{0.46\textwidth}
        \centering \hspace{0mm}\def\svgwidth{82mm}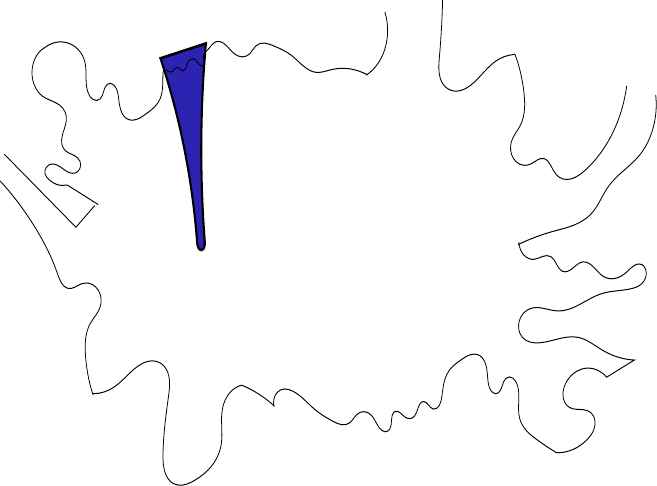
        \caption{}
    \end{subfigure}
    \hfill
    \begin{subfigure}[b]{0.46\textwidth}
        \centering
\hspace{0mm}\def\svgwidth{82mm}
\begingroup%
  \makeatletter%
  \providecommand\color[2][]{%
    \errmessage{(Inkscape) Color is used for the text in Inkscape, but the package 'color.sty' is not loaded}%
    \renewcommand\color[2][]{}%
  }%
  \providecommand\transparent[1]{%
    \errmessage{(Inkscape) Transparency is used (non-zero) for the text in Inkscape, but the package 'transparent.sty' is not loaded}%
    \renewcommand\transparent[1]{}%
  }%
  \providecommand\rotatebox[2]{#2}%
  \newcommand*\fsize{\dimexpr\f@size pt\relax}%
  \newcommand*\lineheight[1]{\fontsize{\fsize}{#1\fsize}\selectfont}%
  \ifx\svgwidth\undefined%
    \setlength{\unitlength}{315.18510233bp}%
    \ifx\svgscale\undefined%
      \relax%
    \else%
      \setlength{\unitlength}{\unitlength * \real{\svgscale}}%
    \fi%
  \else%
    \setlength{\unitlength}{\svgwidth}%
  \fi%
  \global\let\svgwidth\undefined%
  \global\let\svgscale\undefined%
  \makeatother%
  \begin{picture}(1,0.73967401)%
    \lineheight{1}%
    \setlength\tabcolsep{0pt}%
    \put(0,0){\includegraphics[width=\unitlength,page=1]{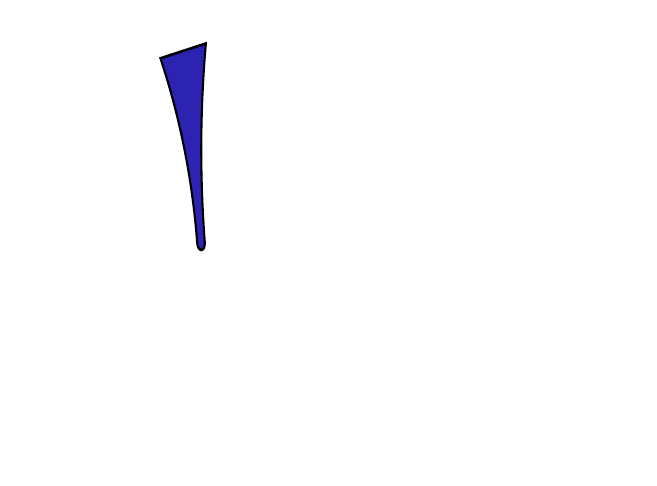}}%
    \put(0.06480509,0.43067786){\color[rgb]{0,0,0}\makebox(0,0)[lt]{\lineheight{1.25}\smash{\begin{tabular}[t]{l}Trace-out\end{tabular}}}}%
    \put(0.37103994,0.54857843){\color[rgb]{0,0,0}\makebox(0,0)[lt]{\lineheight{1.25}\smash{\begin{tabular}[t]{l}Trace-out\end{tabular}}}}%
  \end{picture}%
\endgroup%

        \caption{}
    \end{subfigure} 
    \caption{   (a) A sketch of a situation where our observable part of the universe is embedded in a slow roll  eternally inflating geometry. Our observable part is shaded in blue, together with its past trajectory.  (b) In order to compute the density matrix of the observable region we only need to concentrate on the past lightcone of the observable region, so that the rest of the universe is not relevant.  }
\la{SlowEternal}
\end{figure}

One particular case where the quantum corrections might lead to interesting results is the case of slow roll eternal inflation. More precisely, we can imagine a potential, such as $m^2 \phi^2$, that has a region of eternal inflation and a region of ordinary inflation \cite{Linde:1986fd}. In such a case, we could contemplate the situation where our observable region $\Sigma_{\rm in}$ is part of a very large region $\Sigma$, so large that the universe was in the eternal inflating region in the past. As we discussed in section \ref{HHSection} , the classical solution always exerts a probability pressure for the universe becoming smaller. However, it is  conceivable that some quantum corrections could become important and lead to a local maximum of the probability.  Alternatively, it could be that we can do the functional integral over the small fluctuations without having to talk about an on shell solution, since the geometry is close to an on shell solution. The main quantum integration variable seems to be the number of e-folds since the Euclidean region, see figure \ref{SlowEternal}. 
It is possible, that in analogy to some other situations, such as near extremal black holes \cite{Engelsoy:2016xyb,Jensen:2016pah,Maldacena:2016upp,Stanford:2017thb}, this functional integral can be done exactly over the relevant subset of variables. 

In fact, one expects that bubble nucleation eternal inflation and slow roll eternal inflation should not be conceptually too different. Since we discussed a contribution to the density matrix from the bubble nucleation regime one might naively expect also one in the slow roll eternal regime. 

Understanding these issues seems necessary in order to find out whether an observer selection mechanism along the lines of that proposed in \cite{Hartle:2010vi}  works or not.



\subsection*{Acknowledgments}

We would like to thank R. Bousso, D. Harlow, T. Hertog, L. Iliesiu, D. Jafferis, J. Kudler-Flam, S. Leutheusser, G. Penington, G. Satishchandran, D. Stanford and E. Witten for discussions. V.I. would like to thank Jeff Shen for useful discussions about numerics.    

J.M. is supported in part by U.S. Department of Energy grant DE-SC0009988.  

The work of Y.Z.L is supported by the US National Science Foundation under Grant No. PHY- 2209997

\appendix

\section{Density matrices on finite size slices }

\la{FiniteTrApp}

In this appendix we discuss the generalization of some of the formulas to surfaces that are {\it not} in the asymptotic future or do not involve distances always at superhorizon scales. This means in practice that we will generalize the formulas to finite values of $\tau_r$. 

\subsection{Density matrix on a spherical ball}
\la{DMCapfintau}

The discussion of \nref{DMCap} at finite $\tau_{r}$ is still very similar to the large $\tau_{r}$ one. It is slightly more convenient to expand the fields as
\begin{equation}
\label{homexpfintau}
\varphi^{\pm}=\sum_{l}c_{\ell}^{\pm}Y_{\ell}(\theta)\frac{f_{\ell}^{\pm}(\tau)}{f_{\ell}^{\pm}(\tau_{r})}\,, 
\end{equation}
so that the prefactor of $c_{\ell}Y_{\ell}(\theta)$ is one at $\tau=\tau_{r}$. Again, $c_{\ell}^{+}=c_{\ell}^{-}$ for diagonal components of the density matrix, and the $c_\ell$ are real. For $\theta>\theta_{0}$ we must also impose that $\partial_{\tau}\varphi^{+}-\partial_{\tau}\varphi^{-}=0$. This is slightly different at finite $\tau_{r}$, using
\begin{equation}
i \cosh^{3}\tau_{r} \bigg(\frac{\partial_{\tau}f_{\ell}^{+}(\tau_r)}{f_{\ell}^{+}(\tau_r)}-\frac{\partial_{\tau}f_{\ell}^{-}(\tau_r)}{f_{\ell}^{-}(\tau_r)}\bigg)  =- \frac{2   \ell(\ell+1)(\ell+2)}{|f_{\ell}(\tau_{r})|^{2}}\,.
\end{equation} 
we arrive at the final form of the equations that determine $c_\ell$
\begin{align}
\label{masslesscondfintau}
& \sum_{\ell=0}^\infty c_{\ell}Y_{\ell}(\theta)=\varphi_{b}(\theta)\,,\quad ~~~~~~~\text{for}\,\, 0 \leq \theta\leq\theta_{0}\,,\nonumber\\
& \sum_{\ell=0}^\infty\frac{\ell(\ell+1)(\ell+2)}{1+\frac{\ell(\ell+2)}{\cosh^{2}\tau_{r}}}c_{\ell}Y_{\ell}(\theta)=0\,,\quad \text{ for}\,\,  \theta_{0} < \theta \leq \pi  \,.
\end{align}

This system of equations can be solved numerically as discussed in appendix \ref{Numerical}, but no analytic solution was attempted. 

We get the density matrix for this configuration by computing the first line in the action \nref{WaFu1}. 
\begin{equation} \la{ActFifintau}
\log \rho[ \varphi_b(\theta), \varphi_b(\theta)] \sim -2 \pi^{2} \sum_{\ell}\frac{\ell(\ell+1)(\ell+2)}{1+\frac{\ell(\ell+2)}{\cosh^{2}\tau_{r}}}c_{\ell}^{2}\,,
\end{equation}
where the $c_\ell $ depend on $\varphi_b(\theta) $ through the equations in \eqref{masslesscondfintau}. 

Note that for large $\tau_r$ this is the same expression as we had in \nref{DenMaF}. This is to be expected, since the tracing out procedure is done via saddle point, so that in the end we are evaluating the same action on a particular configuration, namely the solution of \nref{masslesscondfintau}. 

\subsection{Quick review of the spatially homogeneous solution}

\begin{figure}[h]
   \begin{center}
    \includegraphics[width=0.45\textwidth]{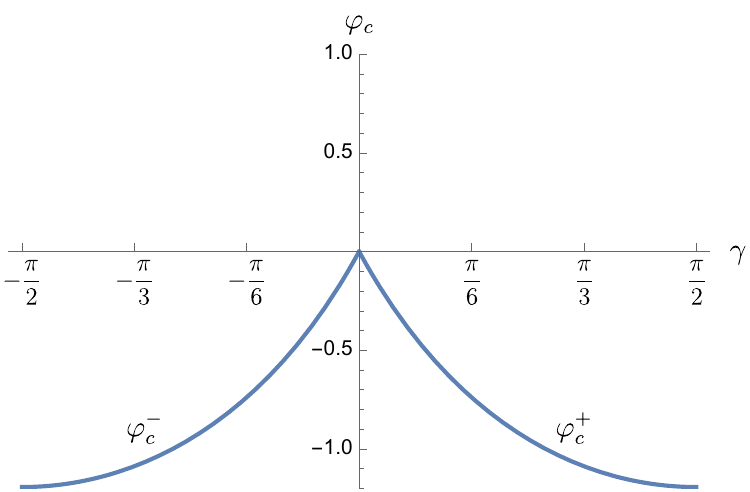}
    \end{center}
    \caption{Profile of the solution $\varphi_c^\pm$ in the Euclidean region for the case $\tau_r=0$. The profiles in the euclidean region  for the case $\tau_r>0$ are the same up to   overall constants (different for $\pm$).    }
    \label{PhicEuclidean}
\end{figure}

The spatially homogeneous solution is the same as in \nref{phic} but we now take into account finite $\tau_{r}$ effects. The first line in \nref{AcSca}   gives again the de-Sitter entropy with cosmological constant $V_r$, $ S_r = 24\pi^2 M_{pl}^4/V_r $, and the final result is
\be \la{Acco}
\log \rho_c  = i(I_+ - I_-) = S_r  + { 2 \epsilon_r M_{pl}^2 \over H_r^2 } 8 \pi^2 \left[ -   \log \cosh \tau_r - { 1 7 \over 12} +   \log 2 +{ 1 \over \cosh^2 \tau_r } \right] ~,
\ee  
where the second term in  comes from inserting \nref{phic} into the action \nref{AcSca} and integrating $\tau$ from $\pm i \pi/2$ to $\tau_r$. 
We first note that if we vary $\tau_r $ in order to search for the most probable solution, we find that the most probable solution sits at $\tau_r=0$. The solution can be viewed as a purely Euclidean sphere, obtained by setting $\tau = i \gamma $,  and a field profile equates to 
\be 
ds^2 = d\gamma^2 + \cos^2 \gamma d\Omega_3^2~,~~~~~~~~~~\varphi^\pm_c = { 1 \over 1 \pm \sin \gamma } - \log(1 \pm \sin \gamma) -1, ~~~~~~~~~{\rm for } ~~~\pm \gamma > 0\,. 
\ee 
This is illustrated in figure \ref{PhicEuclidean}.  Of course, this reproduces the well known fact that the no-boundary proposal tries to have as little inflation as possible. Since there we are putting the future boundary at $\phi_r$ it just simply says we have no inflation and we go directly  into Euclidean space. 

\subsection{Quick review of the spatially inhomogeneous case }
\la{InHomApp}

We can still decompose $\varphi_b$ into spherical harmonics, as in \nref{VarPhivExp}, and $\tilde{\varphi}^\pm$ as in \nref{homexp}. The only difference from subsection \ref{dmsubregion} is that we keep $\tau_r$ finite, which gives rise to the density matrix as follows:
\be 
\log \rho = \log \rho_c  - { 2 \epsilon_r M_{pl}^2 \over H_r^2 }  2 \pi^2 \left[    4   c_0  + \sum_{\ell, m_1,m_2 } { \ell ( \ell +1) (\ell + 2) \over 1 + {\ell (\ell +2) \over \cosh^2 \tau_r } }  |  c_{\ell, m_1,m_2}|^2  \right]\,. \la{FinRho}
\ee 
Note that by setting $\tau_r =0$ we recover the wavefunction in the form discussed in \cite{Chen:2024rpx}.
Setting large values of $\tau_r $ we are exploring the wavefunction on superhorizon scales. In fact, we will be more interested in setting $\tau_r \gg 1$, where we can drop the $\cosh \tau_r$ term in the denominator of \nref{FinRho} to obain a $\tau_r$ independent answer as we expect at superhorizon distances. 

\subsection{The density matrix for a subregion}

\label{dmsubregionfintau}

Once we solve the equations \nref{EqnsTo1} \nref{EqnsTo2}, the final value of the density matrix is given by the same expression as the density matrix evaluated on the whole spatial slice. 
In other words it is given by \nref{FinRho}. 
The only difference is that we are not directly given the values of the $c_{\ell,m_1,m_2}$, we need to find them in terms of $\omega(\vec \Omega) $ in region $\Sigma_{\rm in} $ by solving the problem \nref{EqnsTo1} \nref{EqnsTo2}.  

Let us give more details on how we can determine the solution to \nref{EqnsTo1} \nref{EqnsTo2}. 
We write the field $\varphi = \varphi_c + \tilde \varphi $ as in \nref{phiDec} and expand $\tilde \varphi$ 
  in spherical harmonics,  as in \nref{homexp},   
\begin{equation} \la{Expvftfintau}
\tilde{\varphi}^{\pm}=\sum_{\ell=0}^{\infty}   c_{\ell,m_{1},m_{2}}Y_{\ell,m_{1},m_{2}}(\Omega_{3}) \frac{f_{\ell}^{\pm}(\tau)}{f_{\ell}^{\pm}(\tau_{r})}\,, 
\end{equation}
we have used that, for the diagonal elements of the density matrix,  \nref{EqnsTo2} imply that $\varphi^+ = \
\varphi^- $ on the whole spatial slice, so that     $ c_{\ell,m_{1},m_{2}}^{\pm}$ are equal (and we can drop the $\pm$ index in the $  c$'s).
The term in the trace-out boundary condition \nref{EqnsTo2} involving the derivative becomes 
\be 
 i \cosh^3 \tau_r ( \partial_\tau \tilde \varphi^+ - \partial_\tau \tilde \varphi^+ ) = -  i \cosh^3 \tau_r ( \partial_\tau  \varphi^+_c - \partial_\tau   \varphi^+_c ) = 4\,, 
 \ee 
 after using \nref{DervFc}.
 This implies that at finite $\tau_{r}$ the problem \nref{EqnsTo2} becomes a problem for the coefficients $  c_{\ell,m_1,m_2}$ 
\begin{equation}
\begin{gathered}
\la{FinEqc}
\sum_{\ell,m_{1},m_{2}}  c_{\ell,m_{1},m_{2}}Y_{\ell,m_{1},m_{2}}(\vec \Omega)=\omega (\vec \Omega ), ~~~~~\text{ for } ~~~~~~\vec{\Omega}  \in \Sigma_{\rm in}\,,\\
\sum_{\ell,m_{1},m_{2}}^{\infty} \frac{\ell(\ell+1)(\ell+2)}{1+\frac{\ell(\ell+2)}{\cosh^{2}\tau_{r}}}c_{\ell,m_{1},m_{2}}Y_{\ell,m_{1},m_{2}}(\vec \Omega)=-2 ,~~~~~~~~\text{ for }~~~~~~~~\vec{\Omega} \in \Sigma_{\rm out} \,.
\end{gathered}
\end{equation}

The final action at finite $\tau_r $ is then given by 
\begin{equation}
\begin{gathered}
\label{OnshellactHHfintau}
\log\rho =S_{r}-\frac{8 \pi^{2}\epsilon_{0} M_{pl}^{2}}{H_{r}^{2}}\bigg(2 \log \cosh\tau_{r}+2 c_{0}+\frac{17}{6}-2\log 2+\frac{1}{2}\sum_{\ell,m_{1},m_{2}}^{\infty} \frac{\ell(\ell+1)(\ell+2)}{1+\frac{\ell(\ell+2)}{\cosh^{2}\tau_{r}}}c_{\ell,m_{1},m_{2}}^{2}\bigg)\,.
\end{gathered}
\end{equation}

 
\subsection{The density matrix when we fix the area of a surface}

\label{Finitetauarea}

At finite $\tau_{r}$, the equations 
\eqref{FinEqc} become 
\begin{equation} \la{EqnSefintau}
\sum_{\ell} c_\ell Y_{\ell}\left({ \pi \over 2} \right)=0\,,\quad  ~~~~\sum_{\ell}^{\infty} c_{\ell}\frac{\ell(\ell+1)(\ell+2)}{|f_{\ell}(\tau_{r})|^{2}}Y_{\ell}(\theta)=-2+b \delta\left(\theta-{ \pi \over 2 } \right)\,.
\end{equation}
where $b$ is a constant to be determined. Integrating \nref{EqnSe} against the other harmonics  implies 
\begin{equation} \la{clposfintau}
c_{\ell}=\frac{2 Y_{\ell}(\frac{\pi}{2})}{\ell(\ell+1)(\ell+2)}\bigg(1+\frac{\ell(\ell+2)}{\cosh^{2}\tau_{r}}\bigg) ~~~~~~{\rm for }~~~~~ \ell > 0\,.
\end{equation}
and $c_{0}$ is again determined by requiring that $\varphi=0$ at $\theta=\frac{\pi}{2}$
\begin{equation} \la{clzerofintau}
c_{0}=-\sum_{\ell=1}^{\infty}c_{\ell}Y_{\ell}\bigg(\frac{\pi}{2}\bigg)=2\log 2-\frac{3}{2}-\frac{2}{\cosh^{2}\tau_{r}}\sum_{\ell=1}^{\infty}\frac{Y_{\ell}^{2}\big(\frac{\pi}{2}\big)}{\ell+1}\,.
\end{equation}
Note that this sum is divergent for finite $\tau_r$ (though it is finite if $\tau_r\to \infty $ first). 
This is a UV divergence arises from fixing the local field profile at very short distances. 
Physically, it is reasonable to consider a problem where we only look at distances bigger than the local Hubble scale. The proper distance is related to the angular distance by $ds = H^{-1}_r \cosh\tau_r d\theta$ which means that a Hubble scale cutoff in proper distance is a cutoff on $\ell $
\be 
\label{lcutoff}
\ell_{max } < { \nu \over 2} \cosh \tau_r \,,
\ee 
where $\nu$ is an order one constant that parametrizes the cutoff in Hubble units. After adding a factor $e^{-\ell/\ell_{max} }$ in \nref{clzero} we get 
\begin{equation}
c_{0}=2\log 2-\frac{3}{2}-\frac{1}{\cosh^{2}\tau_{r}}\big(\log(\nu \cosh \tau_{r})-2\big)\,.
\end{equation}
We see that for large $\tau_r $ the second term drops out. 

For the same reason explained around \eqref{eq: fixA c0 eq}, we have
\begin{equation}
0=\int_{\Omega_{3}}\varphi(\partial_{\tau}\varphi^{+}-\partial_{\tau}\varphi^{-})=\frac{16 i}{3}\bigg(2 c_{0}+\sum_{\ell=1}^\infty c_{\ell}^{2}\frac{\ell(\ell+1)(\ell
+2)}{1+\frac{\ell(\ell+2)}{\cosh^{2}\tau_{r}}}\bigg)\,.
\end{equation}
Then,  the final action 
gives
\begin{equation}
\label{fixaactionfintau}
\log \rho = S_r -\frac{8 \pi^{2}\epsilon_{r} M_{pl}^{2}}{H_{r}^{2}}\bigg(2 \log \cosh \tau_{r}+\frac{4}{3}-\frac{\log(\nu \cosh \tau_{r})}{\cosh^{2}\tau_{r}}\bigg)\,.
\end{equation} 

Another interesting solution for the fixed area problem from \nref{FixArea} is the one with $\tau_{r}=0$, which is very different from the limit discussed in the main text. Nevertheless, the solution for $\varphi$ at any $\theta$ and $\tau$ is also quite simplified and is given as follows in spacelike separations from the strip where we fix $\varphi$
\begin{equation} \la{Soltrzerofintau}
\varphi(\theta,\tau)=-\log \nu-\frac{1}{2}\log\bigg(\frac{1-\cosh^{2}\tau \sin^{2}\theta}{4}\bigg)\,,
\end{equation}
where one has to approach the light cone while properly applying the angular momentum cutoff discussed in \nref{lcutoff} to go around the branch cut into large real $\tau$. However, note that $\varphi$ above is perfectly regular in the Euclidean section of the geometry where $\tau=i \tau_{E}$, and is different from the $\varphi_{c}$ solution \nref{phic}. This is because there we were constraining the value of the inflaton to be a constant in the entirety of the $\tau=\tau_r$ surface, where here we instead allows $\varphi$ to be anything it wants away from $\theta=\frac{\pi}{2}$. Therefore, $\varphi$ changes everywhere else in the interval to values that maximize the probability density locally. 

A point about the $\tau_{r}=0$ solution is that it has an enhanced symmetry. To see that it is convenient to define $r$ and $\alpha$ coordinates in terms of which the de Sitter metric is
\begin{equation} \la{StaPaFu}
ds^{2}=-\frac{d\varrho^{2}}{\varrho^{2}-1}+(\varrho^{2}-1)\, d\alpha^{2}+\varrho^{2} d\Omega_{2}^{2}\,,
\end{equation}
with $\varrho=\cosh \tau \sin \theta$ and $\tanh \alpha=-\frac{\tanh \tau}{\cos \theta}$. These coordinates are naturally the static patch coordinates for the right and left wedges of de Sitter. The region $\cosh \tau \sin \theta>1$ where the $\varrho$ direction is timelike is the analogue of an interior. In terms of these coordinates it is clear that \eqref{Soltrzerofintau} is a function of $\varrho$ alone, therefore the usual $SO(3)$ symmetries of the solutions are enhanced to $SO(3) \times SO(1,1)$, since shifts in $\alpha$ are the usual boost isometries in the static patch\footnote{This is clearer to see in embedding space where $\alpha$ acts as a boost}. More than that, $\varrho=1$, where we fix the field, is the bifucartion surface of the killing vector $\partial_{\alpha}$. The continuation of $\varrho$ and $\alpha$ to the Euclidean section of the geometry is done by going to $\varrho<1$ and $\alpha=i \alpha_{E}$, in terms of the metric is
\begin{equation}
ds^{2}=\frac{d\varrho^{2}}{1-\varrho^{2}}+(1-\varrho^{2})\,d\alpha_{E}^{2}+\varrho^{2}\,d\Omega_{2}^{2}\,,
\end{equation}
and the field is
\begin{equation}
\varphi(\varrho)=-\log \nu-\frac{1}{2}\log\bigg(\frac{1-\varrho^{2}}{4}\bigg)\,.
\end{equation}

\section{Numerical and analytic approaches}

\subsection{An numerical approach}
\label{Numerical}
In this appendix, we explain how to numerically find the solution for a massless scalar in de Sitter \eqref{masslesscond} or slow-roll inflation \eqref{FinEqc}, but with SO$(3)$ symmetry. As we explained, the problem reduces to a problem of finding the full profile for $\varphi$ at $\tau = \tau_r$ by solving the trace-out boundary conditions. It is convenient to reproduce those trace-out boundary conditions explicitly here
\begin{align}
& \sum_\ell c_\ell Y_\ell(\theta)=\varphi_b(\theta)\,,\quad \text{for}\,\,\, 0\leq\theta\leq\theta_0\,,\nonumber\\
&  \sum_\ell c_\ell\, \fft{\ell(\ell+1)(\ell+2)}{1+\fft{\ell(\ell+2)}{\cosh^2\tau_r}} Y_\ell(\theta)=\text{const}\,,\quad \text{for}\,\,\, \theta_0<\theta\leq \pi\,,\label{eq: trace-out condition app}
\end{align}
where $\text{const}=0$ for a massless scalar in dS and $\text{const}=-2$ for slow-roll inflation. To solve these equations we choose some integer $n_{\rm max}$ and we run the sum over $\ell $ over $n_{\rm max}$ terms, or up to $\ell = n_{\rm max} -1$. We then disretize the angles $\theta$ into $n_{\rm max}$ points. Then we have $n_{\rm max}$ equations for the same number of variables, the $c_\ell$. Solving these equations we obtain our numerical solutions.

\subsection{An analytic approach}
\label{AnalyticSolution}

In this appendix, we present an analytic approach to obtain the solutions for the trace-out boundary conditions. We start by working on a massles scalar in de Sitter,  fixing it in a slit at the late-time surface of dS, as described in section \ref{DMCap}. Nevertheless, we claim that this strategy is more general and can be applied to analytically solve the density matrix in other cases, such as the fixed curvature density matrix in section $\ref{FixCurv}$, the bubble fluctuation density matrix in section $\ref{sec: bubble}$, as well as other slicing and spacetime configurations.

We consider a massless scalar in pure dS and ignore all the backreaction \eqref{masslesssc}. We decompose the scalar into the angular momentum basis \eqref{vphiExp}. As described in section \ref{DMCap}, our interest is to compute the density matrix on a spherical ball $0 \leq \theta \leq \theta_0$ on the reheating surface $\tau_r$ by fixing $\varphi = \varphi_b(\theta)$ for $0 \leq \theta \leq \theta_0$ and gluing the bra and ket in $\theta_0 < \theta \leq \pi$. This amounts to having the boundary conditions at large $\tau_r$ \eqref{masslesscond}, which we repeat here
\begin{align}
& \sum_\ell c_\ell Y_\ell(\theta)=\varphi_b(\theta)\,,\quad \text{for}\,\, 0\leq\theta\leq\theta_0\,,\nonumber\\
&  \sum_\ell c_\ell\, \ell(\ell+1)(\ell+2) Y_\ell(\theta)=0\,,\quad \text{for}\,\, \theta_0<\theta\leq \pi\,.\label{eq: large time massless scalar in dS}
\end{align}
The strategy to analytically solve this boundary condition is to put it in the form of more standard Dirichlet-Neumann mixed boundary conditions for a single function. The trick is that we first multiply both sides by $\sin\theta$. 
The second step is to invert $\ell(\ell + 2)$, which can be implemented by inverting the differential operator $-\partial_\theta^2-1$. We then transform the boundary condition to
\begin{align}
& F(\theta)=\sin\theta\, \varphi_b(\theta)=\sum_{\ell=1} c_\ell \sin((\ell+1) \theta)\,,\quad \text{for}\,\, -\theta_0\leq\theta\leq\theta_0\,,\nonumber\\
& H(\theta)= \alpha \sin\theta +\alpha^\prime = \sum_{k=0} c_\ell (\ell+1) \sin((\ell+1)\theta)\,,\quad \text{for}\,\, \theta_0<\theta\leq 2\pi-\theta_0\,.
\end{align}
Note that this procedure extends the domain of $\theta$, so we need to carefully design the even/odd properties of the boundary condition. In particular, we extend $\varphi_b(\theta)$ so that it is even under $\theta \rightarrow -\theta$; therefore, $F(\theta)$ is odd. Additionally, we require $H$ to be odd under $\theta \rightarrow 2\pi - \theta$, setting $\alpha' \equiv 0$. Note that we still have $\alpha$ to be determined, which will become clear later as it admits a unique value to ensure the continuity of the first derivative of the solution. 
We can then introduce a variable $\sigma\leq 0$ so that we can define a two dimensional massless field $X(\sigma, \theta)$
\be
\big(\partial_\sigma^2 + \partial_\theta^2\big)X=0\rightarrow X=\sum_\ell c_\ell \sin((\ell+1)\theta) e^{(\ell+1)\sigma}\,,
\ee
with the Dirichlet-Neumann mixed boundary conditions becoming
\begin{align}
& X(\sigma,\theta)\Big|_{\sigma=0}= F(\theta) =\sum_{\ell} c_\ell  \sin((\ell+1)\theta)\,,\quad \text{for}\,\, -\theta_0\leq\theta\leq\theta_0\,,\nonumber\\
& \partial_\sigma X(\sigma,\theta)\Big|_{\sigma=0}= H(\theta)= \sum_{\ell} c_{\ell}(\ell+1) \sin((\ell+1)\theta)\,,\quad \text{for}\,\, \theta_0<\theta\leq 2\pi-\theta_0\,.
\end{align} 
It is worth noting that reconstructing the 4D massless scalar in dS from this 2D field is relatively straightforward by applying a differential operator as in \eqref{eq: fell}. This becomes evident by analytically continuing $\sigma$ to the conformal time $\sigma = -0^+ \pm i\eta$, as defined in \eqref{ConfCo}, so that we have
\be
\varphi^{\pm}(\eta,\theta)= [\cos\eta-\sin\eta \partial_\eta] \fft{X(\pm i\eta,\theta)}{\sin\theta}\,.\label{eq: construct full}
\ee

Since this field is massless and free, this implies that it embodies the 2D conformal symmetry. There then exist 2D conformal transformations such that the slit $|\theta| \leq \theta_0$ gets mapped to the real axis and the complement $\theta_0 < \theta \leq 2\pi - \theta_0$ to the imaginary axis in $\sigma = 0$
\begin{equation}
z^2= -\fft{\sin \fft{(\theta-\theta_0)-i\sigma}{2}}{ \sin \fft{(\theta+\theta_0)-i\sigma}{2}}\,,\quad z=x+i y\,.\label{eq: 2D conf map}
\end{equation}
It is worth noting that at the boundary $\sigma=0$, $z$ is either pure real or pure imaginary, depending on $\theta$, i.e., $z^2|_{\sigma=0}=- \sin (\fft{\theta-\theta_0}{2})/\sin(\fft{\theta+\theta_0}{2})$. It is obvious that $y=0$ for $|\theta|\leq \theta_0$ and $x=0$ for $\theta_0<\theta\leq 2\pi-\theta_0$. In other words, the Dirichlet boundary condition is imposed along the real axis of $z$, while the Neumann boundary condition is imposed along the imaginary axis
\begin{align}
& X(z)= F(x)\,,\quad \text{for}\,\, y=0\,,\nonumber\\
& \partial_x X(z) = \fft{H(y)}{J(y)}\,,\quad \text{for} \,\, x=0\,,\label{eq: mix boundary complex}
\end{align}
where $J(y)$ is the Jacobian coming from transforming $\partial_q$ to $\partial_x$. In our example here, we have
\begin{equation}
J(y)=\fft{1+y^4-2y^2 \cos\theta_0}{4y \sin\theta_0} \,.
\end{equation}
It is also worth noting that $F(x)$ is even in $x$ and $H(y)/J(y)$ is odd in $y$.
The art lies in finding the appropriate conformal transformation to map the regions $\Sigma_{\rm in}$ and $\Sigma_{\rm out}$ of a given Cauchy slice to the real and imaginary axes. The general solution for this type of boundary condition is
\begin{equation}
X(z,\bar{z}) = \int_{0}^{\infty} \fft{dx'}{\pi} \Big(\fft{y}{y^2+(x-x')^2}+\fft{y}{y^2+(x+x')^2}\Big)F(x') - \int_0^{\infty} \fft{dy'}{2\pi}\log\Big(\fft{(y-y')^2+x^2}{(y+y')^2+x^2}\Big) \fft{H(y')}{J(y')}\,. \label{eq: gene analytic sol}
\end{equation}
To determine $\alpha$, it is important to note that the first derivative of the solution \eqref{eq: gene analytic sol} generically develops singularity at $x=y=0$ unless $\partial_y X|_{x=0,y=0}=0$
\begin{equation} \la{ConsCo4}
\lim_{x\rightarrow 0}\partial_\sigma X\Big|_{y=0} \sim \fft{1}{x} \partial_y X\Big|_{x=0,y=0}\rightarrow \partial_y X\Big|_{x=0,y=0}\equiv 0\,.
\end{equation}
This condition guarantees that the first derivative with respect to $\theta$ at $\theta=\theta_0$ for the solution of $\varphi(\tau_r,\theta)$ is continuous.
We do not have a very solid argument for this condition. We observed that it is true numerically. In addition, it makes the field smoother at this point, while still allowing a solution. We suspect that it should arise from the condition that the solution should have a finite action. 
Then \nref{ConsCo4} leads to
\be
\int_0^\infty dx' \fft{F(x')}{x'^2}-\int_0^\infty dy' \fft{H(y')}{ J(y')y'}=0\,,\label{eq: consisency cond}
\ee
which uniquely determines $\alpha$. For \eqref{eq: large time massless scalar in dS} in the context of a massless scalar in dS, we find
\begin{equation}
\alpha=-\sec\fft{\theta_0}{2}\int_{-\theta_0}^{\theta_0} \fft{d\theta}{4\pi} \fft{x(\theta)\sin\theta}{\sin^2\fft{\theta_0-\theta}{2}}\varphi_b(\theta)\,,\label{eq: alpha dS} 
\end{equation}
where we recall
\begin{equation}
x(\theta)=\Big(\fft{\sin\fft{\theta_0-\theta}{2}}{\sin\fft{\theta_0+\theta}{2}}\Big)^{\fft{1}{2}}\,,\quad y(\theta)=\Big(\fft{\sin\fft{-\theta_0+\theta}{2}}{\sin\fft{\theta_0+\theta}{2}}\Big)^{\fft{1}{2}}\,.
\end{equation}
The boundary analytic solution is then given by 
\begin{align}
\varphi(\tau_r,\theta)= \int_{-\theta_0}^{\theta_0}\fft{d\theta'}{2\pi} \fft{y(\theta) \sin\fft{\theta+\theta_0}{2}}{x(\theta') \sin\fft{\theta-\theta'}{2}\sin\fft{\theta_0+\theta'}{2}} \fft{\sin\theta'}{\sin\theta} \varphi_b(\theta') +2\alpha\, y(\theta)\fft{\sin\fft{\theta+\theta_0}{2}}{\sin\fft{\theta}{2}}\,,\quad \theta>\theta_0\,.
\end{align}
We can insert $\alpha$ in \eqref{eq: alpha dS} back and recall $F=\varphi_b \sin\theta$, then complete integration over $(\theta_0,2\pi-\theta_0)$, generically we find the boundary value
\be
\varphi(\tau_r,\theta)=\int_{0}^{\theta_0} \fft{d\theta'}{2\pi} K(\theta,\theta') \varphi_b(\theta')\,,
\ee
where the kernel $K(\theta,\theta')$ is
\be
K(\theta,\theta')=\frac{2 \sin \frac{\theta -\theta _0}{2}  \sin ^2\frac{\theta +\theta _0}{2}  \sin \frac{\theta '}{2} \sin\theta'}{\left(\cos \theta'-\cos \theta\right)\sin \frac{\theta'-\theta_0}{2} \sin ^2\frac{\theta '+\theta _0 }{2}\sin \frac{\theta }{2}}\fft{y(\theta)}{x(\theta')}+ 2\pi\Theta(\theta_0-\theta)\delta(\theta'-\theta)\,. 
\ee
In this 2D language, computing the density matrix \eqref{AcFiFS} of a spherical ball boils down to
\be
\log \rho[\varphi_b(\theta),\varphi_b(\theta)]\sim 4\pi \int_0^{\theta_0} d\theta \varphi_b(\theta) (\partial_\theta^2+1)\partial_\sigma X(\sigma,\theta)\Big|_{\sigma=0}\,.
\ee
Using the solution \eqref{eq: gene analytic sol} with \eqref{eq: alpha dS}, we then obtain the bilinear formulation of the density matrix \eqref{eq: bilinear action} with 
\be
G(\theta,\theta')=  \fft{3\sin\theta \sin\theta' \sin\fft{\theta}{2} \sin\fft{\theta'}{2}}{(\cos\theta_0-\cos\theta)^{\fft{3}{2}}(\cos\theta_0-\cos\theta')^{\fft{3}{2}}\big(\cos\theta-\cos\theta'\big)^4}\mathcal{G}(\cos\theta,\cos\theta';\cos\theta_0)\,,\label{eq: G dS}
\ee
where
\begin{align}
&\mathcal{G}(x_1,x_2;x_0)= \left(x_1^3-\left(9 x_2+2\right) x_1^2+\left(8-3 x_2 \left(3 x_2+4\right)\right) x_1+x_2 \left(\left(x_2-2\right) x_2+8\right)+16\right) x_0^3\nonumber\\
&+3 \left(\left(2 x_2+1\right) x_1^3+\left(x_2 \left(12 x_2+7\right)-4\right) x_1^2+\left(x_2 \left(x_2 \left(2 x_2+7\right)-8\right)-8\right) x_1+x_2 \left(\left(x_2-4\right) x_2-8\right)\right) x_0^2 \nonumber\\
&+3 \left(\left(1-4 x_2 \left(2 x_2+1\right)\right) x_1^3+\left(x_2 \left(7-8 x_2 \left(x_2+1\right)\right)+2\right) x_1^2+x_2 \left(x_2 \left(7-4 x_2\right)+12\right) x_1+x_2^2 \left(x_2+2\right)\right) x_0\nonumber\\
&+x_2^3-x_1 x_2^2 \left(2 x_2+9\right)+x_1^2 x_2 \left(4 x_2 \left(2 x_2-3\right)-9\right)+x_1^3 \left(2 x_2 \left(8 x_2^2+4 x_2-1\right)+1\right)\,.\label{eq: shape G}
\end{align}
Although this function is heavy, it does demonstrate the correct short-distance behavior $G \sim (\cos\theta - \cos\theta')^{-4}$. This behavior $1/x^4$ is reminiscent of the conformal two-point function of scalar operators with the scaling dimension $\Delta = 3$, but averaged over $S^2$ (so that $\int d\Omega_2 x^{-6} \sim x^{-4}$), if interpreted in terms of the dS/CFT \cite{Strominger:2001pn,Strominger:2001gp,Maldacena:2002vr}. The shape of $\mathcal{G}$ \eqref{eq: shape G} is made complicated by tracing out the region $\theta>\theta_0$. For the full sphere $x_0=-1$, it tremendously simplifies $\mathcal{G}|_{\rm full}=16(1+x_1)^2(1+x_2)^2 (x_1 x_2-1)$.

In addition to the boundary value, we would also like to obtain the full bulk fields $\varphi(\tau,\theta)$ from the analytic 2D solution \eqref{eq: gene analytic sol}. This can be easily achieved by the analytic continuation $\sigma=-0^+\pm i\eta$ followed by using \eqref{eq: construct full}. We find it more clear to do the analytic continuation by first separating the ``Euclidean'' solution $X(\sigma,\theta)$ into holomorphic and anti-holomorphic functions
\be
X(\sigma,\theta)= i X_h(z)-i X_{\bar{h}}(\bar{z})\,.
\ee
Using \eqref{eq: gene analytic sol} and \eqref{eq: alpha dS}, we find
\be
X_h(z)= \int_{-\theta_0}^{\theta_0} \fft{d\theta'}{2\pi} \fft{1-x'^2}{x'} \Big(\fft{1}{x'^2-z^2}-\fft{1}{x'^2(1+z^2+2z \sin\fft{\theta_0}{2})}\Big)\varphi_b(\theta')\,.
\ee
The strategy is then to do the analytic continuation for $z$. We find
\be
z^{\pm}=\left(-\fft{\sin\fft{\theta \pm \eta - \theta_0 + i 0^+}{2}}{\sin\fft{\theta \pm \eta + \theta_0 + i0^+}{2}}\right)^{\fft{1}{2}}\,,\quad \bar{z}^{\pm} \left(-\fft{\sin\fft{\theta \mp \eta - \theta_0 - i0^+}{2}}{\sin\fft{\theta \mp \eta + \theta_0 - i0^+}{2}}\right)^{\fft{1}{2}}\,.
\ee
Then the holomorphic and anti-holomorphic functions are properly analytically continued, and the 4D solution is generated by \eqref{eq: construct full}
\be
& \varphi_h^{\pm}(z)=\int_{-\theta_0}^{\theta_0} \fft{d\theta'}{4\pi} \Big(\cos\eta \pm \sin\eta \fft{1+z^4+2z^2\cos\theta_0}{4z\sin\theta_0}\partial_z\Big) \fft{(1-x'^4)z^2 (z+z x'^2+2x'^2 \sin\fft{\theta_0}{2})}{x'^3 (x'^2-z^2)(1+z^2+2z\sin\fft{\theta_0}{2})} \fft{\varphi_b(\theta')}{\sin\theta}\,,\nn\\
& \varphi_{\bar{h}}^{\pm}(\bar{z})=\int_{-\theta_0}^{\theta_0} \fft{d\theta'}{4\pi} \Big(\cos\eta \mp \sin\eta \fft{1+\bar{z}^4+2\bar{z}^2\cos\theta_0}{4\bar{z}\sin\theta_0}\partial_{\bar{z}}\Big) \fft{(1-x'^4)\bar{z}^2 (\bar{z}+\bar{z} x'^2+2x'^2 \sin\fft{\theta_0}{2})}{x'^3 (x'^2-\bar{z}^2)(1+\bar{z}^2+2\bar{z}\sin\fft{\theta_0}{2})} \fft{\varphi_b(\theta')}{\sin\theta}\,,\nn\\
\ee
The full solution then reads
\be
\varphi^{\pm} =i  \varphi_h^{\pm}(z^{\pm})- i \varphi_{\bar{h}}^{\pm}(\bar{z}^{\pm})\,.
\ee
From this analysis, we can immediately see that we have $z^+=-\bar{z}^-$ and $\bar{z}^+=-z^-$ for the trace-out region $\mathcal{M}^{\rm out}$ (which is spacelike to the slit that measured by the density matrix), so that
\be
\varphi^+-\varphi^-\Big|_{\mathcal{M}^{\rm out}}= \int \fft{d\theta'}{\pi} \fft{1-x'^4}{x'^3 \sin\theta_0^3} \sin\fft{\theta_0}{2} \sinh 0^+ \times \varphi_b(\theta') = 0\,.
\ee

As we promised, such a solution is easily generalized to the slow roll inflation, where the inhomogeneous part of the inflaton modifies the trace-out boundary condition. We summarize the boundary conditions for the density matrix in slow-roll inflation scenario below
\begin{align}
& \sum_\ell c_\ell Y_\ell(\theta)=\varphi_b(\theta)\,,\quad \text{for}\,\, 0\leq\theta\leq\theta_0\,,\nonumber\\
&  \sum_\ell c_\ell\, \ell(\ell+1)(\ell+2) Y_\ell(\theta)=-2\,,\quad \text{for}\,\, \theta_0<\theta\leq \pi\,,\label{eq: large time slow roll}
\end{align}
where we take large $\tau_r$ limit. The source term $(-2)$ in the trace-out boundary condition simply modifies $H$
\be
H(\theta)= \alpha \sin\theta - (\theta-\pi)\cos\theta\,,
\ee
while the solution \eqref{eq: gene analytic sol} and the consistency condition \eqref{eq: consisency cond} remain valid. For the particular fixed curvature density matrix discussed in section \ref{FixCurv}, we equivalently have $\varphi_b=\tau(\theta)-\tau_r$. Nevertheless, we recall that the fix curvature density matrix is evaluated on the fixed curvature surface rather than the reheating surface $\tau_r$. Eventually, the spatially inhomogeneous part of the density matrix effectively reduces to a massless scalar in dS, as shown in \eqref{sumofsqract} and \eqref{eq: equiprobFixcurv}, and it thus is captured by \eqref{eq: G dS}.

We verify that the analytic solutions perfectly match the solutions obtained by numerical methods detailed in the previous subsection.

 \section{Comments on the wall motion modes}
 \label{app: wall mode}

In this appendix, we make a few comments on the wall motion modes. 
The bubble solutions break the $SO(5)$ symmetry to $SO(4)$.
This means that we  have some obvious solutions for the equations of motions for the small fluctuations. These are obtained by acting with the corresponding rotations and have the form 
\be 
 \delta \varphi_z \propto \varphi'_{\rm bu}(\chi) \cos \theta  \to \varphi'_{\rm bu}(t) \cosh \rho  ~,~~~~~~~{\rm for } ~~~ l =0 \,.
 \ee 
  There is related $l=1$ mode going like $\sinh \rho Y_{1,m}(\Omega_2) $. 
   These modes are pure gauge modes in the case with dynamical gravity. They rigidly move the whole bubble solution and move the observable region in the same way. 

The physical wall modes arise from the following observation.    
 Note that the profile $\cosh \rho $ obeys \nref{YHarmEq} with $p= 2 i$. This means that we can also consider other solutions with the same value of $p$, but higher angular momenta on the $S^2$, $l\geq 2$. The profile in the $t$ or $\chi$ direction can still be given by $\varphi'_{\rm bu}$. These are the physical wall motion modes. 
 
 More precisely, instead of the spherically symmetric solutions \nref{YHarm} to \nref{YHarmEq} we look for solutions with angular momentum and $p= 2i $
 \be \la{Eig3}
  \nabla^2_{H_3} Y_{2i, l} = 3 Y_{2 i,l} ~~~~~\to ~~~~~{ 1 \over \sinh^2 \rho } { d\over d \rho }\left[   \sinh^2 \rho { d Y_{2i, l }(\rho) \over d\rho } \right] - { l ( l+1) \over \sinh^2 \rho } Y_{2 i , l}(\rho)= 3 Y_{2 i , l}(\rho) \,.
 \ee 
 The solutions regular at the origin are \cite{Yamamoto:1996qq} 
 \be
Y_{2 i , l}(\rho) \propto {  P_{3/2}^{ - l -1/2} ( \cosh \rho ) \over \sqrt{ \sinh \rho  } } \propto  { (\sinh {\rho \over 2} )^l \over 
      ( \cosh {\rho \over 2 } )^{1 + l} }  ~_2F_1\left( - { 3 \over 2} , { 5 \over 2}    , { 3 \over 2 } + l , - \sinh^2 { \rho \over 2 }  \right)\,.
      \ee 

That the final expression for the wall modes in the Left region is 
\be 
\delta \varphi_w \sim  \varphi'_{\rm bu}(t_L) Y_{2 i , l }(\rho) {\cal Y}_{lm}(\Omega_2) \,,
\ee 
where ${\cal Y}_{l,m}(\Omega_2) $ are the usual spherical harmonics. 

 Though these represent flucutations of the scalar, they do not change the value of the scalar curvature of the surface, which is given by a formula similar to \nref{CurvSli} (with a change $6 \to -6$ due to the fact that we start with negative curvature) 
 \be 
 R^{(3) } = {e^{ - 2 \omega } \over \sinh^2 t_r } \left[- 6 - 2 (\nabla \omega)^2 - 4 \nabla^2 \omega \right] ~,~~~~~\to ~~~~~ 
 \delta R^{(3) } \propto \nabla \delta \omega - 3 \delta \omega \,,
 \ee 
 where $\delta \omega $ will be equal to $\delta \varphi$. Then we see that since the wall modes obey \nref{Eig3} we find that $\delta R^{(3)} =0$. 

 They are still physical fluctuations and they can be viewed as tensor modes \cite{Garriga:1996pg,Garcia-Bellido:1995ruf}. 
 
 Of course, they are physically describing the fluctuations of the domain wall. One comment is the following. 
 The wall motion modes, viewed as modes on   $dS_3$ are tachyonic, with a  negative mass squared equal to $-3$ in de-Sitter units. This simply follows from the equation \nref{Eig3} in the $dS_3 $ region.
 The modes are Goldstone modes in the sense of \cite{Watanabe:2013uya}, they arise from the breaking of the SO(5) symmetry to $SO(4)$ and the symmetry algebra is determining their mass. 
  
\section{Spectrum of Bubble fluctuations}
\label{SpecBub}

As explained in section \nref{dmbub},  the problem of determining the diagonal density matrix around a bubble background is obtained by analytically continuing $\varphi$ from the ket to the bra following their connection in \nref{Sheets}, and imposing the two large asymptotics value of the field $\varphi^{\pm}$ to match in $\Sigma_{\text{in}}$. In this whole section when we refer to $\varphi$, we mean its $t$ dependent part in the separation of variables \nref{Expansi}. The asymptotic solution for the field at large Lorentzian time in the left (where we fix the fields) is
\begin{equation}
\label{latebna}
\varphi^{\pm}=c_{\text{B}}^{\pm}(1+O(e^{-2 t_{r}}))+c_{\text{S}}^{\pm}e^{-3 t}(1+O(e^{-2 t_{r}}))\,.
\end{equation}
and the problem we need to solve is to find $c^{+}$ in terms of $c^{-}$.  The connection between these two sets of coefficients is just a change of basis matrix $M$. For the problem of the diagonal density matrix, note that $c_{B}^{+}=c_{B}^{-}=c_{B}$ at leading order in $t_{r}$, so this matrix is effectively fixing $c_{S}^{\pm}$ in terms of $c_{B}$. 

\begin{figure}[h!]
\begin{subfigure}[h]{0.4\linewidth}
    \centering
    \includegraphics[width=0.8\linewidth]{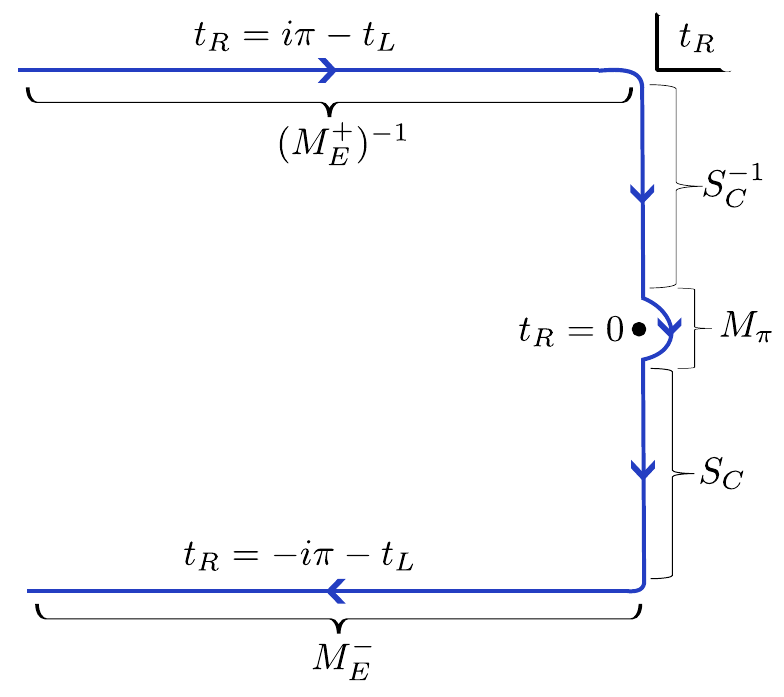}
\caption{}
\end{subfigure}
\begin{subfigure}[h]{0.55\linewidth}
    \centering
    \includegraphics[width=1.1\linewidth]{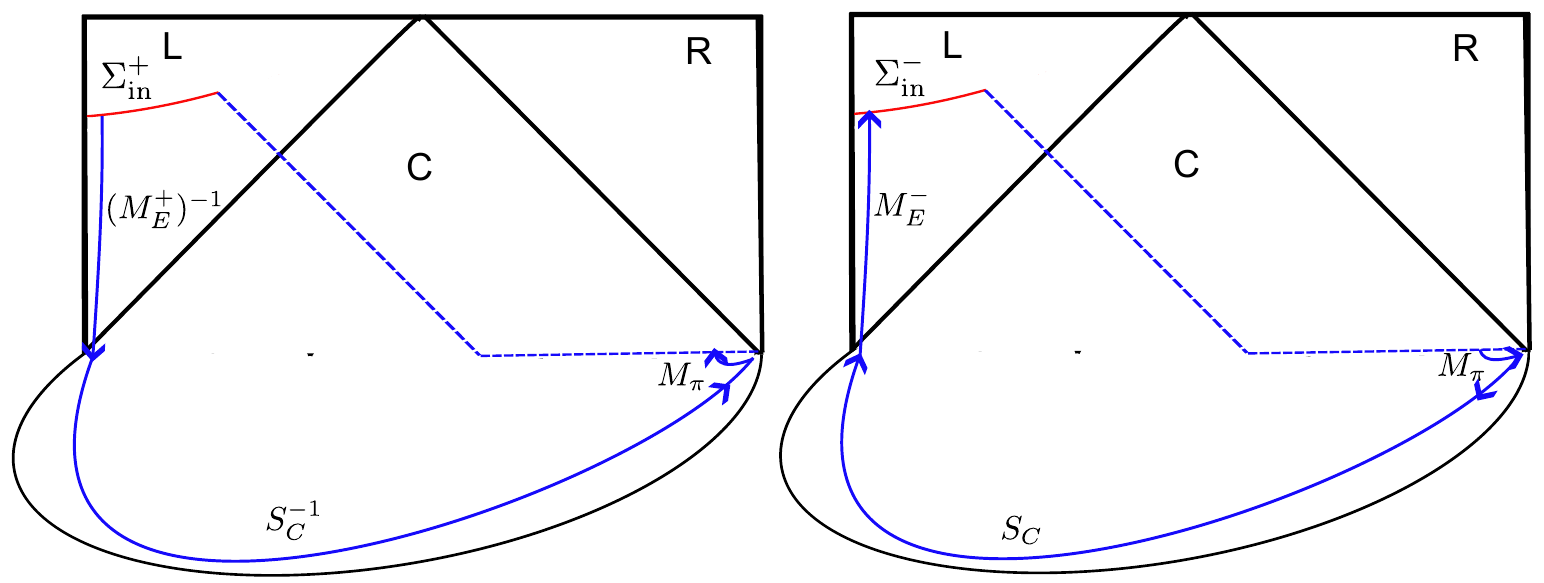}
\caption{}
\end{subfigure}
    \caption{a) Contour of the variable $t_{R}$ as one goes from large times in the ket to large times in the bra. The evolution of the function can be convenient seen as happening in the given intervals, by change of basis matrices as given. b) Representation of the contour in the Penrose diagram of the bra and ket. $M_{\pi}$ is shared between both bra and ket and is obtained by a small contour around the $t_{R}=0$ point.}
\label{contbubble}
\end{figure}

It is convenient to compute $M$ by writing it as a product of consecutive changes of basis, which corresponds to re-expressing the field in terms of simple solutions as we go through different regions in the geometry. The relevant regions are shown in figure \nref{contbubble}, as well as a specific path connecting the two asymptotics regions. The first step is to connect the field basis decomposition from small to large times via $M_{E}^{\pm}$, the $\pm$ index comes from the fact that at small times we express the fields in the basis $f_{pC}^{\dot{\pm} i p}\sim (\chi)^{\dot{\pm} i p}$, so the change of basis depends on the analytic continuation from Lorentzian time which differ in the bra and ket. The continuation is then followed through the central region $C$ via a matrix $S_{C}$, with an extra step connecting the bra and the ket sides $M_{\pi}$. This results in a change of basis matrix
\begin{equation}
M=M_{E}^{-}S_{C}M_{\pi}S_{C}^{-1}(M_{E}^{+})^{-1}\,.
\end{equation}

One of the motivations for introducing such a split is that the change of basis in the central region of the geometry can be seen as a scattering problem that is not sensitive to details of the analytic continuation \cite{Yamamoto:1996qq}. Defining $x=\log \tan \frac{\chi}{2}$, it is convenient to define the basis of solutions in $C$ according to their behaviour for very large $x$, e.g, near $\chi=0$ or $\chi=\pi$, as
\begin{equation}
f_{p C}^{\dot{\pm}} \approx 2^{-1} e^{|x|}e^{\dot{\pm}i p x}(-\text{sgn}(x) \dot{\pm} i p)\,.
\end{equation}

Starting the scaterring from $x=\infty$ and going to $x=-\infty$ one has
\begin{equation}
\varphi|_{\chi \approx \pi} \approx c_{\rm in}^{\dot{+}}f_{p C}^{\dot{+}}+c_{\rm in}^{\dot{-}}f_{p C}^{\dot{-}} \Rightarrow \varphi|_{\chi \approx 0} \approx c_{\rm out}^{\dot{+}}f_{p C}^{\dot{+}}+c_{\rm out}^{\dot{-}}f_{p C}^{\dot{-}} \,,
\end{equation}
and $c_{\rm in}$ and $c_{\rm out}$ are related via
\begin{equation}
\begin{pmatrix}
    c_{\rm out}^{\dot{+}}\\c_{\rm out}^{\dot{-}}
\end{pmatrix}=S_{C}\begin{pmatrix}
    c_{\rm in}^{\dot{+}}\\c_{\rm in}^{\dot{-}}
\end{pmatrix}
\,,\quad
S_{C}=\begin{pmatrix}
\frac{1}{T} && \frac{R}{T} \\
\frac{\bar{R}}{\bar{T}} && \frac{1}{\bar{T}}
\end{pmatrix}\,,
\end{equation}
where $R$ and $T$ are reflection coefficients one defines on the scattering problem, and conservation of Wronskian requires $|R|^{2}+|T|^{2}=1$.

Having posed the problem in this language, the final answer for $M$ should depend only on $p$ and the reflection coefficients. Defining $R=|R| e^{i \delta_{R}}$ and $T=|T|e^{i \delta_{T}}$ one can find $M$ to be
\begin{equation}
\begin{pmatrix}
c_{B}^{-}\\c_{S}^{-}\end{pmatrix}=\begin{pmatrix}
1+\frac{2 \sinh^{2}\pi p}{|T|^{2}}+\frac{2 i |R| \sinh \pi p}{|T|^{2}}\sin \delta  &&-\frac{2 i \sinh(\pi p)}{\mu_{0}|T|^{2}}(\cosh(\pi p)+|R|\cos \delta) \\
\frac{2 i \mu_{0}\sinh(\pi p)}{|T|^{2}}(\cosh(\pi p)-|R|\cos \delta) && 1+\frac{2 \sinh^{2}\pi p}{|T|^{2}}-\frac{2 i |R| \sinh \pi p}{|T|^{2}}\sin \delta
\end{pmatrix}\begin{pmatrix}c_{B}^{+}\\c_{S}^{+}\end{pmatrix}\,,
\end{equation}
with $\mu_{0}=\frac{8 p(p^{2}+1)}{3}$ and $\delta=\delta_{R}-2 \delta_{T}$. Imposing that $c_{B}^{+}=c_{B}^{-}=c_{B}$ one can the solve for the small coefficient as
\begin{equation}
c_{S}^{\pm}=\mp i \frac{\mu_{0}(\sinh \pi p \pm i |R|\sin \delta)}{\cosh(\pi p)+|R|\cos \delta}c_{B}\,,
\end{equation}
and since $\mu$ as defined in \nref{phibigsmall} only refers to the imaginary part of $\varphi$, we have
\begin{equation}
\mu=\frac{1}{1+|R|\frac{\cos \delta}{\cosh \pi p}}\,.
\end{equation}

A similar result was discussed on \cite{Yamamoto:1996qq} in a slightly different context, where the authors approximate the potential by a domain wall with a large mass for fluctuations in one of the sides, in which case $|R|=1$. 

Note than in the example of a pure massless scalar in de Sitter $S_{C}$ is the identity, that is, there is no scattering and $\mu=1$.

\subsection{Analytic Solvable potential}

Taking the potential in \nref{SimPotLin} it is straightforward to compute $\mu$, because there $v''$ is simply proportional to a delta function, namely
\begin{equation}
v''(\varphi)=-6 \delta(\varphi)=-3 \delta\big(\chi-\frac{\pi}{2}\big)\,.
\end{equation}

Therefore away from $\chi=\frac{\pi}{2}$ the perturbations follow the free massless scalar equation of motion, being exactly a combination of $f_{p}^{\dot{\pm}}$. To find the reflection coefficients we assume one of the solutions on the left, with reflection and transmission such that $\varphi$ is continuous at $\chi=\frac{\pi}{2}$ and $\varphi'$ respects a discontinuity following from the delta function in $v''$, that is, take
\begin{equation}
\begin{gathered}
\varphi=f_{p}^{\dot{\pm}}\,,\qquad  \text{for}\quad \chi<\frac{\pi}{2}\,,\\
\varphi=\frac{R}{T}f_{p}^{\dot{\mp}}+\frac{1}{T}f_{p}^{\dot{\pm}}\,,\quad \text{ for}\quad \chi>\frac{\pi}{2}\,.
\end{gathered}
\end{equation}

So one can solve for the scattering matrix to be
\begin{equation}
\begin{pmatrix}
c^{\dot{+}}\\c^{\dot{-}}
\end{pmatrix}\bigg|_{\chi=0}=\begin{pmatrix}1+\frac{3 i p}{2 (p^{2}+1)} && \frac{3 i p}{2(p^{2}+1)} \\
-\frac{3 i p}{2(p^{2}+1)} && 1-\frac{3 i p}{2(p^{2}+1)}\end{pmatrix}\begin{pmatrix}
c^{\dot{+}}\\c^{\dot{-}}
\end{pmatrix}\bigg|_{\chi=\pi}\,.
\end{equation}

In particular this implies that
\begin{equation}
\mu_{\delta} \coloneqq \mu(p)=\frac{1}{1+\frac{9 p^{2}}
{\big(4(1+p^{2})^{2}+ 9 p^{2} \big)\cosh(\pi p)}}\,.
\end{equation}

\begin{figure}[h!]
    \centering
    \includegraphics[width=0.7\linewidth]{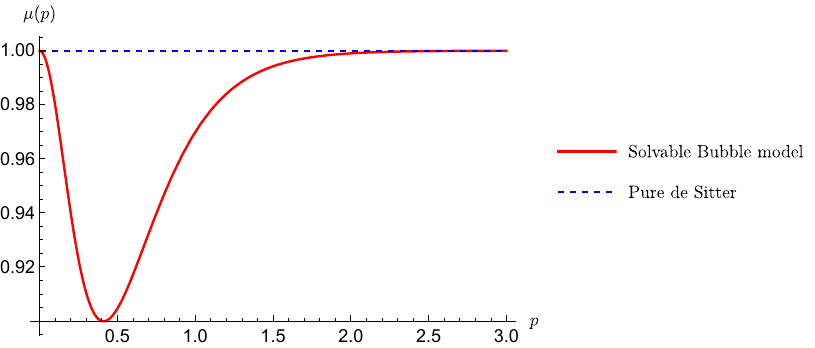}
    \caption{Spectrum of fluctuations, as in \nref{phibigsmall}, of both a massless scalar in pure de Sitter, and one in a bubble with potential given by \nref{SimPotLin}. We see that the difference is small. }
    \label{mucompplot}
\end{figure}

As one can see in figure \nref{mucompplot} the spectrum of fluctuations around this bubble is not very different from the pure de Sitter one.

\bibliographystyle{apsrev4-1long}
\bibliography{GeneralBibliography.bib}
\end{document}